\documentclass[a4paper,11pt]{article}
\usepackage{jheppub} 
\usepackage{lineno}
\usepackage{xcolor}
\usepackage{amsmath}
\usepackage{mathrsfs}
\usepackage{dsfont}
\usepackage{multirow}

\newcommand{\1}{\mathds1}
\newcommand{\Z}{{\mathbb Z}}
\newcommand{\F}{{\mathbb F}}
\newcommand{\C}{{\mathbb C}}
\newcommand{\R}{{\mathbb R}}
\newcommand{\A}{{\mathbb A}}
\newcommand{\Q}{{\mathbb Q}}
\DeclareMathOperator*{\IM}{Im}
\DeclareMathOperator*{\RE}{Re}

\DeclareMathOperator{\aut}{Aut}
\DeclareMathOperator{\spn}{span}
\DeclareMathOperator*{\diag}{diag}
\DeclareMathOperator*{\Tr}{Tr}
\DeclareMathOperator*{\irr}{Irr}

\DeclareMathOperator{\rk}{rk}
\DeclareMathOperator{\defect}{def}
\newcommand{\gapline}{\noindent \texttt{\color{gray}gap> }}

\title{Holographic duality from Howe duality: Chern--Simons gravity as an ensemble of code CFTs}

\author{Anatoly Dymarsky,$^{a}$}
\author{Johan Henriksson,$^{b,c}$ \&}
\author{Brian McPeak$^{d,e}$}
\affiliation[a]{Department of Physics, University of Kentucky, 506 Library Drive, Lexington, KY, 40506}
\affiliation[b]{Universit\'e Paris--Saclay, CEA, Institut de Physique Th\'eorique, 91191, Gif-sur-Yvette, France}
\affiliation[c]{Theoretical Physics Department, CERN, 1211, Geneva, Switzerland}
\affiliation[d]{McGill University, 3600 Rue University, Montréal, QC H3A 2T8, Canada
}
\affiliation[e]{Syracuse University, Crouse Dr, Syracuse, NY 13210}

\addtocontents{toc}{\protect\setcounter{tocdepth}{3}}

\emailAdd{a.dymarsky@uky.edu}
\emailAdd{johan.henriksson@cern.ch}
\emailAdd{bmmcpeak@syr.edu}

\abstract{
We discuss the holographic correspondence between 3d  ``Chern--Simons gravity'' and an ensemble of 2d Narain code CFTs. Starting from 3d abelian Chern--Simons theory, we construct an ensemble of boundary CFTs defined by gauging all possible maximal subgroups of the bulk one-form symmetry. Each maximal non-anomalous subgroup is isomorphic to a classical even self-dual error-correcting code over $\mathbb Z_p\times \mathbb Z_p$, providing a way to define a boundary ``code CFT.'' The average over the ensemble of such theories is holographically dual to Chern--Simons gravity, a bulk theory summed over 3d topologies sharing the same boundary. In the case of prime $p$, the sum reduces to that over handlebodies, \emph{i.e.} becomes the Poincar\'e series akin to that in semiclassical gravity. As the main result of the paper, we show that the mathematical identity underlying this holographic duality can be understood and rigorously proven using the framework of  Howe duality over finite fields. This framework is concerned with the representation theory of two commuting groups forming a dual pair: the symplectic group of modular transformations of the boundary, and an orthogonal group mapping codes to each other. Finally, we reformulate the holographic duality  as an identity between different averages over quantum stabilizer states, providing an interpretation in terms of quantum information theory.
}

\begin{document}

	\makeatletter
	\let\old@fpheader\@fpheader
	\renewcommand{\@fpheader}{  \vspace*{-0.1cm} \hfill CERN-TH-2025-060}
	\makeatother

\maketitle
\flushbottom

\section{Introduction}

The
duality between the 3d Chern--Simons theory and 2d rational conformal field theories \cite{Witten:1988hf} was discovered almost a decade before the AdS / CFT correspondence \cite{Maldacena:1997re, Witten:1998qj, Gubser:1998bc}.
The bulk Chern--Simons theory in  CS / RCFT is not gravitational, and therefore this duality is not a genuine example of holography. However, a new chapter in this story was opened by the recent observation of \cite{Afkhami-Jeddi:2020ezh, Maloney:2020nni} that after the bulk theory is coupled to topological gravity, i.e.~summed over topologies, the duality extends to an \textit{ensemble} of 2d CFTs. The boundary theories in this duality are all possible Narain CFTs.

Recently, it has been appreciated that the Narain ensemble of \cite{Afkhami-Jeddi:2020ezh, Maloney:2020nni} emerges as a limit from the duality between a finite ensemble of Narain ``code  CFTs'' and an abelian 3d 
Chern--Simons with a compact gauge group summed over bulk topologies \cite{Aharony:2023zit}. 
In this paper, we discuss this duality in detail. Code CFTs are defined by even self-dual error correcting codes of length $n$ over $\Z_k \times \Z_k$, which emerge as possible non-anomalous maximal subgroups of 1-form symmetry of the bulk $(U(1)  \times U(1))_k^n $ Chern--Simons theory \cite{Barbar:2023ncl}. We consider boundary CFTs on arbitrary Riemann surfaces of genus $g$, and show that the holographic duality that equates  ``boundary ensemble average = sum over bulk topologies'' originates from a topic within representation theory known as Howe duality, or the theta correspondence \cite{Howe1973Preprint,Howe1979,Howe1989,Howe1989a}.

For prime $k=p$, where $\Z_p=\F_p$ is a field, the bulk sum over topologies reduces to the sum over handlebodies, and can be expressed as a Poincar\'e series, which is a sum over all modular images of the vacuum character
\begin{align}
    \label{eq:abstractSW}
    \sum_{i \in \{ \text{theories} \} } Z_i(\Omega)\propto\sum_{\gamma\, \in \{ \text{modular transformations}\} } \chi_{0}(\gamma\, \Omega).
\end{align}
We will rigorously show that \eqref{eq:abstractSW} holds for all genus, and with non-zero fugacities of the $U(1)^n \times U(1)^n$ symmetry. Our proof 
does not rely on explicitly computing both sides. Instead, we will show that the RHS and LHS are both invariant under the action of two commuting groups, and, using Howe duality for finite fields, that there is a unique such invariant element. This implies that they are proportional.

We shall also discuss complementary pictures, reinterpreting them in our language. The Hilbert space of $(U(1) \times U(1))^n_p$  Chern--Simons theory on a genus-$g$ Riemann surface is isomorphic to $g$ tensor copies of $2n$ level-$p$ qudits. This allows us to give an interpretation of the duality in terms of quantum information theory. Namely, both the path integral of a code CFT at the boundary and the path integral of the bulk theory on any individual handlebody topology are quantum stabilizer states -- hence the holographic duality becomes an identity between different families of stabilizer states.  
We also explain that the ensemble of code CFTs discussed in this paper is the orbifold groupoid \cite{Gaiotto:2020iye} of a given Narain theory, providing a SymTFT interpretation of the boundary ensemble.

\subsection{Ensemble averaging in holography}

Ensemble averaging in quantum gravity is motivated by a sharp conceptual issue with wormholes. The existence of wormhole solutions \cite{Maldacena:2004rf,Arkani-Hamed:2007cpn} of supergravity that connect two asymptotically AdS regions of spacetime suggests a dual description where a conformal field theory on two disconnected manifolds has non-zero correlations, in clear contradiction of locality. This is the essence of the so-called ``factorization puzzle'' -- from the field theory point of view, correlation functions between the two boundaries should factorize. 

Coleman \cite{Coleman:1988cy, Coleman:1988tj} already pointed out in the 1980s that the possible loss of quantum coherence due to wormholes in quantum gravity could be solved by considering an ``ensemble.'' These ideas have taken a much more concrete form in the context of AdS/CFT. This was first observed for JT gravity -- a simple 2d model of gravity plus a dilaton \cite{Jackiw:1984je, Teitelboim:1983ux} -- which was shown to be dual to a random ensemble of matrices, rather than to a system with a single Hamiltonian \cite{Saad:2019lba}. In the same spirit, refs.~\cite{Afkhami-Jeddi:2020ezh, Maloney:2020nni} showed that the partition function computed by averaging over the ensemble of all Narain CFTs matches (the perturbative part of) 3d Chern--Simons theory with non-compact group, summed over all possible handlebody geometries sharing the same boundary. This sum is properly implemented by summing over all of unique images of the vacuum character under the modular group, an observation that was anticipated by \cite{Maloney:2007ud}, where it was noted that the sum over all classical gravity configurations plus their one-loop corrections is a sum over images of the Virasoro vacuum block. 
In the case of the Narain ensemble, the correct starting seed is the vacuum character of $U(1)^n \times U(1)^n$ chiral symmetry. The result, at genus 1, is 
\begin{align}
    \label{eq:PoincareSumIntro}
    Z^{\text{bulk}}(\tau) = \sum_{\gamma \in \Gamma \backslash SL(2, \Z)} \frac{1}{|\eta (\gamma \tau)|^{2n}}\,,
\end{align}
where $\eta$ is Dedekind's eta function. 
The summand with $\gamma = 1$ gives the contribution of thermal AdS. The quotient in the sum is over the subgroup $\Gamma$ which acts trivially on the seed. The RHS of \eqref{eq:PoincareSumIntro} can essentially be taken as the definition of a real-analytic Eisenstein series, leading to  
\begin{align}
\label{Eisenstein}
    Z^{\text{bulk}}(\tau) = \frac{E_{n/2}(\tau)}{\IM(\tau)^{n} |\eta(\tau)|^{2n}} \, 
\end{align}
for the bulk path integral. 

Relating this to an ensemble average requires use of the \emph{Siegel--Weil formula}, which computes the average of the lattice theta functions over all even self-dual lattices. In particular, the real analytic Eisenstein series is the average over lattices of signature $(n,n)$,
\begin{align}
    \label{eq:averNarain}
    \int_{\mathcal{M}_n} d\mu(m) \Theta(\tau, m) \ = \ E_{n /2} (\tau) \IM(\tau)^{-n} \, ,
\end{align}
where $m\in \mathcal M_n$ represent the moduli of the so-called Narain lattices -- even self-dual lattices in $\R^{n,n}$. 

In what follows, we will refer to the sum over modular images of the vacuum character as a Poincar\'e series. The vacuum character and its modular images can be interpreted as a path integral on different 3d spaces, namely solid tori, or genus-one handlebodies. Poincar\'e series have been known to arise in 3d quantum gravity for some time \cite{Dijkgraaf:2000fq,Krasnov:2000zq, Maloney:2007ud, Castro:2011zq, Keller:2014xba, Jian:2019ubz}. Following the work on averaging over Narain moduli space \cite{Maloney:2020nni, Afkhami-Jeddi:2020ezh} a number of generalizations and extensions for various  Narain ensembles have been discussed, including adding fugacities \cite{Datta:2021ftn}, considering rational points or discrete subsets in the Narain moduli space \cite{Raeymaekers:2021ypf, Meruliya:2021utr, Raeymaekers:2023ras, Aharony:2023zit, Ashwinkumar:2021kav, Ashwinkumar:2023jtz,Ashwinkumar:2023ctt}, orbifolds of the Narain moduli space \cite{Benjamin:2021wzr, Kames-King:2023fpa, Forste:2024zjt}, WZW models \cite{Dong:2021wot, Meruliya:2021lul}, and via top-down constructions in string theory \cite{Heckman:2021vzx}.

\subsection{Code CFTs}
\label{sec:codeIntro}

Classical error-correcting codes -- used to transmit messages over noisy channels -- have long been known to have rich connections to other branches of mathematics, including invariant theory \cite{Sloane1977,Nebe2006book}, sphere packings \cite{LeechSloane1971,Conway1999}, and modular forms \cite{Runge1996}. In the early 1990s, binary codes were also connected to chiral CFTs via a lattice construction \cite{Dolan:1989kf, Dolan:1994st}. Recently this topic has received renewed interest due to a number of new code-CFT constructions \cite{Gaiotto:2018ypj, Dymarsky:2020qom, Dymarsky:2020bps, Dymarsky:2021xfc,  Henriksson:2021qkt, Henriksson:2022dnu, Yahagi:2022idq, Angelinos:2022umf,  Henriksson:2022dml, Kawabata:2023nlt,  Furuta:2023xwl, Alam:2023qac, Kawabata:2023usr, Kawabata:2023iss, Kawabata:2023rlt,Ando:2024gcf,Ando:2025hwb} including codes over other fields, non-chiral CFTs, and fermionic CFTs.

An additive code ${\mathcal C}$ is a collection of elements (called codewords) in some abelian group $\mathscr D$. The simplest case is binary codes, for which $\mathscr D=\{0,1\}^n$. Via ``Construction A'' of Leech and Sloane \cite{LeechSloane1971}, a binary code can be used to define a lattice by embedding it in $(\Z/\sqrt{2})^n\subset \mathbb R^n$. When the code is doubly-even and self-dual (type II), the corresponding lattice will be even and self-dual, and can therefore be used to define a chiral CFT \cite{Dolan:1989kf, Dolan:1994st}. The torus partition function of a chiral CFT based on doubly-even self-dual code is given by
\begin{align}
\label{Zchiral}
    Z_{\mathcal C}(\tau) = \frac{ W_{\mathcal C}(\theta_3(2\tau), \theta_2(2\tau) )}{\eta(\tau)^{2n}},
\end{align}
where $ W_{\mathcal C}(x, y)=\sum_{c\in \cal C} x^{n-|c|} y^{|c|}$, is the so-called enumerator polynomial of the code $\mathcal C$, (see \cite{Dymarsky:2020qom} or \cite{Henriksson:2021qkt} for a definition and overview), and $\theta_i$ are Jacobi theta functions. Enumerator polynomials of doubly-even self-dual codes are invariant under 
\begin{align}
\label{eq:MacWilliamsIds}
    S: (x,y)\mapsto \left(\tfrac{x+y}{\sqrt2},\tfrac{x-y}{\sqrt2}\right), \qquad T: (x,y)\mapsto (x,iy)\,.
\end{align}
We refer to invariance under these transformations as the MacWilliams identity \cite{Macwilliams1963}. 
These ensure invariance under modular $S$ and $T$ transformation of the 2d CFT torus partition function \eqref{Zchiral}.\footnote{Modulo the gravitational anomaly, which gives rise to phases.} 
Although listing all individual codes  quickly becomes intractable (for type II codes, this has been accomplished for $n\leqslant 40$ \cite{Betsumiya2012}, see \cite{Harada2015} for a database), it is still possible to study statistical properties of the ensemble of all codes. Importantly, the average enumerator polynomial of all doubly-even self-dual codes was derived by Pless and Sloane in 1975 \cite{Pless1975},
\begin{equation}
\label{eq:typeIIaverage}
    \langle  W_{\mathcal C}(x,y) \rangle =\frac1{1+2^{2-n/2}}\left[
    x^n+\left(\tfrac{x+y}{\sqrt2}\right)^n+\left(\tfrac{x+iy}{\sqrt2}\right)^n+\left(\tfrac{x-y}{\sqrt2}\right)^n+\left(\tfrac{x-iy}{\sqrt2}\right)^n+y^n
    \right].
\end{equation}
Recently, two of us noted that this average is precisely of the ``Poincar\'e series'' form -- up to an overall coefficient, the average enumerator polynomial is a sum over ``modular transformations'' generated by~\eqref{eq:MacWilliamsIds} acting on $x^n$ \cite{Henriksson:2022dml}. The six terms in \eqref{eq:typeIIaverage} represent the six unique images of $x^n$ under all possible combinations of $S$ and $T$. Furthermore, \cite{Henriksson:2022dml} noted that the combinations $\theta_3(2\tau) / \eta(\tau)$ and $\theta_2(2\tau) / \eta(\tau)$ are precisely the characters of $SU(2)_1$ chiral algebra, with the seed $x^n$ corresponding to $SU(2)_1^n$ vacuum character. 
This led to a conjecture that the average over corresponding code CFTs is holographically dual to $SU(2)_1^n$ Chern--Simons theory.\footnote{The example here is for codes related to chiral theories with $(c,\bar c)=(n,0)$. A more general average  for codes relevant for $c=\bar c$ CFTs, was discussed for CFTs in \cite{Henriksson:2022dml}, see also \cite{Bannai2020,Nebe2023} in the mathematical literature} 

The equality ``ensemble average = Poincar\'e series'' for certain code-based ensembles is known to go beyond the torus partition function. This was confirmed by computing the genus 2 and genus 3 partition functions in low-$n$ cases where every code is known \cite{Henriksson:2022dml}. In the present paper, we will consider $\mathscr D=(\F_p \times \F_p)^n$ with prime $p$. These are related to lattices inside $\mathbb{R}^{n,n}$ which define Narain CFTs. We prove that the equality between code averaging and Poincar\'e series holds at all genera.

\subsection{Bulk TQFT}

In \cite{Aharony:2023zit,Barbar:2023ncl,Dymarsky:2024frx}, the ``3d Chern--Simons gravity'' -- CS theory summed over topologies -- was constructed and shown to be dual to the ensemble of CFTs based on codes over $\F_p\times \F_p$. Let us give a brief overview of this picture.

The $U(1)^n \times U(1)^n$ 3d Chern--Simons theory is a topological theory. The Hilbert space ${\mathcal H}^g$ of this theory placed on a genus-$g$ Riemann surface $\varSigma$ is finite dimensional, with the basis vectors being certain RCFT conformal blocks \cite{Witten:1988hf}. In the abelian case, there is a continuous variety of boundary conditions, called embeddings in \cite{Barbar:2023ncl}, and the corresponding blocks are not holomorphic \cite{Belov:2005ze, Aharony:2023zit}. A path integral of this theory on $\varSigma$ times an interval $[0,1]$ would calculate a transition amplitude between two states in  ${\mathcal H}^g$. When the theory admits gapped (topological) boundary conditions \cite{Kapustin:2010hk,Kaidi:2021gbs}, imposing such a b.c.~at one of the boundaries would define a state at the other boundary, the corresponding wavefunction being the  partition function of a certain 2d CFT on $\varSigma$.

In the case of abelian Chern--Simons theory with a compact gauge group, topological boundary conditions correspond to even self-dual codes $\mathcal C$ \cite{Barbar:2023ncl}. Chern--Simons boundary wavefunctions are then related to partition functions of code CFTs.
A way to construct a given boundary theory comes from  gauging a particular maximal non-anomalous subgroup of 1-form symmetry in the bulk \cite{Benini:2022hzx}, which renders the bulk theory trivial. The Hilbert space of this theory on any $\varSigma$ is one-dimensional and embedded in the Hilbert space of the original theory. The path integral of this theory defines a particular state $\Psi_\mathcal{C}$, where $\mathcal{C}$ is the code defining the gauging, and it is equal to the partition function of the code CFT: $\Psi_\mathcal{C}=Z_\mathcal{C}$. This picture suggests that the ensemble-averaged CFT partition function can be understood as a particular state in the original Hilbert space ${\mathcal H}^g$, a superposition of different states $\Psi_\mathcal{C}$ corresponding to  different maximal gaugings (topological boundary conditions), or, equivalently, codes $\mathcal{C}$. 
The construction above has a suggestive interpretation in terms of quantum information theory. The finite-dimensional Hilbert space ${\mathcal H}^g$ can be associated with the space of qudits, and each state  $\Psi_\mathcal{C}$ is a stabilizer state associated with a self-dual stabilizer code of the  Calderbank--Shor--Steane (CSS) type, associated with the classical code $\mathcal{C}$. We develop this picture in section~\ref{sec:quantumStabiliserCodes} below. 

The bulk  Chern--Simons theory coupled to topological gravity by summing over all handlebody topologies gives rise to the Poincar\'e series, \textit{i.e.}~the sum over modular images of the vacuum character.   
In case of the genus-one boundary, the vacuum character is the path integral of the CS theory placed on thermal AdS, a handlebody geometry with a particular cycle topologically trivial in the bulk. All other handlebodies are related to thermal AdS by modular transformations. 
The resulting sum is the Hartle--Hawking wavefunction of ``CS gravity.'' In fact this definition would call for a sum over \emph{all} bulk topologies, which is the prescription in the general case \cite{Dymarsky:2024frx}, but in certain cases such as those discussed in this paper, this sum simplifies to include only handlebodies, or in fact a finite sum over equivalence classes of handlebodies -- see section~\ref{sec:greduction} for a discussion.

\subsection{Howe duality}
\label{sec:introHowe}

The duality between the code ensemble and Chern--Simons gravity is  non-trivial. The aim of this paper is to show that the duality is a consequence of a rich mathematical structure known as \emph{Howe duality} or the \emph{theta correspondence}. Here we will show that a version of Howe duality implies the equality ``boundary ensemble average = bulk Poincar\'e series'' for the code ensemble, where the average is over a discrete set of boundary code CFTs and the bulk theory is Chern--Simons theory at finite level. 

The central statement of Howe duality concerns a symplectic group $\mathcal{S}=Sp(2N, F)$ over a field $F$, and ``dual pairs,'' which are a pair of subgroups of $\mathcal{S}$ which are mutually centralizing and act reductively on a space $V$. The study of dual pairs was introduced by Howe \cite{Howe1973Preprint,Howe1979,Howe1989,Howe1989a}. 
Howe duality says that a certain representation of $\mathcal S$ has a particular structure when restricted to the pair of commuting subgroups. 
In the present case we will have $F=\F_p$ -- the field with $p$ elements ($p$ prime) -- and the dual pair is formed by a group 
$O(n, n,F)$ which acts transitively on the set of codes, and the symplectic group $Sp(2g,F)$ of modular transformations/MacWilliams identities.

The precise statements of Howe duality take different forms depending on the type of field $F$ (local, global, finite). The local case has already found application in physics \cite{Howe:1987tv,Rowe:2012ym,Basile:2020gqi}, while the global case leads to the Siegel--Weil formulas, which are behind the average over the whole Narain moduli space \cite{Siegel1951} discussed in \cite{Afkhami-Jeddi:2020ezh,Maloney:2020nni}, and the average over Euclidean even unimodular lattices \cite{Siegel1935}. The theory for the case of finite fields, leading to our proof, was developed in \cite{Aubert1996}, and the specific statement that we need was only recently proven in \cite{Pan2019,Ma2022},\footnote{We are grateful to Anne-Marie Aubert for discussions and for providing us with these references.} see also \cite{Pan2019b,Pan2020}. Our proof does not require that we evaluate both sides of the duality explicitly, and as such it applies for all values of $p $, $n$ and $g$.

\paragraph{Outline of the proof.} Consider a vector space $\mathcal X$ over $\C$ whose basis vectors are labeled by $g$-tuples of ``codewords'' $c\in \F_p^{2n}$,
\begin{equation}
    \mathcal X=\mathrm{span}_{\C}\{\psi_{c_1,\ldots,c_g},c_i\in \F_p^{2n}\}.
\end{equation}
For every $g$, each code $\cal C$ has an associated vector $\sum_{c_i\in \cal C}\psi_{c_1,\ldots,c_g}$ in this space, which fully determines the genus-$g$ partition function of the corresponding code CFT. 
Individual terms in the Poincar\'e series are also vectors in this space. 
To connect with the picture above, the space $\mathcal X$ is isomorphic to the Hilbert space ${\mathcal H}^g$ of $(U(1)\times U(1))^n_{p}$ CS theory on a genus-$g$ Riemann surface, and to the Hilbert space of $2gn$ qudits.
The groups $Sp(2g,\F_p)$ and $O^+(2n,\F_p)=O(n,n,\F_p)$ both
act linearly on $ \mathcal X$. 
The group actions commute and $O^+(2n,\F_p)\times Sp(2g,\F_p)$ forms a reductive dual pair inside a larger symplectic group $Sp(4ng,\F_p)$, which also acts linearly  on $\mathcal X$. It is straightforward to show that both the ensemble average and the Poincar\'e series  are invariant under both subgroups.

The space $\mathcal X$ can be identified with the \emph{oscillator representation} $\omega$, which is the central object in Howe duality. 
Howe duality dictates the decomposition of the oscillator representation into representations of the two commuting subgroups:
\begin{equation}
\label{eq:HoweIntro}
\mathcal X\cong   \omega= \sum_{\pi\in \irr G} \pi \otimes \Theta(\pi)\,,
\end{equation}
where $\pi$ are irreps of $G=O^+(2n,\F_p)$, and $\Theta(\pi)$ are representations of $H=Sp(2g,\F_p)$. For finite fields, $\Theta(\pi)$ is not necessarily irreducible, so the theta correspondence is not one-to-one. 
Understanding the map $\Theta(\pi)$ for all representations is a subject of ongoing mathematical research, but precise statements, conjectured in \cite{Aubert1996} and recently proven in \cite{Pan2019,Ma2022}, are available for the singlet and other ``simple'' representations $\pi$. These imply that the sum \eqref{eq:HoweIntro} contains a unique term (one-dimensional vector space) that is invariant under both $G$ and $H$. Since both the ensemble average and the Poincar\'e series are invariant, they are proportional.

In appendix~\ref{sec:Plessproof} we discuss a more direct proof, for genus $g\leq 2$, using a method analogous to that of Pless and Sloane \cite{Pless1975}.
The shortcomings of this approach is that it relies on an explicit evaluation of the ensemble average and  the Poincar\'e series, and extending it to higher genus quickly becomes prohibitively complicated.

\subsection*{Outline and comments on notation}

In section~\ref{sec:ensembles and duality}, we introduce the ensemble under consideration as the set of maximal gaugings of the abelian 3d TQFT, and explain how the group theory relevant to Howe duality arises from the multiplication law of Wilson line operators in the bulk. We then show how this ensemble is the same as the ensemble of 2d Narain CFTs  parametrized by codes over $\F_p\times \F_p$. Finally, we show how the space of Chern--Simons  wavefunctions  on a Riemann surface of genus $g$ is isomorphic to Hilbert space of  $g$ copies of $2n$ level-$p$ qudits. In section~\ref{sec:proof}, we discuss Howe duality for finite fields and show that it implies that the average over  code CFTs is equal to  Poincar\'e series of the vacuum block.
In section~\ref{sec:discussion} we conclude and outline several interesting future directions.

Throughout this paper, we use $k$ as a general integer, and specify $k = p$ when it is prime (we take $p$ an odd prime); $n$  denotes the length of the code (CFT central charge $c=\bar c=n$), so that codewords are elements of ${\mathscr D}=\F_p^{2n}$, and $g$  is the genus of the Riemann surface $\varSigma$. We will use $\psi_{c_1, \ldots, c_g}$ to denote a term in the enumerator polynomial corresponding to codewords $c_1$, \ldots, $c_g$. The capital letter $\Psi_{c_1,\ldots, c_g}$ will be reserved for the ``codeword block'' (bulk TQFT wave-function) which is a function of the modular parameters (period matrix) $\Omega$ of the $\varSigma$, and fugacities $\xi, \bar \xi$ of the $U(1)^n \times U(1)^n$ symmetry.

\section{Ensemble of code-based CFTs and Chern--Simons gravity} \label{sec:ensembles and duality}

Let us now define the ensemble of 2d theories in question, and discuss it from several different perspectives.

\subsection{Bulk Chern--Simons theory}

The role of ``Chern--Simons gravity'' as the bulk dual of the code CFT ensemble has recently been worked out in detail \cite{Aharony:2023zit, Barbar:2023ncl, Dymarsky:2024frx}. We will review this connection, and introduce the necessary elements: the Heisenberg--Weyl group of Wilson lines and the symplectic group that acts as its outer automorphisms. 

\subsubsection[$(U(1) \times U(1))_k$ Chern--Simons theory on a torus]{$\boldsymbol{(U(1) \times U(1))_k}$ Chern--Simons theory on a torus}
\label{onefieldg1}
For simplicity let us start  with the level-$k$ ``$AB$ theory:''  a single pair of gauge fields $A$ and $B$ on a 3d manifold $\mathcal{M}$ whose boundary is a torus, $\varSigma=\partial \mathcal M$. Within our general story this corresponds to the case of $n =g= 1$. The 3d action is
\begin{align}
    S \ = \ \frac{k}{4 \pi} \int_\mathcal{M} (A \wedge d B + B \wedge d A)\, .
    \label{CStheory}
\end{align}
This model is known as $\Z_k$ gauge theory; for $k=2$ it is the topological quantum field theory describing the low-energy limit of the toric code \cite{Kitaev:1997wr}. In what follows we will be mainly interested in the case when $k$ is an odd prime.
The bulk theory must be invariant under large gauge transformations
\begin{align}
    A  \to A + \omega_A \, , \qquad B \to B + \omega_B \, ,
\end{align}
with $\omega_{A,B}$ being canonically-normalized cohomologies in $\mathcal M$.
For any real number $r$, we can diagonalize the action by defining
\begin{align}
    A_{\pm} = \sqrt{\frac{k}{2}} \left( \frac{A}{r} \pm B r \right) ,
\end{align}
leading to (proper quantization of this theory requires additional boundary terms)
\begin{align}
    S = \frac{1}{4 \pi} \int_\mathcal{M} \left( A_+ \wedge d A_+ - A_- \wedge d A_-\right).
\end{align}
The quantization of this theory is reviewed in \cite{Aharony:2023zit} following \cite{Bos:1989wa, Elitzur:1989nr}, see also \cite{Belov:2005ze}. The first important element is the wavefunction, defined by a path integral on the handlebody (solid torus) with possible Wilson loop insertions, and  fixed boundary conditions for the fields $A$ and $B$ at $\partial M$. For level-$k$ ``$AB$ theory'' placed on a torus, the Hilbert space ${\mathcal H}$ is a linear combination of $k^2$ wavefunctions, known as non-holomorphic blocks in \cite{Gukov:2004id,Belov:2005ze} (and codeword blocks in \cite{Aharony:2023zit}), given by 
\begin{align}
\label{psiab}
    \Psi_{\alpha,\beta}(\tau, \xi, \bar \xi) = \frac{1}{|\eta(\tau)|^2} \sum_{n, m} e^{i \pi (\tau p_L^2- \bar \tau p_R^2) + 2 \pi i (p_L \xi - p_R \bar \xi) + \frac{\pi}{2 \tau_2} (\xi^2 + \bar \xi^2)},
\end{align}
where
\begin{align}
\label{eq:pLpRgenus1}
    p_{L, R} &= \sqrt{\frac{k}{2}} \left(\frac{n + \alpha / k}r \pm (m + \beta/k) r \right) \,, 
\qquad \alpha, \beta \in \Z_k\,, 
    \qquad n, m \in \Z\,.
\end{align}
Here $r$ is the embedding parameter, controlling boundary conditions at $\partial\mathcal M$ and $\tau = \tau_1 + i \tau_2$ is the modular parameter of the boundary torus. We also introduced the fugacities $\xi,\bar \xi$, which  are related to the boundary values of the gauge fields,
\begin{align}
    \xi = \frac{\tau_2}{i \pi} (A_+)_{\bar{z}} \, , \qquad \bar \xi = \frac{\tau_2}{i \pi} (A_-)_z\,,
\end{align}
where we used complex coordinate on the torus and  $A = A_z dz + A_{\bar z} d \bar z$, in a gauge where there is no component in the direction transverse to the boundary.

\begin{figure}
    \centering
    \includegraphics[scale=0.6]{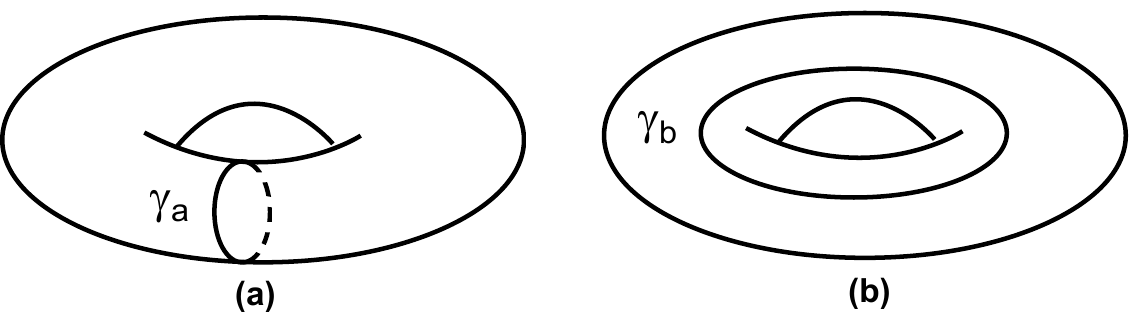}
    \caption{The two Wilson loops that wrap a genus-1 surface. The $\gamma_a$ loop in (a) is contractible in the bulk so it acts as the identity on the vacuum, while $\gamma_b$ in (b) is non-contractible so it acts non-trivially on the vacuum.
    }
    \label{fig:wilsonloops}
\end{figure}

The Chern--Simons theory is topological. The only non-trivial observables are Wilson loop operators.
When $\varSigma=\partial \mathcal{M}$ is a torus, a conventional choice of basis cycles is $\gamma_a$, $\gamma_b$, as shown in figure~\ref{fig:wilsonloops}. The Wilson loop operators wrapping the cycles on $\varSigma$ are 
\begin{align}
\label{WL}
   W[a, b] \ = \ \exp \left( i \oint_{\mathcal{\gamma}_a} (a_2\,  A + a_1\, B) +  i  \oint_{\mathcal{\gamma}_b} ( b_2\, A + b_1\, B ) +i{\pi\over k} (a_1 b_2+a_2 b_1) \right)  .
\end{align}
The pure phase factor is necessary to make $W[a, b]$ invariant under the shifts of $a,b$  mod $k$, which is confirmed by its action 
on the wavefunctions \eqref{psiab}  \cite{Aharony:2023zit},
\begin{align}
\label{WLaction}
W[a,b]\, \Psi_{(\alpha,\beta)} = e^{\frac{2\pi i  }k(\alpha\,  a_2 + \beta\, a_1) } \Psi_{(\alpha + b_1, \beta + b_2)}\,,
 \end{align} 
and the multiplication rule
\begin{align}
\label{multi}
   W[a, b] W[a', b'] =  W[a+a',b+b'] e^{-{2\pi i \over k}(a',b)},
\end{align}
where \begin{align}
    \label{eq:H5innerproduct}
 (a,b) = a_i  \eta_{ij}  b_j\,, \qquad \eta = \begin{pmatrix}
        0 & 1 \\
        1 & 0
    \end{pmatrix}.
\end{align}
From here follows the commutation relation 
\begin{align}
    \label{eq:N=2Heisenberg}
    W[a,b] W[a', b'] = e^{\frac{2 \pi i}{k} ((a,b') - (a',b))} W[a', b'] W[a, b]\,  .
\end{align}
The commutator~\eqref{eq:N=2Heisenberg} is the multiplication law for the $N = 2$ Heisenberg--Weyl group $H_5$ written in terms of different generators, see appendix \ref{app:HeisenbergReview} for details. The fact that the Wilson loops of an abelian Chern--Simons theory form the Heisenberg--Weyl group has been  already pointed out in \cite{Moore, Belov:2005ze}. This observation plays a key role in what follows.

The action of $W[a,b]$ on ${\mathcal H}$ can be derived starting from the definition of $\Psi_{0,0}:= |0\rangle$ as a state evaluated by a path integral on the handlebody with shrinkable $a$-cycle, and therefore defined via $W[a,0]|0\rangle=|0\rangle$. All other basis states can be defined using $|b\rangle:=W[0,b]|0\rangle$, implying  \eqref{WLaction} from \eqref{multi}.

\subsubsection{General case: $(U(1) \times U(1))^n_k$ theory on a genus-$g$ Riemann surface}
\label{generalCS}

We now consider the case of a CS theory with $2n$ gauge fields placed on $\mathcal M$ with a genus-$g$ boundary $\varSigma=\partial \mathcal M$. We choose the standard basis of cycles $\gamma_\kappa$, $\kappa=1,2,\dots 2g$ on $\varSigma$ with the symplectic intersection 
\begin{eqnarray}
    \gamma_\kappa \cap \gamma_{\kappa'} =\omega_{\kappa\kappa'}=\left(\begin{array}{cc}
    0 & -  I_{g\times g}\\
    I_{g\times g}& 0\end{array}\right).
\end{eqnarray}
The Chern--Simons action is 
\begin{align}
\label{CSaction}
    S \ = \ \frac{K_{ij}}{4 \pi} \int d^3x\,  A^i \wedge d A^j,
\end{align}
where $K_{ij}$ is the Gram matrix of an even lattice $\Lambda$ of signature $(n,n)$ \cite{Belov:2005ze}. 
In what follows we focus on the special case of ($ n$ copies of) level-$k$ AB theory,
\begin{eqnarray}
\label{etamatrix}
K_{ij}=k\,\eta_{ij},\qquad 
    \eta = \begin{pmatrix}
        0 & \mathbb{I}_n \\
        \mathbb{I}_n & 0
    \end{pmatrix}.
\end{eqnarray} The lattice $\Lambda$ can be chosen (using the ``rigid embedding'') to be 
\begin{eqnarray}
    \Lambda=\{\sqrt{k}\vec{v}\,|\, \vec{v}\in \Z^{2n}\}\,. 
\end{eqnarray}
However any $O(n,n,{\mathbb R})$ transformation of $\Lambda$ will give an algebraically equivalent construction  (although for a different family of Narain theories) \cite{Angelinos:2022umf,Aharony:2023zit}.

For $g=1$, the boundary conditions of the gauge fields $A^i$ are parametrized by a vector $(\xi ,\bar\xi)$ of length $2n$, which changes under large gauge transformations as follows
\begin{align}
    \begin{pmatrix}
        \xi + \bar \xi^* \\
        \xi - \bar \xi^* \\
    \end{pmatrix}  \to \begin{pmatrix}
        \xi + \bar \xi^* \\
        \xi - \bar \xi^* \\
    \end{pmatrix} + \sqrt{2} \Lambda(\vec n + \vec m \tau) \, , \quad \vec n, \vec m \in \Z^{2n} \, ,
\end{align}
where we have used $\Lambda$ here to denote  the generating matrix of the lattice $\Lambda$, and star denotes complex-conjugate.
The $g=1$ wavefunctions of this theory, parameterized by elements of the discriminant group\footnote{$\Lambda^*$ denotes the dual lattice.} $c\in {\mathscr D}=\Lambda^*/\Lambda=(\Z_k \times \Z_k)^n$ are
\begin{align}
\begin{split}
    & \Psi_c \ = \ \frac{\Theta_c}{|\eta(\tau)|^{2n}} \, , \\
    &\Theta_c(\xi, \bar \xi, \tau) \  = \ \sum_{\vec \ell \in \Z^{2n}} e^{i \pi (\tau p_L^2- \bar \tau p_R^2) + 2 \pi i (p_L \cdot \xi - p_R \cdot \bar \xi) + \frac{\pi}{2 \tau_2} (\xi^2 + \bar \xi^2)}\,, \\
    & \qquad \begin{pmatrix}
        p_L + p_R \\
        p_L - p_R
     \end{pmatrix} = \sqrt{2} \Lambda(\vec \ell + (g_\Lambda)^{-1} \vec c ) \,  , \qquad g_\Lambda = \Lambda^T \eta \Lambda \, .
\end{split}
\end{align}
From the point of view of the code CFT construction, $c$ are the possible ``codewords'' and ${\mathscr D}$ is the ``dictionary'' -- the collection of all possible words.
One of the main results of \cite{Aharony:2023zit,Barbar:2023ncl} is that these  wavefunctions in the bulk are the same as the ``code blocks,'' defined in terms of code constructions of CFTs. They can be thought of as non-holomorphic ``conformal blocks'' of the general Narain CFTs \cite{Gukov:2004id,Belov:2005ze}.

We will need the higher-genus generalization of the wavefunctions, forming a basis in the Hilbert space ${\mathcal H}^g$ of the CS theory on $\varSigma$ of genus $g$. These are defined by $g$-tuples of vectors $(c_1,  \ldots , c_g)$, $c_i\in \Z_k^{2n}$. 
From the bulk point of view these are the path integrals on a 3d handlebody geometry $\mathcal M$  ending on $\partial M=\varSigma$, with the Wilson line insertions of charges $c_i$ wrapping $g$ non-shrinkable cycles. 
The genus-$g$ handlebody is characterized by a modular parameter $\Omega$, see e.g.~\cite{Maloney:2020nni}.
Written explicitly, while taking $\xi=\bar \xi=0$, the wavefunctions are
\begin{align}
    &\Psi_{c_1 c_2  \ldots  c_g} \ = \ \frac{\Theta_{c_1 c_2  \ldots  c_g}}{\Phi(\Omega)^{n}} \, , \\
    \label{Theta}
    & \Theta_{c_1c_2  \ldots  c_g}(\Omega) \ = \ \sum_{\vec v_1,  \ldots  \vec v_g} e^{i \pi \vec{v}_I (\eta \RE (\Omega)_{IJ} + i \IM(\Omega)_{IJ}) \vec{v}_J} \, , \\
    & \qquad  \vec{v}_I = \mathcal{O} ( m_I \sqrt{k}+c_I / \sqrt{k} \, ) , \qquad m_I \in \Z^{2n} \, , \qquad  \mathcal{O} \in O(n, n, \mathbb{R}).  \label{Odef}
\end{align}
The full version including the fugacities can be found  in appendix~\ref{app:MacWilliamsHigherGenus}. 

A genus-$g$ wavefunction is a sum over $g$-tuples of vectors $\vec v_I$, with the  index $I$ running from $1,\ldots,g$. In addition, each $\vec{v}_I$ is an $2n$-dimensional vector in the lattice $\Lambda^*$. In \eqref{Theta}, the (implicit) vector indices of $\vec{v}_I$ and $\vec{v}_J$ are contracted with $\eta$ or with each other. A fixed orthogonal matrix $\mathcal{O}\in O(n,n,\mathbb{R})$ specifies the ``embedding'' of codes into the Narain moduli space. The sum in \eqref{Theta}  is performed over all integer-valued vectors $m_I$. The denominator $\Phi(\Omega)$ stands for the one-loop contribution of fluctuations of gauge fields in the bulk, generalizing $|\eta(\tau)|^2$ for $g>1$. This factor would be the same for any Chern--Simons theory with the same signature of matrix   $K_{ij}$ \cite{Belavin:1986cy,DHoker:1986eaw,Zograf1987,Zograf1987b,Sarnak1987,Krasnov:2000zq,McIntyre:2004xs,Gaberdiel:2010jf}.  

A general cycle on $\varSigma$ can be decomposed into a sum of the basis cycles $\gamma^I_a,\gamma^I_b$. A general Wilson line wrapping a cycle $\gamma \subset \varSigma$ can be written as follows
\begin{equation}
\label{WLg}
    W[a,b] = \exp \bigg( i \sum_{I,j,j'} a_{I}^j \, \eta_{jj'} \oint_{\gamma^I_a} A^{j'} dx +i \sum_{I,j,j'} b_{I}^{j}\, \eta_{jj'} \oint_{\gamma^I_b} A^{j'} dx +i{\pi \over k}\sum_{I,j,j'} a_I^j \eta_{jj'}b_I^{j'} \bigg).
\end{equation}
Here $I$ runs from $1$ to $g$ indexing cycles, and indices $j,j'$ run from $1$ to $2n$, indexing gauge fields $A^j=\{A_1,  \ldots , A_n, B_1,  \ldots , B_n\}$. As in the case of $g=n=1$, the phase factor in \eqref{WLg} is introduced to make it well defined for $a,b\in \Z_k^{2n}$.

A short calculation shows that
\begin{align}
    \label{eq:Heisenberg}
      W[a, b] W[a', b'] \ = \exp \left(-\frac{2 \pi i}{k} (a',b) \right)  W[a+a', b+b']\,, 
\end{align}
where
\begin{align}
   (a,b) =  \sum_{i,j=1}^{2n}\sum_{I=1}^g a_{I}^i \,  \eta_{ij} \, b_{I}^j \, .
\end{align}

Generalizing the $g = 1$ case of section \ref{onefieldg1},
the ``vacuum'' state $|0\rangle:= \Psi_{0,\ldots, 0}$ is the CS path integral on the handlebody with shrinkable cycles $\gamma_a^I$. It  can be defined by requiring that any Wilson loop wrapping shrinkable cycles acts trivially, 
\begin{align}
    W [a, 0] | 0 \rangle = | 0 \rangle \, .
\end{align}
The rest of the basis of the Hilbert space ${\mathcal H}^g$  is created by  Wilson lines acting on the vacuum, 
\begin{align}
    W [0, b] | 0 \rangle = |b\rangle := |b_1,\dots ,b_g\rangle:= \Psi_{ b_1,  b_2,  \ldots , b_g} \, .
\end{align}
Since the theory is topological, Wilson lines can be pulled into the bulk of $\mathcal M$.
Thus these wavefunctions \eqref{psiab}, and their higher-genus analogs \eqref{Theta}, are the CS path integrals on the handlebody  
with shrinkable cycles $\gamma_a^I$ and Wilson lines with charges $b_I$ wrapping the non-shrinkable cycles $\gamma_{b}^I$. Additivity of charge also implies $W[0,b] |c\rangle=|b+c\rangle$.
From \eqref{eq:Heisenberg}  we readily find
\begin{eqnarray}
    W[a,0] |c\rangle=e^{{2\pi i\over k}(a,c)}|c\rangle,
\end{eqnarray}
which is sufficient to specify the action of $W[a,b]$ on ${\mathcal H}^g$.

We can see that ${\mathcal H}^g$ has a structure of the tensor product of $g$ copies of $\mathcal H$, and the group of all Wilson line operators \eqref{WLg} acting on ${\mathcal H}^g$ is the Heisenberg--Weyl (HW) group $H_{2N + 1}(\F_p)$ with $2N = 4gn$. We also note that the Heisenberg--Weyl group is often called the (generalized) Pauli group in the quantum information theory literature.

The commutator of two Wilson lines, as follows from \eqref{eq:Heisenberg}, involves the alternating symplectic form $(a, b') - (a',b)$. The automorphisms of the Heisenberg--Weyl group are the transformations which preserve this form, namely,  symplectic transformations.\footnote{More accurately, the symplectic group acts as the set of outer automorphisms of the HW group, and the HW group itself (mod its center) acts as its set of inner automorphisms. The full set of automorphisms is the semidirect product $Sp(2N) \ltimes \mathrm{HW}$. Also relevant is the Clifford group, the normalizer of the generalized Pauli (HW) group inside the group of all unitarity operators acting on ${\mathcal H}^g$. It is essentially the semidirect product $Sp(2N) \ltimes \mathrm{HW}$ up to phases   \cite{Gross2006}. See also \cite{Heinrich2021} for a nice overview.} We discuss this in more detail in section \ref{sec:qudits} below.

\subsection{Maximal gaugings, code CFTs, and CS gravity}

The ensemble of boundary theories is defined by the set of all topological boundary conditions  of the Chern--Simons theory \cite{Kapustin:2010hk}, or equivalently, the set of all maximal non-anomalous subgroups of its one-form symmetry. Let us now explain in detail how this works.

\subsubsection{Maximal non-anomalous gaugings}

We start by considering the group of one-form symmetries  in Chern--Simons theory,  generated by Wilson lines. For the $U(1)$ theory at level $k$, the group of one-form symmetries is $\Z_k$, and the symmetry has a `t Hooft anomaly -- it cannot be gauged \cite{Gaiotto:2014kfa}, see also \cite{Shao:2023gho}. 

The Chern--Simons theory described in the previous section is a bit more complicated. It has a 1-form abelian symmetry group $\mathscr D = (\Z_k \times \Z_k)^{n}$. 
The Wilson lines are parametrized by charges $c=({\alpha},{\beta})\in {\mathscr D}$. We call the charge even if it satisfies $ \alpha\cdot \beta+\beta\cdot\alpha =0\ ({\rm mod}\,\, 2k)$. 
Wilson lines with even charges are bosonic, i.e.~have trivial self-braiding. 
A subgroup of charges ${\mathcal C}\subset {\mathscr D}$  has to be non-anomalous to be gauged, meaning all lines 
with $c\in {\mathcal C}$ should be bosonic and have trivial mutual braiding \cite{Gaiotto:2014kfa}. From \eqref{eq:Heisenberg} one can see that  mutual braiding of two lines with charges $c$ and $c'$ will be trivial if and only if $ \alpha\cdot \beta '+\beta\cdot\alpha'  =0\ ({\rm mod}\,\, k)$.\footnote{The non-trivial condition arises when considering the braiding between a line on an $\gamma_a$ cycle with a line on the conjugate $\gamma_b$ cycle, giving the commutator $[W[ c,0],W[0, c']]$. Then the condition $ \alpha\cdot \beta '+\beta\cdot\alpha'=0\ ({\rm mod}\,\, k)$ follows from \eqref{eq:Heisenberg}.} 

From the code perspective, the abelian subgroup  ${\mathcal C}\subset {\mathscr D}$ is an additive code, and  the condition the symmetry subgroup is non-anomalous  corresponds to the code ${\mathcal C}$ being self-orthogonal. However, this is automatically true of any even code, defined as a linear subgroup of ${\mathscr D}$ consisting of even charges, so every even code defines a non-anomalous subgroup of 1-form symmetry. When a self-orthogonal subgroup is maximal, \emph{i.e.}\ it can not be extended by adding more  codewords (charges), it becomes a Lagrangian subgroup of ${\mathscr D}$. In the language of codes, we say the code is self-dual.

Gauging a maximal non-anomalous subgroup specified  by $\mathcal C $ would define a trivial theory that evaluates the same boundary wave-function $\Psi_{\mathcal C}$ on $\varSigma$ for  any 3d manifold $\mathcal M$ with $\varSigma=\partial \mathcal M$. This is the state defined by the topological boundary conditions specified by $\mathcal C$.

To construct $\Psi_{\mathcal C}$ explicitly when $\varSigma$ is a torus, we can take $\mathcal M$ to be the handlebody with shrinkable $\gamma_a$-cycle. 
Gauging the subgroup $\mathcal C $ 
is equivalent to inserting Wilson lines with all possible charges $c \in \mathcal{C}$, and wrapping all possible cycles. Since $\gamma_a$ are shrinkable, corresponding Wilson lines have no effect on the wavefunction, $|0\rangle= W[c,0]|0\rangle$. Hence $\Psi_{\mathcal C}$ as a state in $\mathcal H$ is given by 
\begin{equation}
\label{psic}
Z_{\mathcal C}:=\Psi_{\mathcal C} = \sum_{ c\in \mathcal C}\Psi_{c} = \sum_{c\in \mathcal C}W[0,c]|0\rangle\,.
\end{equation}
It is straightforward to see that after gauging  the resulting theory is trivial: all Wilson lines with charges in $\mathcal C$ act trivially,
\begin{eqnarray}
\label{stabWL}
    W[c,c']\, \Psi_{\mathcal C} =\Psi_{\mathcal C}\,,\qquad \forall\,\, c, c'\in {\mathcal C}\,.
\end{eqnarray}
In fact, this condition can be used to define a maximal non-anomalous subgroup, and the Hilbert space of the new theory after gauging can be defined, as a subspace of $\mathcal H$, as the one-dimensional space of wavefunctions satisfying \eqref{stabWL}.

Generalizing the construction of $\Psi_{\mathcal C}$ on $\varSigma$ of a higher genus is straightforward: 
\begin{equation}
\Psi_{\mathcal C} = \sum_{ c_1,\dots, c_g\in \mathcal C}\Psi_{c_1,\dots,c_g} \in {\mathcal H}^g.
\end{equation}

\subsubsection{Ensemble of gaugings} 

A Lagrangian subgroup ${\mathcal C}\subset {\mathscr D}$ -- an even self-dual code -- defines a boundary state $\Psi_{\mathcal C}$. As we will see below, this wavefunction is the path integral of a code CFT associated with $\mathcal C$. Now, the ensemble of all even self-dual codes  of length $n$ over $\F_p\times \F_p$ is the same as the ensemble of possible maximal gaugings of one-form symmetry of the CS theory. This leads to a natural notion of the ensemble of code CFTs -- an ensemble of all possible maximal gaugings of the bulk theory.  

We would like to compare our setup with that of \cite{Benini:2022hzx}. There the logic was that a gravitational theory should have no global symmetries \cite{Banks:2010zn,Harlow:2018jwu,Harlow:2018tng,Belin:2020jxr}, hence to formulate the holographic duality, the one-form symmetry in the bulk  must be gauged. Namely one should select a particular maximal non-anomalous subgroup of one-form symmetry, and obtain  a ``gravitational'' theory upon gauging. 
As we mentioned above, the resulting bulk theory is topologically trivial. It therefore yields a factorized wavefunction when $\varSigma$ is not connected, thus solving  the factorization puzzle.
Each topologically trivial theory obtained after maximal gauging is holographically dual to a particular Narain CFT, namely code CFT defined in terms of ${\mathcal C}$. The holographic duality between a given Narain theory and a topologically trivial theory, which can be described by the $k=1$ Chern--Simons theory \eqref{CStheory}, was first pointed out in \cite{Gukov:2004id}, and discussed in detail in \cite{Aharony:2023zit}.   

From the point of  view  of the bulk theory there is no canonical choice of a maximal subgroup $\mathcal C$. In fact for a prime $k=p$ all maximal non-anomalous subgroups are mapped to each other by the action of the orthogonal group $O^+(2n, \F_p)$ -- a symmetry of Chern--Simons theory \eqref{CStheory} implemented by surface operators in the bulk. Our approach is to average over an ensemble of all possible gaugings, such that all of them are on the same footing. 

It is interesting to note that the ensemble of code CFTs defined above has another interpretation as the orbifold groupoid -- it is the ensemble of all possible orbifolds of a given Narain theory.  
The theory is specified by the choice of $\mathcal{O} \in O(n, n, \mathbb{R})$ from \eqref{Odef} and the symmetry that is ``orbifolded'' is the shift symmetry $(\Z_p \times \Z_p)^n \subset U(1)^n \times U(1)^n$. Orbifolding different subgroups leads to an ensemble of Narain theories, while the bulk Chern--Simons theory \eqref{CStheory}, which is the $\Z_p^n$ gauge theory, is the SymTFT of this global symmetry. The resulting construction is the orbifold groupoid of \cite{Gaiotto:2020iye}. 

\subsubsection{Chern--Simons gravity}

Independently from the above, we define a ``gravitational'' bulk theory by promoting bulk TQFT to ``CS gravity'' -- Chern--Simons theory summed over all possible bulk topologies sharing the same boundary.
When $k=p$ is prime, the sum reduces to classes of handlebodies; we discuss this in section \ref{sec:greduction}. 
As we show below, the CS gravity is dual to the entire ensemble of all code CFTs (possible maximal gaugings of 3d CS theory): 
\begin{equation}
\label{LargeEq}
\begin{matrix}\text{CS with global}\\\text{1-form symmetry}\end{matrix} \quad \overset{\text{maximal gaugings}}{\longrightarrow} \quad \left\{ \begin{matrix}
    Z_{{\mathcal C}_1}
    \\
    \vdots 
    \\ Z_{{\mathcal C}_{\mathcal N}}
    \end{matrix} \right.\quad 
    \overset{\text{average}}{\longrightarrow}  \quad \text{ensemble average}=\text{CS gravity}.
\end{equation}
Proving the rightmost equality, understood at the level of boundary states, with help of the Howe duality is the main result of this paper. 

Unlike the original CS theory, the CS gravity has no one-form symmetry. This is because  Wilson lines do not commute with the topology-changing Wheeler–DeWitt Hamiltonian. Furthermore it is invariant under the action of $O^+(2n,\F_p)$, meaning that corresponding global symmetry of the CS theory is gauged in CS gravity \cite{Dymarsky:2024frx}.
Thus our proposal for gravitational bulk theory  is consistent with the ``no global symmetries'' expectation of  \cite{Banks:2010zn,Harlow:2018jwu,Harlow:2018tng,Belin:2020jxr}.

\subsection{Code CFTs}
We have mentioned above that a Lagrangian subgroup ${\mathcal C}\subset {\mathscr D}$ can be understood as an even self-dual error-correcting code, while corresponding boundary wavefunction $\Psi_{\mathcal C}$ is a path integral of a code CFT on $\varSigma$. To see this in detail we briefly  review  the connection between codes over $\F_p\times \F_p$ and CFTs, developed in \cite{Dymarsky:2020qom,Angelinos:2022umf,Yahagi:2022idq,Aharony:2023zit}, building on much earlier works \cite{Dolan:1994st,Dolan:1989kf}. For simplicity  we take $k = p$ to be prime.

\subsubsection{Brief review of codes}
\label{codesdef}

An additive  error-correcting code $\mathcal C$ is a subset of vectors $c$ (codewords) inside an abelian group ${\mathscr D}=G^n$, closed under addition. The group $G$ defines an ``alphabet.'' Here we will consider even self-dual codes over $G=\F_p\times \F_p$, so that ${\mathscr D}=(\F_p\times \F_p)^n$, where $\F_p$ is the field with $p$ elements. We take $p>2$, \textit{i.e.}~it is an odd prime. A code is additive if $c,c'\in \mathcal C$ implies that $c+c'\in \mathcal C$, and a code is even if 
\begin{equation}
(c,c)=0\ (\text{mod } 2p) \label{evenness}
\end{equation}
for all $c\in \mathcal C$, where the inner product is defined by 
\begin{equation}
\label{innderproduct}
(c,c')=\alpha\cdot  \beta'+ \alpha'\cdot  \beta \ (\text{mod } p),
\end{equation} 
and $c=(\alpha,\beta)$. For $p=2$ this would be the Hermitian inner product introduced in \cite{Calderbank:1996aj}.

A code is self-dual if it is equal to its dual $\mathcal C^\perp$, defined as the set of all $c' \in {\mathscr D}$ such that $(c',c)=0$ for all $c\in \mathcal C$. So we see that the condition for a code to be even is the same as for the Wilson lines with the charges belonging to the  subgroup ${\mathcal C}\subset {\mathscr D}$ to have trivial self-braiding, and that a code is self-dual if it corresponds to a maximal non-anomalous subgroup. The number of even self-dual codes of length $n$ over $\F_p\times\F_p$ is \cite{Aharony:2023zit}
\begin{equation}
\label{eq:numcodes}
    \mathcal N_{\mathrm{codes}}=\prod_{i=0}^{n-1}(p^i+1)\,.
\end{equation}

In the theory of classical  codes \cite{MacWilliams1977,Nebe2006book}, one defines various ``enumerator polynomials'' that record partial or complete information about the code. Here we quickly recall the definitions. The \emph{full weight enumerator} at genus $g$ of a code $\mathcal C$ is defined as
\begin{equation}
\label{eq:CWEdef}
 W^{(g)}_{\mathcal C}=\sum_{c_1\in \mathcal C}\sum_{c_2\in \mathcal C}\cdots \sum_{c_g\in \mathcal C} \psi_{c_1,c_2,\ldots ,c_g}\,,
\end{equation}
where $c=(\alpha,\beta)^T$ are the codewords seen as column vectors. 
It is a vector in a complex vector space $\mathcal X$  of dimension $p^{2ng}$ with (formal) basis elements $\psi_{M}$ for all possible $2n\times g$ matrices $M$ valued in $\F_p$. Clearly, the full weight-enumerator contains full information about the code sufficient to construct the code itself. 

In coding literature, one also defines the \emph{complete weight enumerator}. For genus one, it is a polynomial in $p^2$ variables $x_{\alpha,\beta}$ defined by mapping $\psi_{(\alpha,\beta)}\mapsto \prod_i x_{\alpha_i,\beta_i}$. The map from full weight-enumerator to complete weight-enumerator loses partial  information about the code as it forgets about the order of the components of the vector $c=(\alpha,\beta)$. In a code CFT construction with rigid embedding the code-blocks $\Psi_c$ factorize to a product over $i$ with each factor depending only on the pair $(\alpha_i,\beta_i)$. Thus the complete weight-enumerator can be used to determine the CFT partition function, with vanishing fugacities, by substituting each $x_{\alpha,\beta}$ with an appropriate lattice sum \cite{Angelinos:2022umf,Yahagi:2022idq}.\footnote{Analogous statements are more straightforward  for the code CFT construction based on the codes  with the sign-definite inner product, e.g.~conventional binary codes with Hamming weight relevant for chiral CFTs \cite{Dolan:1989kf,Dolan:1994st,Henriksson:2021qkt}.  When formulas for ensemble averages are written in terms of complete weight-enumerators, the sum can be reorganized as a sum over equivalence classes of codes, the contribution from each equivalence class $[\mathcal C]$ will be weighted by $\frac{1}{|\aut[\mathcal C]|}$ where $\aut [\mathcal C]$ is the automorphism group.}

The language of enumerator polynomials is helpful, because the averaged code CFT partition function can be expressed in terms of averaged complete (or full) weight enumerators. 
In our examples in section~\ref{sec:example} and appendix~\ref{app:examples}, we will give explicit expressions for some codes and their enumerator polynomials.

\paragraph{MacWilliams identities.}
The MacWilliams identity \cite{Macwilliams1963} is a relation between the enumerator polynomial of a code and that of its dual. For self-dual code both polynomials are the same, leading to an algebraic identity  satisfied by code's enumerator polynomials. 
The standard MacWilliams identity for binary codes, see \eqref{eq:MacWilliamsIds} in the introduction, generalizes to self-dual codes over $\F_p\times \F_p$ as follows. A genus-1 full weight-enumerator of a self-dual code is  invariant under 
\begin{align}
\label{eq:MacWilliamsS}
  S: \qquad  \psi_c \mapsto \frac1{p^n}\sum_{c' \in {\mathscr D}} e^{-\frac{2\pi i}p (c,c')}\,\psi_{c'}\,.
\end{align}
There is also an identity due to evenness: the full weight-enumerator of an even code is invariant under 
\begin{align}
\label{eq:MacWilliamsT}
   T: \qquad   \psi_c \mapsto
     e^{\frac{i \pi}p (c,c)}\,\psi_c\,.
\end{align}
Historically only invariance under \eqref{eq:MacWilliamsS}, due to self-duality, is called the MacWilliams identity, but we will use the term ``MacWilliams identities'' to refer to both relations, as well as their higher-genus generalizations.
The action of $T$ and $S$ generates a representation of the modular group $SL(2,\Z)$, in fact its quotient $SL(2,\F_p)$.
Full weight-enumerators at higher genus $g$ are invariant under generalizations of \eqref{eq:MacWilliamsS}--\eqref{eq:MacWilliamsT}, which generate $Sp(2g,\F_p)$, a finite quotient of the modular group $Sp(2g,\Z)$. These are discussed in appendix~\ref{app:MacWilliamsHigherGenus}.

\paragraph{Orthogonal group.}
Codewords $c\in {\mathscr  D}$ are vectors in $\F_p^{2n}$. One can consider the group $O^+(2n,\F_p):=O(n,n,\F_p)$ acting on $\F_p^{2n}$, that leaves the inner product \eqref{innderproduct} invariant.\footnote{The are two inequivalent orthogonal groups $O^\pm(2n,\F_p)$ on $\F_p^{2n}$ that preserve two different quadratic forms, see appendix~\ref{app:groupsFiniteFields} for more details.} 
This group naturally maps codes to codes while preserving their property of being even and/or self-dual. 
It acts transitively on the set of all even self-dual codes, i.e.~maps any given code to any other one. The action on $\F_p^{2n}$ can be lifted to the vector space of full-weight enumerators $\mathcal X$,
\begin{equation}
\label{eq:actionOgroup}
   \pi({\mathsf g}):\quad \psi_{c_1,\ldots,c_g} \mapsto \psi_{\mathsf gc_1,\ldots,\mathsf gc_g}\,, \qquad \mathsf g\in O^+(2n,\F_p)\,.
\end{equation}
The group has $|O^+(2n,\F_p)|=2(p^n-1)p^{n(n-1)}(p^2-1)(p^4-1)\cdots (p^{2n-2}-1)$ elements. 

Consider a simple code ${\mathcal C}_0\ni (\alpha,0)$ with all possible $\alpha \in \F_p^n$. Its stabilizer inside $O^+_{2n}$  is given by 
$\Gamma_0=\{\left(\begin{smallmatrix}A&C\\0 &D\end{smallmatrix}\right)\}$, with all elements understood to be in $\F_p$. 
Hence the coset\footnote{See (4.46) of \cite{Aharony:2023zit} and also (4.36) of \cite{Dymarsky:2020qom} for the binary case.}
\begin{equation}
O^+_{2n}\, / \, \Gamma_0
\end{equation}
describes the space of all codes and 
has $\mathcal N_{\mathrm{codes}}$ elements. Accordingly  the set of full weight-enumerators of all codes is given by
\begin{equation}
    \{\pi({\mathsf g})\,  W_{\mathcal C_0}|\, \mathsf g\in O^+_{2n}\, / \, \Gamma_0\}.
\end{equation}
Summing $\pi({\mathsf g})\,  W_{\mathcal C_0}$ over the whole group and not just the coset simply results in a different overall coefficient. 

\subsubsection{Code CFTs}
\label{sec:codeCFTs}
Code-based 2d CFTs were originally constructed from binary error correcting codes in \cite{Dolan:1989kf,Dolan:1994st} and more recently in \cite{Yahagi:2022idq,Angelinos:2022umf,Aharony:2023zit} for the class of codes considered here. 
In short, one considers a Narain CFT \cite{Narain:1986am} constructed from a lattice of Narain type -- even self-dual lattice in $\R^{n,n}$ defined by the code via Construction A of Leech and Sloane \cite{LeechSloane1971}. 
The path integral of the resulting  CFT is specified by the code as follows. Starting from the code full weight enumerator, one promotes the formal variables $\psi_M$ for $M=(c_1,\dots,c_g)$ from \eqref{eq:CWEdef} to the codeword blocks $\Psi_M$ \eqref{Theta},\footnote{For non-zero fugacities, $\Psi_M$ is given by \eqref{Psifugacity}. We also note,  the CFT path integral is different from the CFT partition function by a universal fugacity-dependent factor, see \cite{Kraus:2006nb} and \cite{Aharony:2023zit} for details. In what follows we do not make a distinction, while always working with a path integral whenever fugacities are non-zero.}
\begin{equation}
    Z_{\mathcal C}=W_{\mathcal C}(\Psi)\,.
\end{equation}
As a result, the partition function becomes a sum over code blocks
\begin{equation}
\label{codeCFTZ}
    Z_{\mathcal C}(\Omega, \xi, \bar \xi)=\sum_{c_1\in \mathcal C}\sum_{c_2\in \mathcal C}\cdots \sum_{c_g\in \mathcal C} \Psi_{c_1,c_2,\ldots, c_g}(\Omega, \xi, \bar \xi)\,.
\end{equation}

At this point one can readily see that the space $\mathcal X$ of enumerator polynomials at genus $g$ is isomorphic to ${\mathcal H}^g$ -- the Hilbert space of CS theory on a genus-$g$ surface $\varSigma$, which is  equivalent to the tensor product of $g$ copies of ${\mathcal H}:= {\mathcal H}^1$,
\begin{equation}
    \psi_M \in {\mathcal X} \quad \leftrightarrow \quad \Psi_M \in {\mathcal H}^g.
\end{equation}
The transformations associated with the MacWilliams identities, equations \eqref{eq:MacWilliamsS} and \eqref{eq:MacWilliamsT}, for $g=1$, become the action of the generators $S,T$ of the modular group  acting on the boundary torus $\varSigma$,
\begin{eqnarray}
\label{modulargroup}
    T |c\rangle =e^{{i\pi\over p}(c,c)}|c\rangle, \quad S|c\rangle ={1\over k^n}\sum_{c' \in {\mathscr D}} e^{-{2\pi i\over p}(c,c')}|c'\rangle\,.
\end{eqnarray}
The action of the modular group on Wilson lines can be readily deduced from here,   
\begin{eqnarray}
\label{modularW}
    T\, W[a,b]\, T^\dagger = e^{-{i\pi \over p}(b,b)}W[a+b,b]\,,\qquad S\, W[a,b]\, S^\dagger =e^{{2\pi i\over p}(a,b)}W[-b,a]\,.
\end{eqnarray}The action of the modular group on ${\mathcal H}^g$ for higher $g$ are discussed in detail in the appendix~\ref{app:MacWilliamsHigherGenus}.

Finally, the orthogonal group $O^+(2n)$,  has the following interpretation  in terms of Chern--Simons theory. It is a group of global symmetries implemented by surface operators in the bulk. Its action on ${\mathcal H}^g$ is given by
\begin{eqnarray}
\label{orthogonal}
    \pi({\mathsf g}) |b_1, \ldots ,b_g\rangle &=& |{\mathsf g} b_1,\ldots, {\mathsf g} b_g\rangle,\\
    \pi({\mathsf g}) \, W[a_1,\ldots, a_g,b_1,\ldots, b_g]\, \pi({\mathsf g})^\dagger &=& W[{\mathsf g} a_1,\ldots ,{\mathsf g} a_g, {\mathsf g} b_1,\ldots, {\mathsf g} b_g]. \label{OW}
\end{eqnarray}

For even self-dual  codes over $\F_p\times \F_p$ of length $n$, the averaged full weight enumerator at genus $g=1$ takes the following form \cite{Aharony:2023zit},
\begin{equation}
    \label{eq:genus-1-average}
    \overline { W} \ = \ \frac{1}{1 + p^{1-n}} \left( \psi_{0} + p^{-n}  \sum_{c \in {\mathscr D}} \psi_{c} +  p^{-n} \sum_{c \in {\mathscr D}} e^{\frac{\pi i }{p}(c, c)} \psi_{c} +  \ldots   +  p^{-n} \sum_{c \in {\mathscr D}} e^{\frac{(p-1)i \pi  }{p}(c, c)}\psi_{c} \right) 
\end{equation}
We give a direct derivation of this formula in Appendix \ref{sec:Plessproof}. Following \cite{Aharony:2023zit} 
one can show that, up to an overall coefficient, it agrees with the Poincar\'e series of $\psi_{ 0}$, by rewriting it as follows 
\begin{equation}
    \overline { W}=\frac{1}{1 + p^{1-n}}\bigg(1 +\sum_{k=0}^{p-1} T^kS\bigg)\psi_{ 0}\,.
\end{equation} 
To recognize Poincar\'e series, one needs to take into account that $\psi_{ 0}$ is invariant under $\Gamma_0(p)\subset SL(2,\Z)$ and the coset $\Gamma_0(p)\backslash SL(2,\Z)$
consists of $p+1$ equivalence classes with the representatives $1$ and $S T^k$ for $0\leq k<p$. Note the inverse order of $T^k$ and $S$; the modular group acts on vectors in $\mathcal X$ as an anti-representation. The generalization of this result to higher $g$, even for $g=2$ is non-trivial. Accordingly, our proof below in section \ref{sec:proof} is not relying on the explicit form of $\overline{W}$.

\subsection{Qudit formulation}
\label{sec:qudits}

We have seen in section \ref{generalCS} that the Wilson line operators of the Chern--Simons theory on a genus-$g$ Riemann surface $\varSigma$  form the Heisenberg--Weyl group. We will see below that the Hilbert space ${\mathcal H}^g$ of this theory is a finite-dimensional irreducible representation of the  Heisenberg--Weyl group. This is in line with the axiomatic approach to quantum mechanics, where the Hilbert space is often defined, or constructed, as the irrep of the  Heisenberg--Weyl group. In what follows we show that ${\mathcal H}^g$ can be understood as the Hilbert space of $p$-dimensional \emph{qudits}, giving a quantum information theoretic interpretation to the model of  holographic duality discussed in this paper. 

As before, by $p$ we denote an arbitrary odd prime,\footnote{Although $p=2$ would be natural from a quantum computing perspective, many mathematical statements require more care in that case, see \cite{Gurevich2012,Gross:2021ixc,Heinrich2021}.} and consider $N$ qudits of dimension $p$.
The $N$-qudit Hilbert space ${\mathcal H}_N$ is a direct product of the $1$-qudit Hilbert spaces, which is defined in the following way. It has a basis of $p$ elements $|0\rangle,\ldots,|p-1\rangle$, upon which $\hat X$ and $\hat Z$ operators act as follows \cite{Gottesman:1998se},
\begin{equation}
\label{eq:quditOpsDef}
    \hat X|r\rangle=|r+1\rangle\,, \qquad \hat Z|r\rangle = \zeta^r|r\rangle\,,
\end{equation}
where $\zeta=e^{2\pi i/p}$ is a $p$-th root of unity. Next, we consider an ordered product of the $\hat X$ and $\hat Z$ operators, acting on  ${\mathcal H}_N$,
\begin{equation}
    \hat Z^{ a}\hat X^{ b}|r_1,\dots,r_N\rangle
    :=  
    (\hat Z^{a_1}\hat X^{b_1}|r_1\rangle)\otimes\cdots\otimes 
    (\hat Z^{a_N}\hat X^{b_N}|r_N\rangle)\,.
\end{equation}
Here $a,b\in \F_p^N$ are vectors with $N$ components.
Operators of this type, along with the phases which are $p^\mathrm{th}$-roots of unity, form the \textit{generalized Pauli group}, which is isomorphic to the Heisenberg--Weyl group over finite field $\F_p$, see \cite{Heinrich2021}.\footnote{This group is more commonly called the Pauli group in the coding/information theory literature, and the Heisenberg or Heisenberg--Weyl group in  mathematical physics literature. } 
Following \cite{Gross:2021ixc},
we define \emph{Weyl operators},
\begin{equation}
\label{Wtaudef}
    \mathsf W_{ a, b}=\tau^{-a\cdot b}\,\hat Z^{a}\hat X^{b},\qquad \tau=-e^{i\pi\over p}.
\end{equation}
We note that $\tau=\zeta^{2^{-1}}$ where $2^{-1}=(p+1)/2$ is the inverse element of $2$ in $\F_p$.
It is convenient to define {\it phase space} variables ${\rm x}  =(a, b)$ and ${\rm y}  =(a', b')$, where ${\rm x,y}\in\F_p^{2N}$, and a symplectic  inner product 
$[{\rm x},{\rm y}]:= a\cdot  b'- b\cdot  a'$. Then it is easy to check that the Weyl operators satisfy the following commutation relation defining 
 the Heisenberg--Weyl group
\begin{equation}
\label{eq:Weyl-qudits}
    \mathsf W_{\rm x}\mathsf W_{\rm y} = \tau^{[\rm x,\rm y]}\mathsf W_{\rm x+\rm y}=\zeta^{[\rm x,\rm y]}\mathsf W_{\rm y}\mathsf W_{\rm x}.
\end{equation}
Defining further $({\rm x},z):= {\mathsf W}_{\rm x}\, \zeta^z$ for $z\in \F_p$, we readily find
\begin{equation}
    ({\rm x},z) ({\rm y},z')=({\rm x+y},z+z'+2^{-1}[{\rm x},{\rm y}]),
\end{equation}
which is precisely the group law of the finite-field Heisenberg--Weyl group, c.f.~\eqref{eq:finiteHeisenberg}, 

Considering $g=1$ and $N=2n$, we find that the space of full-weight enumerators $\mathcal X$, which is isomorphic to the Hilbert space ${\mathcal H}$ of Chern--Simons theory on a torus $\varSigma$, 
is isomorphic to ${\mathcal H}_N$,
\begin{equation}
    \psi_{c}\in \mathcal X \ \quad \leftrightarrow\ \quad   \Psi_{ c}
    \in \mathcal {\mathcal H}\  \quad \leftrightarrow \quad \ |c\rangle\equiv \hat X^{ c}|0\rangle\in \mathcal H_N, 
\end{equation}
The generalization to arbitrary $g,n$ and $N=2gn$ is straightforward, 
\begin{equation}
    \Psi_{c_1,\ldots, c_g}\in \mathcal {\mathcal H}^g \ \leftrightarrow\ |c_1,\ldots ,c_g\rangle= \mathsf W_{\rm x}|0\rangle\in \mathcal H_N,\quad {\rm x}=(0,b),\quad b=(c_1,\ldots, c_g)\in \F_p^N.
\end{equation}
There is also an isomorphism between Weyl operators and Wilson lines \eqref{WL},
\begin{equation}
\mathsf W_{\rm x},\quad {\rm x}=(a,b)\quad  \leftrightarrow \quad \tau^{-a\cdot b} W[ \eta\, a, b], \label{WOWL}
\end{equation}
such that $W[a,b] \leftrightarrow X^{\eta a}Z^b$. 
Here by $\eta\, a$ we mean $\eta_{ij} a^j_I$, where $1\leq i,j\leq 2n$ and $1\leq I\leq g$.
In particular, we find that the Wilson line \eqref{WLg} with charges $a_I$ and $b_I$ and without the phase factor $\tilde{W}[a,b]:=W[a,b]e^{-{2\pi i\over p}(a,b)}$ corresponds to $(-1)^{(a,b)}{\mathsf W}_{\eta a, b}$ . This expression is not well defined if charges are understood mod $p$. At the same time, as follows from the definition \eqref{WLg},  under the modular group it transforms without any additional phases, c.f.~\eqref{modularW}
\begin{eqnarray}
\label{modularWtilde}
    T\, \tilde{W}[a,b]\, T^\dagger = \tilde{W}[a+b,b],\qquad S\, \tilde{W}[a,b]\, S^\dagger =\tilde{W}[-b,a].
\end{eqnarray}
We will see the counterpart of this statement in terms of ${\mathsf W}_{\rm x}$ shortly. 

The quantum mechanical perspective is useful for working out how the modular and orthogonal groups are embedded inside the larger symplectic group $Sp(2N)=Sp(4ng)$ of the outer automorphisms of the Heisenberg--Weyl (generalized Pauli) group. 
A subgroup of the unitary group $U(N)$ acting on ${\mathcal H}_N$ which, under the adjoint action, maps elements of the generalized Pauli group to each other, is called the Clifford group ${\bf C}_N$. (An alternative definition  of the Clifford group restricts the overall phase, see \cite{Gross2006}.) 
For any $U$ from the Clifford group we have 
\begin{eqnarray}
    U\, W_{\rm x}\, U^\dagger = e^{{2\pi i \over p} f({\rm x})}\, W_{\rm y}, \label{SpwC}
\end{eqnarray}
where necessarily ${\rm y}=h\, {\rm x}$, for some symplectic 
\begin{equation}
\label{eq:hOnWform}
   h=\begin{pmatrix}
        A&B\\C&D
    \end{pmatrix}\in Sp(2N,\F_p), 
\end{equation}
which is the same for all $\rm x$, and $f$ is some function defined mod $p$. Equation \eqref{SpwC} defines a map from the Clifford group to the space of pairs $(h,f)$.  There is a subgroup of the Clifford group which is mapped to the symplectic  group and  trivial $f({\rm x}) = 0$. The latter is called the Egorov condition in the literature \cite{Gurevich2016}. This map is surjective, 
and can be reversed, defining $U_h\in U(N)$ for each $h\in Sp(2N)$ up to an overall phase.
This defines a representation of the symplectic group $Sp(2N)$ acting on Weyl operators 
\begin{equation}
\label{eq:hOnW}
   U_h\, {\mathsf W}_{\rm x}\, U_h^\dagger = \mathsf W_{h\, \rm x}.
\end{equation}

It is easy to see that $U_h$ is a projective representation of  $Sp(2N)$ on ${\mathcal H}_N$, $U_h U_h' =c(h,h') U_{hh'}$.
In fact a stronger statement holds: one can trivialize $c(h,h')$ to define a representation  $U_h$ of $Sp(2N,\F_p)$ inside ${\bf C}_N \subset U(N)$,
namely
\begin{equation}
    U_h U_{h'} =U_{h h'}.
\end{equation} 
This  representation can be  constructed explicitly by specifying  the  generators of $Sp(2N,\F_p)$ \cite{Neuhauser2002, Gurevich2016}. 
For any matrix $A\in GL(N,\F_p)$ we define
\begin{eqnarray}
\label{SA}
S_A=\left(\begin{array}{cc}
(A^{-1})^T & 0\\
0 & A
\end{array}\right),\qquad U_A\, |b\rangle =\left(\!\tfrac{\det A}p\!\right)\,|A\, b\rangle,
\end{eqnarray}
where $b\in F_p^N$ is understood to be a vector of length $N$, and $\left(\!\tfrac{\det A}p\!\right)$ is the Legendre symbol of the determinant of $A$.\footnote{For non-zero $x\in \F_p$, the Legendre symbol $\left(\!\tfrac{x}p\!\right)$ is equal to 1 if $x$ is a square, and $-1$ if it is not. If $A$ implements a permutation $\sigma$, we have simply $\left(\!\tfrac{\det A}p\!\right)=(\mathrm{sign}\,\sigma)^{\frac{p-1}2}$ for odd primes $p$.}
Then for any $\F_p$-valued and symmetric $T$,
\begin{eqnarray}
\label{ST}
S_T=\left(\begin{array}{cc}
1_N & T\\
0 & 1_N
\end{array}\right),\qquad U_T\, |b\rangle =\tau^{b\cdot T\,b}\,|b\rangle.
\end{eqnarray}
And finally we have 
\begin{eqnarray}
\label{SJ}
S_J=\left(\begin{array}{cc}
0 & -1_N\\
1_N & 0
\end{array}\right),\qquad U_J\, |b\rangle ={i^{N{p-1\over 2}}\over p^{N/2}}\sum_{b'\in \F_p^N} \zeta^{-b\cdot b'}\,|b'\rangle.
\end{eqnarray}
While the  combinations of $S_A,S_T$ and $S_J$ generate the whole $Sp(2N)$, $U_A,U_T, U_J$ generate a faithful representation acting on ${\mathcal H}_N$.
The phase factor in the definition of $U_A$ \eqref{SA} is necessary for consistency once the phase factor in  $U_J$ \eqref{SJ} is introduced: $S_J^2$ should be the same as $S_A$ with $A=-1$. Abolishing these phase factors would result in $U_A,U_T, U_J$ generating a projective representation -- a multiple cover of $Sp(2N)$.

Since ${\mathcal H}^g$ and ${\mathcal H}_N$ with $N=2gn$ are isomorphic, we can discuss how the modular and the orthogonal groups, \eqref{modulargroup} and \eqref{orthogonal}, are embedded inside the representation of $Sp(2N)$. 
We first consider the case of $g=1$ when $N=2n$. It is straightforward to see that the modular group is embedded into $Sp(2N,\F_p)$ as follows
\begin{equation}
\label{modinsideSP}
    h=\left(\begin{array}{cc}
       \eta  &  0 \\
       0  &  1_{2n}
    \end{array}\right)\left(
    \left(\begin{array}{cc}
       a  &  b \\
       c  &  d
    \end{array}\right) \otimes 1_{2n} \right)
    \left(\begin{array}{cc}
       \eta  &  0 \\
       0  &  1_{2n}
    \end{array}\right)\in Sp(4n,\F_p),\qquad \left(\begin{array}{cc}
       a  &  b \\
       c  &  d
    \end{array}\right)\in SL(2,\F_p).
\end{equation}
We first note that the equations \eqref{SpwC} (with $h$ given by \eqref{modinsideSP}) and \eqref{modularWtilde} are compatible, both being phase-free, because $(-1)^{(b,b)}=1$ for any $b\in \F_p^{2n}$. Next, we discuss how $U_h$ is acting on ${\mathcal H}^g$.
We first consider the ``$S$'' transformation with $a=0,b=-1,c=1,d=0$. The corresponding $Sp(2N,\F_p)$ matrix can be written as follows $S_\eta S_J$
where $S_\eta$ is the transformation $S_A$ \eqref{SA} with $A=\eta\in GL(2,\F_p)$. Using \eqref{SA} and \eqref{SJ} we readily find action on the Hilbert space ${\mathcal H}_N$ to be the Fourier transform 
\begin{eqnarray}
    S_\eta S_J |b\rangle ={1\over p^n} \sum_{b'\in \F_p^2} e^{-{2\pi i\over p}(\alpha \alpha'+\beta \beta')}  |\eta b'\rangle, 
\end{eqnarray}
where $b=(\alpha,\beta)$ and $b'=(\alpha',\beta')$. Since $|\eta b\rangle=|\beta',\alpha'\rangle$ we recognize this to be the same as \eqref{modulargroup}. Notice that the phase factors in \eqref{SA} and \eqref{SJ} cancel each other. 

Similarly taking $a=1,b=1,c=0,d=1$ we find corresponding $Sp(4n,\F_p)$ transformation to be given by \eqref{ST} with $T=\eta$. Taking into account the value of $\tau=-e^{i\pi \over p}$ we find this also to match \eqref{eq:MacWilliamsT} because $(-1)^{(b,b)}=1$. This proves that $U_h$ with $h\in SL(2,\F_p)$ given by \eqref{modinsideSP} coincides with the action of the modular group on ${\mathcal H}$. The generalization to higher $g$ is straightforward; modular group $Sp(2g)$
is a subgroup of $Sp(4gn)$ with the action on ${\mathcal H}^g \cong {\mathcal H}_N$ generated by \eqref{SA}--\eqref{SJ}.

Embedding of the orthogonal group $O^+(2n,\F_p)$ inside $Sp(2N,\F_p)$ is also straightforward. For simplicity we focus on $g=1$,
\begin{eqnarray}
\label{oinsidesp}
    h_{\mathsf g}=\left(\begin{array}{cc}
       ({\mathsf g}^{-1})^T  &  0 \\
        0 & {\mathsf g}
    \end{array}\right).
\end{eqnarray}
A  quick comparison of \eqref{SA} with $A={\mathsf g}$ and 
\eqref{orthogonal} reveals they differ by a phase factor $(-1)^{\frac{p-1}2}$ (or $(-1)^{\frac{p-1}2g}$ at general genus), which is non-trivial unless $p=1$ (mod $4$). For this reason a care should be taken applying the results of Howe duality when $p=3$ (mod $4$) -- we give more details in appendix~\ref{sec:p3mod4}.

\subsubsection{Relation to quantum stabilizer codes} 
\label{sec:quantumStabiliserCodes}

The code CFT path integrals \eqref{codeCFTZ}, as well as path integrals of the Chern--Simons theory on individual handlebody geometries, in terms of qudit Hilbert space ${\mathcal H}_N$ have a natural interpretation as quantum stabilizer states. This was first pointed out in \cite{Dymarsky:2024frx},\footnote{The connection to CSS stabilizer states was independently made by A.~Barbar.} and is explained in more detail below. A relation between finite-field theta correspondence and quantum stabilizer codes was also noted  in \cite{Montealegre-Mora2021}.

A stabilizer group is an abelian subgroup $S$ of the Pauli group that does not contain any nontrivial multiple of the identity operator. 
As noted early on in \cite{Calderbank:1996aj} there is a close relation between quantum stabilizer codes and classical additive symplectic codes \cite{Calderbank:1996hm,Rains:1997uh,Gottesman:1997zz,Ketkar:2006fcv,Haah:2016ntz}. This connection is manifest in the ``phase space'' formalism, as follows. We can define the stabilizer group, as reviewed by \cite{Gross:2021ixc}, in the following way:
\begin{equation}
S=\{e^{2\pi if({\rm x})/p}\, \mathsf W_{\rm x}\,|\,{\rm x}\in {\mathscr C}\} \, .
\end{equation}
Here ${\mathscr C} \subset \F_p^{2N}$ is an abelian group which is self-orthogonal (isotropic), meaning it is contained in its dual, ${\mathscr C} \subset {\mathscr C}^\perp$. The dual here is with respect to  symplectic form,
$[\rm x,\rm y]=0$ for any $\mathrm x$ and $\mathrm y\in {\mathscr C}$, where
\begin{eqnarray}
\label{sympinner}
    [{\rm x,\rm y}]:=a\cdot b'-a'\cdot b,\qquad {\rm x}=(a,b),\quad {\rm y}=(a',b'),\quad {\rm x,y}\in \F_p^{N} \oplus \F_p^{N}.
\end{eqnarray}
The condition that ${\mathscr C}$ is isotropic is necessary for $S$ to be abelian. We also require $f(0)=0$.
Ultimately, ${\mathscr C}$ is a classical additive self-orthogonal symplectic code.

A stabilizer group $S$ defines a projector
\begin{equation}
\label{projector}
    \hat P_S=\frac1{|S|}\sum_{\mathsf W \in S} \mathsf W
\end{equation}
on the code subspace of dimension $p^N/|S|$. When the group $S$ is maximal, i.e.~when ${\mathscr C}={\mathscr C}^\perp$ is self-dual, the corresponding quantum code is one-dimensional and is conventionally referred to as the stabilizer state
\begin{equation}
   |S\rangle={\Psi_S  \over |\Psi_S|} ,\quad  \Psi_S \propto \frac1{|S|}\sum_{\rm x \in {\mathscr C}} e^{2\pi if({\rm x})/p}\, \mathsf W_{\rm x}|0\rangle.
\end{equation}
Comparing this with the genus-one code CFT path integral states $\Psi_{\mathcal C}$ \eqref{psic}  we readily find that corresponding stabilizer group is defined by 
\begin{eqnarray}
\label{CSS}
    {\mathscr C}={\mathcal C}^*\oplus {\mathcal C},\quad f(\rm x)=0,
\end{eqnarray}
where ${\mathcal C}^*=\{\eta\,c\,|\, c\in {\mathcal C}\}$ is dual to $\mathcal C$ in the sense of conventional (Euclidean) scalar product mod $p$. It is easy to see that ${\mathscr C}$ defined by \eqref{CSS} is self-dual with respect to \eqref{sympinner}, and hence \eqref{CSS} is  a Calderbank--Shor--Steane (CSS) self-dual stabilizer code  \cite{Calderbank:1995dw,Steane:1995vv}, defined in terms of the classical code $\mathcal C$. The generalization to arbitrary genus is straightforward, 
\begin{eqnarray}
\label{CSS2}
    {\mathscr C}={\mathcal C}^*\oplus \dots \oplus {\mathcal C}^*\oplus {\mathcal C} \oplus \dots \oplus {\mathcal C},\quad f(\rm x)=0,
\end{eqnarray}
where each $\mathcal C$ and ${\mathcal C}^*$ appear $g$ times. 

Remarkably, each term appearing in the Poincar\'e series is a stabilizer state as well. To see that, we start with $g=1$  and the ``vacuum'' state 
$|0\rangle\in {\mathcal H}_N$, which is the state evaluated by the Chern--Simons path integral on a solid torus with the shrinkable $\gamma_a$-cycle
$|0\rangle \in {\mathcal H}$. As was pointed out in \cite{Salton:2016qpp}, it is a stabilizer state. In our notations it has  
\begin{equation}
    {\mathscr C}=\F_p^{N}\oplus 0,\qquad f(\rm x)=0.
\end{equation}
Any modular transformation $\gamma \in SL(2,\Z)$ of this state is a stabilizer state with 
\begin{equation}
\label{bulkCSS}
    {\mathscr C}=\{(\eta a,b)\,|\, (a,b)=(\tilde{a},0) \gamma^T\, \, ({\rm mod}\, \, p),\, \, {\tilde a}\in  \F_p^{N}\},\quad  f(\rm x)=0.
\end{equation}
Each self-dual code ${\mathscr C}$ \eqref{bulkCSS} can be written in terms of a symplectic code ${\mathcal L}$ \cite{Dymarsky:2024frx},
\begin{equation}
    {\mathscr C}=(\eta \otimes 1)({\mathcal L}\otimes \F_p^{N})\subset \F_p^{N}\oplus \F_p^{N},
\end{equation}
where ${\mathcal L}\subset \F_p^2$ is an additive classical symplectic self-dual code of length $2g=2$, such that for any 
$(a,b)\in {\mathcal L}$ and $(a',b')\in {\mathcal L}$,
\begin{equation}
a b'-a'b\,\, ({\rm mod}\,\,  p)=0,
\end{equation}
and $\mathcal L$ can not be increased by adding more elements. When $p$ is prime, all such codes are the $SL(2,\F_p)$ images of ${\mathcal L}_0=(a,0)$, $a\in \F_p$. For example we can consider 
\begin{eqnarray}
    \gamma=T^k S=\left(\begin{array}{cc}
    k & -1\\
    1 & 0\end{array}\right)
\end{eqnarray} 
such that the corresponding code is ${\mathcal L}_k\ni(a k, a)$ for all $a\in \F_p$. Accordingly the corresponding isotropic self-dual code \eqref{bulkCSS} is given by 
\begin{eqnarray}
{\mathscr C}=\{(\eta ka,a)\,|\, a\in \F_p^{N}\},    
\end{eqnarray}
and  \eqref{projector} is the projector  on the state $p^{-n}\sum_{c\in \mathscr D} e^{{i\pi\over p}k(c,c)} |c\rangle $, in full agreement with \eqref{eq:genus-1-average}.

The generalization to arbitrary $g$ and $\gamma \in Sp(2g,\F_p)$ is straightforward, 
\begin{equation}
    {\mathscr C}=\{(\eta a_1,..., \eta a_g, b_1,..., b_g)\,|\, (a_1,..., a_g, b_1,..., b_g)\!=\!(\tilde{a}_1 ,..., \tilde{a}_g,0,..., 0) \gamma^T\, \, ({\rm mod}\, \, p),\, \, {\tilde a}_i\in  {\mathscr D}\}.
    \label{topology}
\end{equation}
Each such self-orthogonal ${\mathscr C}$ can be defined in terms of a classical symplectic self-dual code ${\mathcal L}$ of length $2g$, which form an orbit under $Sp(2g,\F_p)$.

Going back to the holographic identity \eqref{eq:abstractSW}, which relates the averaged boundary CFT path integral to bulk CS theory summed over different handlebody topologies, we find that it is (up to an overall coefficient) an equality between two families of stabilizer states 
\begin{eqnarray}
\label{holrelcodes}
    \sum_{S_{\mathcal C} }|S_{\mathcal C}\rangle \propto  \sum_{S_{\mathcal L}  }|S_{\mathcal L}\rangle
\end{eqnarray}
Here $S_{\mathcal C}$ is a set of CSS stabilizer codes \eqref{CSS2} describing code CFTs, while $S_{\mathcal L}$ is a family of stabilizer codes \eqref{topology} describing path integrals over different handlebody topologies. This provides a quantum information theoretic interpretation of the holographic duality \eqref{eq:abstractSW}.

The holographic relation \eqref{holrelcodes}, written explicitly as a relation  between two families of quantum stabilizer codes, and its proof with help of Howe duality in next section, resonate with the quantum information equalities of \cite{Gross:2021ixc}. It would be very interesting to place both results within the same mathematical framework.

\section{Howe duality and a proof of the averaging formula}
\label{sec:proof}

Howe duality is a relation between representations of two Lie groups $G$ and $H$ over a field $F$, embedded as commuting subgroups inside a larger symplectic group $\mathcal{S}$. In particular, this relation is a pairing or correspondence between the representations of $G$ and $H$ inside $G \times H \subset \mathcal S$, but considering different types of fields $F$ leads to different precise statements. 
This correspondence was first introduced by Howe \cite{Howe1973Preprint,Howe1979,Howe:1987tv,Howe1989,Howe1989a} based on earlier work by Siegel and Weil on modular forms \cite{Siegel1935,Siegel1951,Weil1964,Weil1965} and Segal and Shale on the quantization of the boson field \cite{Segal1959,Segal1961, Shale1962}.

The central statement of Howe duality concerns a certain representation $\omega$, the ``oscillator representation,'' of $\mathcal S$, specifically how it decomposes (branches) under the pair of commuting subgroups. This decomposition takes the form
\begin{equation}
\label{eq:HoweSec3}
    \omega|_{G\times H}=\sum_i \pi_i \otimes r_i
\end{equation}
namely a pairing of representations $\pi_i$ of $G$ and $r_i$ of $H$, both labeled by a joint index $i$. In section~\ref{sec:HoweDualityGeneral} we will give a general statement and in section~\ref{sec:finiteFieldAndProof} we will discuss the finite-field case relevant for our proof. This is followed by an explicit example in section~\ref{sec:example}. 
Some introductory (yet quite challenging) references for the general presentation are \cite{Prasad1993,Prasad1996,KudlaCastle,Li2021}. 

Howe duality is similar to Schur--Weyl duality, which, at least at the level of examples, may be more familiar to the reader. Consider the tensor product of $g$ copies of a vector space $V$, which we may take to be $\C^n$. On this space the symmetric group $S_g$ acts by permuting the copies, and the matrix group $GL(V)=GL(n,\C)$ acts within each copy. These groups commute, and Schur--Weyl duality is a statement about the  decomposition
\begin{equation}
\label{eq:Sch-W}
    \C^n\otimes \ldots\otimes \C^n = \sum_{\lambda} \pi_\lambda \otimes \rho_\lambda\,.
\end{equation}
Here $\pi_\lambda$ are representations of $S_g$, and $\rho_\lambda$ are representations of $GL(n,\C)$. They are both labeled by Young tableaux with $g$ boxes. For example, in the case of just two factors, we have $ V\otimes V=S^2V+ \Lambda^2V$, where $S^2V$ is the symmetric part (upon which $S_2$ acts trivially), and $\Lambda^2V$ is the antisymmetric part (upon which $S_2$ acts by a sign). Here $\pi_\lambda$ are the two one-dimensional irreps of $S_2$ and $\rho_\lambda$ are $n(n\pm 1)/2$ dimensional irreducible representations of $GL(n,\C)$.\footnote{Schur--Weyl duality has no direct relation to averages discussed in this work. However, see \cite{Maxfield:2023mdj} for a setting where Schur--Weyl duality is central to characterizing the Hilbert space in a toy model of quantum gravity.}

\subsection{Generalities}
\label{sec:HoweDualityGeneral}

Howe duality applies when $\mathcal S$ acts on a space $W$ and when its subgroups $G$ and $H$ form a \textit{dual pair}, which requires that they be 1)  reductive groups and 2) mutually centralizing inside $\mathcal S$. A group $G$ acts reductively on $W$ if it breaks $W$ into a finite sum of $G$-invariant subspaces. The centralizer of a subgroup $G \subset \mathcal S$ is the set of elements in $\mathcal S$ which commute with all $g\in G$. Mutually centralizing means that $G$ is the centralizer of $H$ and \emph{vice versa}.

Howe gave a classification of reductive dual pairs \cite{Howe1979,Howe:1987tv,Howe1989a}:
\begin{align}
\label{eq:ListHowedualities}
    O\times Sp\subset Sp,\qquad  
    U\times U \subset Sp,\qquad     GL\times GL\subset Sp.
\end{align}
The relevant case for our purposes will be the pair $G\times H=O(2n,F)\times Sp(2g,F)\subset \mathcal S=Sp(4gn,F)$, and the field in question will be $F=\F_p$ for prime $p$. However, in this subsection we keep our discussion general and defer the specialization to $\F_p$ to section~\ref{sec:finiteFieldAndProof}.

A central role is played by a special projective representation of $\mathcal S$ called the \emph{oscillator representation} $\omega$, also known as the Weil representation or the metaplectic representation. The name ``oscillator representation'' was chosen by Howe \cite{Howe1973Preprint} in direct analogy with physics, where the basis of this representation is a Fock space created by powers of  $N$ different creation operators $a^\dagger_i$ acting on a vacuum state, 
\begin{equation}
\label{eq:OmegaAsFockSpace}
    \omega=\mathrm{span}_\C\left\{\big(a_1^\dagger\big)^{k_1}\big(a_2^\dagger\big)^{k_2}\cdots \big(a_N^\dagger\big)^{k_N}|0\rangle \right\}.
\end{equation}
When the groups are defined over $\R$, each creation operator can act an arbitrary number of times, rendering $\omega$ infinite-dimensional. For a finite field $\F_p$, there is a direct relation between the Fock space picture and the various formulations described in the previous section: 
\begin{equation}
\label{eq:equivalence-ahat-X}
    a_i^\dagger\, \leftrightarrow\, W[0,\mathbf e_i]\,\leftrightarrow \,\mathsf W_{0,\mathbf e_i} \, \leftrightarrow\, \hat X_i
\end{equation}
where $\mathbf e_i$ is a $2n\times g$ matrix with zeros except a single $1$ at a position determined by $1\leq i\leq 2gn$ (say row by row), and $\hat X_i$ is the operator $\hat X$ of \eqref{eq:quditOpsDef} acting on the $i^{\text{th}}$ qudit. In the case of $F=\F_p$, each ``creation operator'' can only act $p$ times, and $\omega$ has dimension $p^{2ng}$. The oscillator representation $\omega$ is then isomorphic as a vector space to $\mathcal X$ introduced above.

The oscillator representation $\omega$ is an irreducible representation of the Heisenberg--Weyl group $H_{2ng}$. It also transforms projectively under the group of (outer) automorphisms of $H_{2ng}$, which is $Sp(4ng, F)$  -- these are the canonical transformations of quantum mechanics. In the case $F=\R$, $\omega$ is a projective representation of $Sp(4ng, F)$ but a linear representation of the double cover $\widetilde{Sp}$ of the symplectic group, which is called the metaplectic group, and the pair $G\times H$ lifts to a pair of subgroups $\tilde G\times \tilde H$ of $\widetilde{Sp}$.\footnote{For finite fields, it is possible to write down a genuine (non-projective) representation by including suitable phases, but this is impossible for $\R$. For the field $\C$, the oscillator representation is a genuine representation of $Sp(4gn,\C)$ \cite{Howe1979}.} A review of the details of this construction can be found in appendix~\ref{app:HeisenbergReview}, where we give an introduction to the oscillator representation for both $\R$ and $\F_p$.

When restricting to $\tilde G\times \tilde H$, the oscillator representation $\omega$ ceases to be irreducible, and Howe duality makes precise statements about the decomposition of $\omega$ under the pair of commuting subgroups
\begin{equation}
\label{eq:HoweGeneral}
    \omega|_{\tilde G\times\tilde H}=\sum_i \pi_i(\tilde G)\otimes \rho_i(\tilde H).
\end{equation}
Equation~\eqref{eq:HoweGeneral} says that $\omega$ decomposes as a sum of representations $\pi_i$ of $\tilde G$ and $\rho_i$ of $\tilde H$, indexed by a joint label $i$. In case of  $F=\R$, both $\pi_i$ and $\rho_i$ are irreducible representations and the decomposition is multiplicity-free. Over finite fields, we do not have such a bijection between irreps, but it is still possible to make precise statements about this decomposition, as we discuss below. 

\subsection{Finite-field Howe duality}
\label{sec:finiteFieldAndProof}

We now focus exclusively on the case of $G \times H =O^+(2n,\F_p)\times Sp(2g,\F_p)$ inside $\mathcal S=Sp(4gn,\F_p)$. The groups $G$ and $H$ act on the space  $\mathcal X$ of the code variables $\psi_{c_1,\ldots, c_g}$ as described in section \ref{sec:codeCFTs}.
The group actions commute -- morally, 
$G$ acts on the left and $H$ on the right on the  $2n\times g$ matrix of labels $(c_1,\dots,c_g)$.\footnote{This argument doesn't extend to all generators of $H=Sp$. To check that the whole group $H$ commutes with $G$ (which acts by \eqref{eq:actionOgroup}) is straightforward using the generators given in Appendix~\ref{app:MacWilliamsHigherGenus}.}
These groups form a reductive dual pair inside  $Sp(2g,\F_p)$, and since the space $\mathcal X$ is isomorphic to the oscillator representation, we can study representations of  $G\times H$ on $\mathcal X$ using finite-field Howe duality.\footnote{There is an ambiguity with sign for $p=3$ (mod $4$) and odd $g$, also encountered in section~\ref{sec:qudits} above. We discuss this in appendix~\ref{sec:p3mod4}. 
We also remark that in the case $n=1$, $p=3$, the groups $O^+\times Sp$ do not form a reductive dual pair, as these groups are not mutually centralizing inside $Sp(4g,\F_3)$, see p.~15 of \cite{Moeglin1987}. However explicit computations confirm that this case conforms to the general pattern (at least for $g=1$, $g=2$), in particular it has a single vector (up to rescaling) invariant under both groups. 
}

In finite-field Howe duality, the statement \eqref{eq:HoweGeneral}  takes the following form: 
\begin{equation}
\label{eq:HoweFinite}
    \omega|_{G\times H}=\sum_{\pi\in \irr(G)} \pi\otimes \Theta(\pi)
\end{equation}
where $\Theta(\pi)$  is not necessarily an irrep, meaning that the correspondence between irreps is not one-to-one.\footnote{Some work has gone into understanding if there is a natural choice for a canonical representative irrep in $\Theta(\pi)$ which makes the correspondence one-to-one, see \cite{Gurevich2016,Aubert2016,Pan2020}.} 
Since all involved spaces are finite, it is always possible to write the decomposition as \eqref{eq:HoweFinite} (if one allows for $\Theta(\pi_i)=0$ for some $\pi_i$). The non-trivial part is therefore to characterize the map $\Theta$. A conjecture about $\Theta$ for a subclass of irreps, so-called unipotent representations, that includes the singlet (trivial representation) was given in \cite{Aubert1996} and proven in \cite{Pan2019,Ma2022}. 
We give more details on this below, but first we formulate our main argument.

\subsubsection{Proof of equality ``Ensemble average equals Poincar\'e series''}

Consider now two vectors $\phi_\text{A}$ and $\phi_\text{P}$ in $\mathcal X$, defined as the boundary ensemble average and the bulk Poincar\'e series respectively,
\begin{equation}
    \phi_{\text A}=\frac1{\mathcal N}\sum_{C} \psi_C, \qquad     \phi_{\text P}=\frac1{\mathcal N'}\sum_{h\in \Gamma\backslash Sp} U_h \psi_0
\end{equation}
It is easy to see that $\phi_\text{A}$ and $\phi_\text{P}$ are both invariant under both $H$ and $G$.\footnote{The invariance of $\phi_{\text A}$ under $G$ and $\phi_{\text P}$ under $H$ are immediate since they are group averages. The invariance of $\phi_{\text A}$ under $H$ follows from modular invariance of each $\psi_C$. Finally, the invariance of $\phi_{\text P}$ under $G$ follows from commutativity of actions of $G$ and $H$ and the fact that $\psi_0$ is invariant under $G$.} They are also both non-zero.\footnote{Neither $\phi_{\text A}$ nor $\phi_{\text P}$ can vanish. Each term contributing to $\phi_{\text A}$ has a positive scalar product with $\psi_0$; each term contributing to $\phi_{\text P}$ has a positive scalar product with $\psi_{\mathcal C}$ for any code $\mathcal C$.} 
Then we invoke the following lemma, proven below.
\paragraph{Lemma.} 
\textit{There is a one-dimensional subspace $V_{\mathrm{inv}}\subset \mathcal X$ that is is invariant under both $G$ and $H$. }\\

\noindent The lemma immediately proves our main statement, since both $\phi_\text{A}$ and $\phi_\text{P}$ lie in $V_{\mathrm{inv}}$. Thus we have the proportionality
\begin{equation}
\phi_\text{A}\propto \phi_\text{P}.
\end{equation}
The proportionality constant can be fixed by demanding that the vector $\psi_0$ appears with the same unit coefficient on both sides. 

The lemma follows if the image of the singlet of $G$ maps under $\Theta$ to a representation which decomposes into irreps of $H$ as a sum which contains the singlet precisely once: 
\begin{equation}
\label{eq:singlet-singlet-overview--0}
    \Theta(\mathbf 1_O) = \mathbf 1_{Sp} \oplus [\text{other irreps}]. 
\end{equation}
Below we shall characterize further the $\Theta$ map and verify that this is indeed the case.

\subsubsection{Theta map and proof of the lemma} 
\label{sec:HoweFinite}

We now describe the details about the map $\Theta$ defined above. Our presentation closely follows references \cite{Pan2019,Pan2019b,Pan2020}. In particular reference \cite{Pan2019} proves a conjecture from \cite{Aubert1996}, also proven independently in \cite{Ma2022}, which confirms that
\begin{equation}
\label{eq:singlet-singlet-overview}
    \Theta(\mathbf 1_O) = \mathbf 1_{Sp} \oplus \underbrace{\text{other irreps}}_{\min(n,g)\, \text{terms}}\, ,
\end{equation}
\emph{i.e.} that the singlet of $O(2n,\F_p)$ is mapped to a total of $1+\min(n,g)$ irreps, with the singlet of $Sp(2g,\F_p)$ appearing once.

We now introduce some definitions following \cite{Pan2020}, which will allow us to explain \eqref{eq:singlet-singlet-overview} from the statements of \cite{Aubert1996,Pan2019,Ma2022}. 
An important subclass of representations of both orthogonal and symplectic groups over $\F_p$ ($p$ odd) are so-called unipotent representations\footnote{The definition of unipotent is complicated, see Definition~7.8 of \cite{Deligne1976}; for us it suffices to know that this is a subclass of representations which includes the singlet of the respective groups.} and can be labeled by ``symbols.'' This follows older results by Lusztig \cite{Lusztig1977}. A symbol is a pair of collections of strictly decreasing numbers (or empty set),
\begin{equation}
\Lambda=  \begin{pmatrix}
      K\\L
  \end{pmatrix}   =\begin{pmatrix}
        k_1,k_2,\ldots,k_m \\l_1,l_2,\ldots, l_n
    \end{pmatrix}\,,
\end{equation}
where $m,n\in \{0\}\cup \mathbb N$ and $k_i>k_{i+1}$, $l_i>l_{i+1}$. A symbol is reduced if $0\notin K\cap L$, that is not both $K$ and $L$ have zero as the last element. We only consider reduced symbols. 

The rank and the defect of the symbol is defined by
\begin{equation}
    \rk\begin{pmatrix}
        K\\L
    \end{pmatrix}=\sum_{k\in K}k+\sum_{l\in L}l-\left\lfloor \frac{m+n-1}2\right\rfloor, \qquad \defect\begin{pmatrix}
        K\\L
    \end{pmatrix}=m-n\,.
\end{equation}
The classification of Lustzig \cite{Lusztig1977} in the language of \cite{Pan2020} implies that the unipotent irreducible representations are those with
\begin{align}
   \rk\Lambda=N,\qquad \defect\Lambda=
   \begin{cases}
       1\text{ (mod $4$),}\quad & Sp(2N,\F_p),
       \\
       0\text{ (mod $4$),}\quad & O^+(2N,\F_p),
       \\
       2\text{ (mod $4$),}\quad & O^-(2N,\F_p).
       \\
   \end{cases}
\end{align}
The dimension of the irrep corresponding to the symbol $\Lambda=\left(\begin{smallmatrix}
    K\\L
\end{smallmatrix}\right)$ is given by (same formula for $Sp$ and $O$)
\begin{equation}
\label{eq:dimensionIrrep}
    \dim \rho_\Lambda=\frac{G_p}{2^{\lfloor \frac{m+n-1}2\rfloor}}\frac{\Delta(K,p)\Delta(L,p)}{\Theta(K,p)\Theta(L,p)}\frac{\Pi(\Lambda,p)}{p^{\left(\begin{smallmatrix}
    m+n-2\\2
\end{smallmatrix}\right)}p^{\left(\begin{smallmatrix}
    m+n-4\\2
\end{smallmatrix}\right)}\cdots }\,,
\end{equation}
where the exponents of $p$ in the last denominator are binomial coefficients, and we have the following definitions:
\begin{align}
\nonumber
    \Delta(K,p)&=\prod_{k,k'\in K, k>k'}(p^k-p^{k'})\,, & \Theta(K,p)&=\prod_{k\in K}\prod_{h=1}^k(p^{2h}-1)\,,
    \\
    \Pi(\Lambda,p)&=\prod_{k\in K}\prod_{l\in L} (p^k+p^l)\,.
\end{align}
Finally, $G_p$ denotes the order of the group $G$ divided by its maximal factor $p^m$. See appendix~\ref{app:groupsFiniteFields} for expressions for the order of the groups.

Unipotent representations map to unipotent representations under the $\Theta$ map. To give an explicit description of how, we change labels from symbols to bipartitions. A bipartition is a pair of Young tableaux, equivalently a pair of integer collections $[\kappa_1,\ldots,\kappa_m]$, $\kappa_i\geqslant \kappa_{i+1}$. Restricting to the case of $\defect\Lambda=1$ for $Sp$ and $\defect\Lambda=0$ for $O^+$, we can map symbols to bipartitions by
\begin{equation}
    \Upsilon: \begin{pmatrix}
        k_1,k_2,\ldots,k_m \\l_1,l_2,\ldots, l_n
    \end{pmatrix}\mapsto \begin{pmatrix}
        [k_1-(m-1),k_2-(m-2),\ldots, k_m]
        \\
        [l_1-(n-1),l_2-(n-2),\ldots, l_n]        
    \end{pmatrix}.
    \label{eq:Upsilon}
\end{equation}
The bipartitions we consider have a total of $n$ boxes for $O^+(2n)$ and $g$ boxes for $Sp(2g)$.

For the irreps labeled by bipartitions $(\kappa,\lambda)$, we use the notation $\pi_{\kappa,\lambda}$ in the case of $O^+$ and $r_{\kappa,\lambda}$ in the case of $Sp$. With this notation, the theta map takes the following explicit form for unipotent representations\footnote{The statement of \eqref{eq:ThetaMapUnipotent} can equivalently be described in terms of Weyl groups, see \cite{Aubert1996}.}
\begin{equation}
\label{eq:ThetaMapUnipotent}
\Theta(\pi_{\kappa,\lambda})=\sum_{r\in I} r, \qquad I=\{r_{\kappa',\lambda'}| \lambda'\preceq \kappa \text{ and } \lambda\preceq \kappa'\},
\end{equation}
where the symbol $\preceq$ is defined by $\eta\preceq \mu $ iff $\mu_1\geq \eta_1\geq \mu_2\geq \eta_2\geq \mu_3 \geq\cdots$ and $\eta_i$ denotes the $i$th row of the Young tableau $\eta$. The form \eqref{eq:ThetaMapUnipotent} for the $\Theta$ map was conjectured by \cite{Aubert1996} and proven by \cite{Pan2019,Ma2022}.\footnote{There is also a similar formula for $O^-$, which we will not need here.}

Let us now apply this consideration to the singlet (trivial) irrep. In terms of bi-partitions, the trivial irrep of $O^+(2n,\F_p)$ has $(\kappa,\lambda)=([n],\emptyset)$, where $[n]$ denotes the Young tableau with a single row with $n$ boxes -- see proposition 2.9 of \cite{Pan2020}. By checking the conditions in \eqref{eq:ThetaMapUnipotent}, we then find
\begin{equation}
\label{eq:detailsThetaOfId}
\Theta(\pi_{[n],\emptyset})=\sum_{k=0}^{\min(n,g)} r_{[g-k],[k]}
\end{equation}
We see that it contains $1+\min(n,g)$ terms. The singlet $r_{[g],\emptyset}$ appears once, proving our lemma.\footnote{We thank Anne-Marie Aubert for indispensable correspondence.
}

\subsection{Explicit example: $p=3$, $n=1$, $g=1$}
\label{sec:example}

Having discussed the general theory, we are ready to give an example with all the details spelled out. We will consider the simplest case, $p=3$, length $n=1$ and genus $g=1$. In appendix~\ref{app:examples}, we give four additional examples: increasing each of the parameters (thus $[png]=[511], \, [321], \, [312]$), and finally the case $[png]=[322]$, where the sum \eqref{eq:detailsThetaOfId} contains three terms instead of two as in the other examples. 

The involved classical Lie groups over finite fields are finite groups, so we can make use of techniques for manipulations with finite groups such as character tables. For some manipulations it is convenient to use the computer algebra package GAP \cite{GAP}, which automatizes many operations for finite groups, including computations over finite fields. In appendix~\ref{app:GAP} we explain the relevant manipulations -- here we focus on the results. 

For $[png]=[311]$, the space $\mathcal X$ is
\begin{equation}
\label{eq:Xspacep3}
\mathcal X=\spn_\C\{\psi_{00},\psi_{01},\psi_{02},\psi_{10},\psi_{11},\psi_{12},\psi_{20},\psi_{21},\psi_{22}\}.
\end{equation}
Here we write the indices as row vectors, however to conform with the notation in the rest of the paper, they should be viewed as a matrices with a single column vector: 
$
\psi_{\alpha\beta}\leftrightarrow \psi_{\left(\begin{smallmatrix}\alpha\\\beta\end{smallmatrix}\right)}
$.

At this length, there are two codes with three codewords each. Their full-weight enumerators are
\begin{equation}
\label{eq:codesP3N1}
 W_{\mathcal C_1}=\psi_{00}+\psi_{10}+\psi_{20}\,, \qquad  W_{\mathcal C_2}=\psi_{00}+\psi_{01}+\psi_{02}\,, \qquad 
\end{equation}
and the average is thus
\begin{equation}
\label{eq:averForm1}
    \overline{ W}=\psi_{00}+\frac12\left(\psi_{01}+\psi_{02}+\psi_{10}+\psi_{20}\right).
\end{equation}
We should compare this with the formula for the average as given by \eqref{eq:genus-1-average}. Putting $n=1$, and $p=3$, we get
\begin{align}
\nonumber
\label{eq:averForm2}
    \overline{ W}=\frac{1}{1+1}\big(&\psi_{00}+\frac13[\psi _{00}+\psi _{01}+\psi _{02}+\psi _{10}+\psi
   _{11}+\psi _{12}+\psi _{20}+\psi _{21}+\psi
   _{22}]
   \\
   \nonumber
   &+\frac13[\psi _{00}+\psi _{01}+\psi _{02}+\psi _{10}+\zeta\psi
   _{11}+\zeta^2\psi _{12}+\psi _{20}+\zeta^2\psi _{21}+\zeta\psi
   _{22}]
   \\&+\frac13[\psi _{00}+\psi _{01}+\psi _{02}+\psi _{10}+\zeta^2\psi
   _{11}+\zeta\psi _{12}+\psi _{20}+\zeta\psi _{21}+\zeta^2\psi
   _{22}]\big),
\end{align}
where $\zeta=e^{2\pi i/3}$. Expanding this expression, it agrees with \eqref{eq:averForm1} due to cancellation of phases.  

Next, we consider the representation theory of the involved groups. We begin with the modular (symplectic) group $\mathrm{SL}(2,\F_3)=\langle S,T\rangle $, which has $24$ elements and 7 irreducible representations of dimensions $1,1,1,2,2,2,3$.
The character table is given in table~\ref{tab:charTabSL2}, left. 
The representation of $\mathrm{SL}(2,\F_3)$ on our space $\mathcal X$ can be constructed from the representation of the generators $S$ and $T$. Using the MacWilliams identities \eqref{eq:MacWilliamsS}--\eqref{eq:MacWilliamsT}, we find straightforwardly the representation matrices
\begin{align}
\label{eq:genus1repS}
\rho(S)&=\frac13\begin{pmatrix}
1 & 1 & 1 & 1 & 1 & 1 & 1 & 1 & 1 \\
 1 & 1 & 1 & \zeta ^2 & \zeta ^2 & \zeta ^2 & \zeta  & \zeta  & \zeta  \\
 1 & 1 & 1 & \zeta  & \zeta  & \zeta  & \zeta ^2 & \zeta ^2 & \zeta ^2 \\
 1 & \zeta ^2 & \zeta  & 1 & \zeta ^2 & \zeta  & 1 & \zeta ^2 & \zeta  \\
 1 & \zeta ^2 & \zeta  & \zeta ^2 & \zeta  & 1 & \zeta  & 1 & \zeta ^2 \\
 1 & \zeta ^2 & \zeta  & \zeta  & 1 & \zeta ^2 & \zeta ^2 & \zeta  & 1 \\
 1 & \zeta  & \zeta ^2 & 1 & \zeta  & \zeta ^2 & 1 & \zeta  & \zeta ^2 \\
 1 & \zeta  & \zeta ^2 & \zeta ^2 & 1 & \zeta  & \zeta  & \zeta ^2 & 1 \\
 1 & \zeta  & \zeta ^2 & \zeta  & \zeta ^2 & 1 & \zeta ^2 & 1 & \zeta
\end{pmatrix},
\\\label{eq:genus1repT}
\rho(T)&=\diag(1,1,1,1,\zeta ,\zeta ^2,1,\zeta ^2,\zeta ),
\end{align}
From squaring the modular $S$-matrix we can confirm $\rho(S)\rho(S)$ acts as charge conjugation $\psi_v\mapsto \psi_{-v}$, consistent with $S^2=C$. With these matrices at hand, we confirm that
 $   \overline{ W}=\frac12(\psi_{00}+\rho(S)\psi_{00}+\rho(T)\rho(S)\psi_{00}+\rho(T)^2\rho(S)\psi_{00}).
$
This agrees term by term with \eqref{eq:averForm2}. Finally, using the character table, and some additional information extractable from GAP,\footnote{Specifically, as described in appendix~\ref{app:GAP}, by simple commands GAP gives representatives of the conjugacy classes. Then we can determine these the form of these representatives in the representation at hand, and finally \eqref{eq:XdecompSL2} follows from standard computations with the character table.} we find that decomposition of the entire space $\mathcal X$ under $SL(2,\F_3)$ is
\begin{equation}
\label{eq:XdecompSL2}
    \mathcal X=2r_1+ r_2'+ \bar r_2'+ r_3\,.
\end{equation}

\begin{table}
\centering 
{\small
\begin{tabular}{|r|c|c|c|c|c|c|c|}
\hline
  $SL_{2,3}$   & $ [1] $ & $ [4] $ & $ [4]' $ & $ [1]' $ & $ [4]'' $ & $ [4]''' $ & $ [6]    $ \\ \hline
    $ r_1$& $  1 $ & $ 1 $ & $ 1 $ & $ 1 $ & $ 1 $ & $ 1 $ & $ 1 $ \\ \hline
$r_1'$&$  1 $ & $ \zeta ^2 $ & $ \zeta  $ & $ 1 $ & $ \zeta  $ & $ \zeta ^2 $ & $ 1 $ \\ \hline
$\bar r_1'$ &$  1 $ & $ \zeta  $ & $ \zeta ^2 $ & $ 1 $ & $ \zeta ^2 $ & $ \zeta  $ & $ 1 $ \\ \hline
$r_2$ &$  2 $ & $ 1 $ & $ 1 $ & $ -2 $ & $ -1 $ & $ -1 $ & $ 0 $ \\ \hline
$r_2'$ &$  2 $ & $ \zeta  $ & $ \zeta ^2 $ & $ -2 $ & $ -\zeta ^2 $ & $ -\zeta  $ & $ 0
   $ \\ \hline
$\bar r_2'$ & $ 2 $ & $ \zeta ^2 $ & $ \zeta  $ & $ -2 $ & $ -\zeta  $ & $ -\zeta ^2 $ & $ 0
   $ \\ \hline
$r_3$ & $ 3 $ & $ 0 $ & $ 0 $ & $ 3 $ & $ 0 $ & $ 0 $ & $ -1$
\\\hline
\end{tabular}\quad 
\begin{tabular}{|r|c|c|c|c|}
\hline
   $O_{2,3}^+$  & $ I $  & $ \left(\begin{smallmatrix}0&2\\2&0\end{smallmatrix}\right) $ & $ \left(\begin{smallmatrix}0&1\\1&0\end{smallmatrix}\right) $ & $ 2I $ \\ \hline
  $ \pi_1$ & $ 1  $  &  $ 1  $  &  $ 1  $  &  $ 1  $  \\ \hline
    $ \pi_1'$ &$
 1  $  &  $ -1  $  &  $ -1  $  &  $ 1  $  \\ \hline
    $ \pi_1''$ & $
 1  $  &  $ 1  $  &  $ -1  $  &  $ -1  $  \\ \hline
    $ \pi_1'''$ & $
 1  $  &  $ -1  $  &  $ 1  $  &  $ -1 $
\\\hline
\end{tabular}
}
\caption{Character tables for $SL(2,\F_3)$ (left) and $O(1,1,\F_3)\cong O^+(2,\F_3)\cong \Z_2\times \Z_2$ (right). We denote a conjugacy class with $k$ elements by $[k]$, and a $d$-dimensional representation by $r_d$ resp. $\pi_d$. Primes and bars are added to avoid repeated labels. }\label{tab:charTabSL2}
\end{table}

Next we consider the orthogonal group here is
$O_2^+=O(1,1,\F_3)$, which contains four elements and is isomorphic to the Klein four-group or $\Z_2\times \Z_2$. Its character table is given in table~\ref{tab:charTabSL2}, right. $O_2^+$ acts on the indices of $\psi_{\alpha\beta}$ according to \eqref{eq:actionOgroup} and it is easy to confirm that this action is transitive on the set of codes (each code can be mapped to any other). By an exercise with the character table, we find that the oscillator representation decomposes as
\begin{equation}
    \mathcal X=4\pi_1+\pi_1'+2\pi_1''+2\pi_1'''.
\end{equation}

The larger symplectic group that acts on $\mathcal X$ is $\mathrm{Sp}(4,\F_3)$. It has $51\,840$ elements and 34 irreps. Four of these are ``minimal'' (see appendix~\ref{app:groupsFiniteFields}) of dimension $4,4,5,5$, the other non-trivial irreps have dimensions between $6$ and $81$. Our space $\mathcal X$, which is isomorphic to the oscillator representation, is here a direct sum of one irrep with dimension $5$ (charge-conjugation even) and one irrep with dimension $4$ (charge-conjugation odd) under $Sp(4,\F_3)$.

Considering simultaneously the representation theory of both subgroups,\footnote{This does not require any representation theory of the larger symplectic group $\mathcal S=Sp(4,\F_3)$, just the subgroup $G\times H=O(2,\F_3)\times SL(2,\F_3)$. For this computation, we use that the character table of a direct product is the ``outer product'' of the character tables of the two factors.} we arrive at the following decomposition:
\begin{align}
\mathcal X|_{G\times H}&=\pi_1\otimes (r_1+ r_3)+ \pi_1' \otimes r_1  
\nonumber\\
&\quad +\pi_1''\otimes r_2'+\pi_1'''\otimes \bar r_2',
\label{eq:XDecompP3}
\end{align}
where the first line is the charge-conjugation-even subspace and the second line is the charge-conjugation-odd subspace. Relating this special example to our proof, we note that the decomposition \eqref{eq:XDecompP3} has only one term that is a singlet under both groups: $\pi_1\otimes r_1$. Since we know explicitly that both expressions $\phi_{\mathrm P}$, $\phi_{\mathrm A}$ are invariant under $G\times H$, we must have that both $\phi_{\mathrm P},\phi_{\mathrm A}\in \pi_1\otimes r_1$, i.e. that they must be proportional.

Next, we can compare this result with the predictions for the theta map for unipotent representations, \eqref{eq:ThetaMapUnipotent}. In terms of bipartitions, the relevant irreps are $\pi_{\emptyset,[1]}$, $\pi_{[1],\emptyset}$, $r_{\emptyset,[1]}$ and $r_{[1],\emptyset}$
and using \eqref{eq:ThetaMapUnipotent} we see that
\begin{equation}
    \Theta\left(\pi_{\emptyset,[1]}\right)=r_{[1],\emptyset}\,, \quad \Theta\left(\pi_{[1],\emptyset}\right)=r_{[1],\emptyset}+r_{\emptyset,[1]}\,.
\end{equation}
We also compute the dimension of these irreps by inverting the map $\Upsilon$ from symbols to bipartitions, \eqref{eq:Upsilon}, and then using the dimension formula \eqref{eq:dimensionIrrep}. We find
\begin{equation}
    \dim\pi_{\emptyset,[1]}=1, \quad \dim\pi_{[1],\emptyset}=1, \qquad \dim r_{\emptyset,[1]}=3, \quad \dim r_{[1],\emptyset}=1.
\end{equation}
Comparing with \eqref{eq:XDecompP3} we conclude that the unipotent irreps of $O^+(2,\F_3)$ are $\pi_{[1],\emptyset}=\pi_1$ and $\pi_{\emptyset,[1]}=\pi_1'$ and of $SL(2,\F_3)$ are $r_{[1],\emptyset}=r_1$ and $r_{\emptyset,[1]}=r_3$.

We finish this section with a comment on modular invariants. We see that the following two spaces,
\begin{align}
    \pi_1\otimes r_1 &\leftrightarrow
\spn\{2\psi_{00}+\psi_{01}+\psi_{02}+\psi_{10}+\psi_{20}\},
    \\
    \pi_1'\otimes r_1 &\leftrightarrow
    \spn\{\psi_{01}+\psi_{02}-\psi_{10}-\psi_{20}\},
\end{align}
are invariants of $SL(2,\F_3)$, one of which is even and one is odd under $\psi_{\alpha\beta}\mapsto \psi_{\beta\alpha}$. 
We also note that the space of modular invariants can be identified as the space spanned by the enumerators:
\begin{align}
\mathrm{inv}_H(\mathcal X)&=\spn\{ W_{\mathcal C_1}, W_{\mathcal C_2}\} \cong \spn\{ W_{\mathcal C_1}+ W_{\mathcal C_2}, W_{\mathcal C_1}- W_{\mathcal C_2}\}
\\&
=\spn\{2\psi_{00}+\psi_{01}+\psi_{02}+\psi_{10}+\psi_{20},\psi_{01}+\psi_{02}-\psi_{10}-\psi_{20}\} =
\pi_1+ \pi_1'.
\nonumber
\end{align}
This idea that the enumerator polynomials of codes span all invariants of the MacWilliams identities (generally denoted Clifford groups) is an important concept in the study of error-correcting codes \cite{Nebe2006book}. However, it is known that codes cannot be used to generate all modular forms (at large enough genus), even in the chiral case, as discussed in \cite{SalvatiManni1986,Runge1996,Oura2008}. 

\section{Discussion}
\label{sec:discussion}

This paper has two main goals. The first is to unify recent literature about ensembles of CFTs based on codes over $\F_p \times \F_p$ of length $n$. We reviewed the construction of  Chern--Simons theory behind these ensembles and explained that each individual code CFT is holographically dual to a trivial TQFT in the bulk emerging after gauging of a particular maximal non-anomalous subgroup of the one-form symmetry, as outlined in  \cite{Barbar:2023ncl, Benini:2022hzx, Aharony:2023zit}. We have also shown that the Hilbert space ${\mathcal H}^g$ of Chern--Simons theory on a genus-$g$ Riemann surface  is equal to the tensor product of $g$ Hilbert spaces of $2n$ level-$p$ qudits. Gauging a maximal subgroup of one-form symmetry projects the Hilbert space to a one-dimensional ``code subspace'' -- which, in terms of the qubits, is defined by a CSS self-dual quantum stabilizer code. 
In this language, a state of the Chern--Simons theory on any  particular handlebody is also a quantum stabilizer state, providing a quantum information theory interpretation of the holographic duality \eqref{eq:abstractSW}.  

Our second goal is to understand how and why the holographic duality arises between the Chern--Simons theory summed over handlebody topologies and the ensemble of boundary code CFTs. We linked it to a rich mathematical structure known as Howe duality, or the theta correspondence, which relates representations or automorphic forms of the dual pairs of groups. In our case such a pair is provided by the symplectic group acting by modular transformations on the Riemann surface  where the 2d CFT lives, and the orthogonal group acting on the moduli space of the code CFTs. From the Chern--Simons point of view, the orthogonal group is a symmetry of the bulk action but not the boundary conditions.
In CFT terms, there is a duality between the conformal manifold (moduli space of the target space) and the moduli space of the worldsheets. 
Bulk wavefunctions are the functions on the product of these moduli spaces.  

The role of Howe duality in this paper is to organize the representation theory of these functions. Using Howe duality for finite fields in section~\ref{sec:finiteFieldAndProof} we proved the equality between the boundary ensemble average and the bulk Poincar\'e series, by showing there is only one vector in the Hilbert space ${\mathcal H}^g$ invariant under both groups. The equality follows after fixing the normalization on both sides, using the vacuum character (\emph{i.e.}\ demanding that there is a unique vacuum). 

Our work opens  a host of interesting questions, which we discuss below.\footnote{For further explorations, we would like to reference some relevant literature not cited elsewhere in the paper. This includes work covering anyons and finite fields in physics \cite{Frohlich:1988di,Kitaev:2005hzj,Gibbons:2004dij,Lu:2012dt,Burnell:2021reh},
lattices and codes \cite{Montague1994,Bachoc2006,Keller:2017iql,Musin2019,Buican:2021uyp,Kao2022,Buican:2023ehi}, 
dual pairs and Howe duality \cite{Rowe2005,Rowe:2011zz,Roberts:2016hpo,EpequinChavez2018}, and 
averages and ensembles of CFTs \cite{Marolf:2020xie,Collier:2022emf,BoyleSmith:2023xkd, Romaidis:2023zpx,Padellaro:2023blb,Rayhaun:2023pgc,Saidi:2024zdj,Sammani:2025aat}.
We believe that there are many more connections across these topics to uncover.}

\subsection{Genus reduction and the sum over all topologies}
\label{sec:greduction}

First we would like to place our results in the context of \cite{Dymarsky:2024frx}, which has shown that any 3d TQFT summed over all possible topologies is holographically dual to a probabilistic ensemble of boundary CFTs, defined by maximal gaugings of the bulk theory. The proof there is rigorous for any abelian Chern--Simons theory, although there are many subtleties for more complicated theories \textit{e.g.}~those with a continuous spectrum, such as Virasoro TQFT \cite{Collier:2023fwi, Collier:2024mgv,Takahashi:2024ukk}. The general picture is that when the boundary Riemann surface has a very large genus $g$, the holographic duality trivializes and only handlebodies contribute to the bulk sum. All other non-trivial topologies then emerge via genus reduction. An important point is that genus reduction unavoidably yields singular topologies, which can be understood as handlebodies with Wilson line insertions. 

What our paper shows is that in certain simpler cases, such as the case of prime $p$ of this paper, the contribution of all possible topologies, including the singular ones, is proportional to a sum that only includes handlebodies, and hence merely renormalizes the overall coefficient. This statement applies to the sum, not to individual topologies. The genus reduction can be formulated in quantum mechanical terms, as a particular linear map of a state from ${\mathcal H}^g$ to  ${\mathcal H}^{g-1}$, see \cite{Dymarsky:2024frx}. What our results imply, and which would be difficult to see directly, is that when $p$ is a prime the genus reduction maps the sum over the handlebodies of genus $g$ to the sum over the handlebodies of genus $g-1$. 

That only handlebodies contribute to the bulk sum  has a clear interpretation in the context of semiclassical gravity. It is well known  that the solid tori are the only solutions of classical 3d gravity with negative cosmological constant and a torus boundary, and hence the only saddles of the bulk path integral in the semiclassical limit \cite{Maloney:2007ud}. The resulting semiclassical gravity torus partition function was shown in \cite{Maloney:2007ud} to have some negative degeneracies. These can potentially be attributed to missing contributions from the singular topologies, which unavoidably emerge via genus reduction \cite{Dymarsky:2024frx}. Using the framework of TQFT gravity, it would be very interesting to formulate a  model of quantum gravity with a semiclassical limit, such that the bulk sum is dominated by the handlebody contributions. Chern--Simons gravity satisfies this criterion but is not a genuine theory of gravity. However more complicated models of TQFT gravity, \textit{e.g.}~based on a symmetric orbifold of $n\gg 1$ theories, are expected to admit a gravity limit \cite{Pakman:2009zz,Haehl:2014yla,Belin:2014fna}. It therefore would be  very interesting to see if the property that only handlebodies contribute to the bulk sum can be preserved in these more complicated cases.

\subsection{$U(1)$ gravity}

An interesting question is whether our ensemble over a discrete set of lattice theories gives rise to the average over the whole Narain moduli space dual to ``$U(1)$-gravity'' of \cite{Afkhami-Jeddi:2020ezh, Maloney:2020nni} in the limit $p\rightarrow \infty$.\footnote{With the exception of the case of $n=1$.} 
Strictly speaking, at this point the relation to the average over the whole moduli space is conjectural. Although it is known that in the large-$p$ limit the number of code CFTs becomes infinite,
it is an open question to show that they will densely and homogeneously -- with respect to Zamolodchikov metric -- populate the Narain moduli space. Evidence in support of the limit was given in \cite{Aharony:2023zit}, which evaluated  the code CFT averaged torus partition function to become the Eisenstein series \eqref{Eisenstein}. A related check can be found in appendix~\ref{app:gap}, where  we show that the gap of $U(1)$ primaries in the large-$p$ limit approaches $\Delta^* = 1 / (2\pi e )$ expected for $U(1)$-gravity \cite{Afkhami-Jeddi:2020ezh}.

It is also an open question to show that the higher genus averaged partition function would approach and converge to the higher genus generalization of the Eisenstein series, matching the higher-genus analog of the Siegel--Weil formula. In fact, the latter can be understood as a consequence of Howe duality over global fields, providing a unifying framework for both finite $p$ of this work and the continuous case of \cite{Afkhami-Jeddi:2020ezh, Maloney:2020nni}. We outline the connection in the next subsection, but leave the task of developing a detailed connection to future work.

\subsection{Howe duality for other fields}
\label{eq:SiegelFormulas}

The mathematical literature on Howe duality goes beyond just finite fields. Other primary cases of fields are \textit{local fields} and \textit{global fields}. Local fields are complete with respect to a valuation and include $\R$, $\C$, and the $p$-adic numbers $\Q_p$. Global fields have multiple inequivalent valuations and include the rationals $\Q$, number fields (finite extensions of $\Q$, function fields over finite fields $\F_q(t)$ and finite extensions of these function fields.
Here we will comment on the manifestation of Howe duality for global and local fields respectively, and mention various connections to physics.

\paragraph{Local fields.}
A local field $F$ is the one which is complete with respect to a metric, and the classic examples include $\R$ and $\C$, and the $p$-adic numbers $\Q_p$. Howe's original work considered local fields, where the oscillator representation is precisely the infinite-dimensional Fock space \eqref{eq:OmegaAsFockSpace}. This is a representation of the metaplectic group $\widetilde{Sp}$, with subgroups $\tilde G\times\tilde H$. We review some fundamental definitions for the local field case in appendix~\ref{app:HeisenbergReview}, including the construction of the Heisenberg-Weyl group and oscillator representation.

A key statement of Howe duality over local fields is that the decomposition \eqref{eq:HoweGeneral} is multiplicity-free, meaning that it forms a one-to-one map between irreps of the two groups. Howe proved the duality \eqref{eq:HoweGeneral} so-called archimedean fields (e.g. $\R$ and $\C$) \cite{Howe1989}. The progress on the mathematical side for non-archimedean local fields is nicely reviewed in the talk \cite{Gan2014}, with relevant theorems proven in \cite{Waldspurger1990,Minguez2008,Gan:2014vza,Gan2017}. 

The theta correspondence can be thought of as a specific example of a more general principle -- indeed, there are examples in physics with two groups acting on a physical system and exhibiting a special relationship among their representations \cite{Howe:1987tv,Rowe:2012ym,Basile:2020gqi}. Here typically one of the groups is a symmetry group, and another is a ``dynamical group.'' Representations of the two groups are in one-to-one correspondence and can be indexed by a joint label. A concrete example is the central potential problem, \emph{e.g.} the hydrogen atom. In this case a basis of eigenfunctions can be taken to be $\psi_{nlm}$, where $(n,l)$ are labels for the radial problem furnishing irreps of the dynamical group $SU(1,1)$,
and $(l,m)$ are labels for the spherical problem, furnishing irreps of the symmetry group $SO(3)$. The common label (corresponding to $i$ in \eqref{eq:HoweSec3}) is $l$, which determines the rotational irrep with $2l+1$ states labeled by $m$, but also a highest-weight irrep of $SU(1,1)$, with infinitely many states labeled by $n\geqslant l+1$ \cite{Rowe:2012ym}.
More examples of this kind can be found in \cite{Rowe:2012ym,Basile:2020gqi}, demonstrating that the dual pair correspondence is common in physics.

\paragraph{Global fields.}

For global fields, including $\Q$ and its finite extensions, Howe duality manifests in the theory of automorphic forms -- functions on spaces defined by a group $G$ with certain transformation properties under a subgroup $\Gamma \subset G$, plus other conditions on smoothness/fall-off. 
Howe duality becomes a correspondence between the automorphic forms of the two groups in the dual pair, which takes the form of a map between automorphic representations of the two groups. A general treatment requires introducing the so-called \textit{adeles} $\A$ of a local field $F$\footnote{This is the (restricted) direct product of all completions of $F$ with respect to each of its valuations. If $F = \Q$, then an element of $\A=\A_\Q$ is an infinite tuple consisting of elements in $\R$, $\Q_2$, $\Q_3$, $\Q_5$, \ldots where $\Q_p$ are the $p$-adics. There is an additional restriction that all but a finite number of the $p$-adic elements are integers. The motivation of introducing the adeles is that the rationals inside the adeles form a discrete subset, just like the integers inside $\R$. Statements including the familiar cosets $G(\Z)\backslash G(\R)$ can then be rephrased as statements for $G(\Q)\backslash G(\A)$, where the advantage is now that $\Q$ is a field, with nicer algebraic properties than the ring $\Z$.
}
and the structure of a pairing between representations is formulated as a ``theta lift,'' which relates automorphic forms of the two commuting subgroups.

Using these, the Howe duality is phrased as results for the \textit{theta lift},
\cite{Prasad1993,Prasad1996,Li2021}
\begin{equation}
    \label{eq:thetaLift}
    \theta_{\varphi}[\phi](\mathsf h)=\int_{G(\Q)\backslash G(\A)}\theta(\mathsf h,\mathsf g,\varphi)\phi(\mathsf g)d\mathsf g,
\end{equation}
which maps and automorphic form $\phi$ of $G$ into an automorphic form $\theta_{\varphi}[\phi]$ of $H$. This is given as an integral over an adelic quotient against a kernel $\theta(\mathsf h,\mathsf g,\varphi)$, defined with the help of a Schwarz function $\varphi$.

\paragraph{Lattice averages and Siegel--Weil formulas.}

A key application of the theta lift \eqref{eq:thetaLift} is the case of $(G,H)=(O,Sp)$ and constant $\phi(\mathsf g)=1$, where the resulting theta lift is a generalized Eisenstein series, as proven in \cite{Weil1965,Kudla1988,Kudla1988b}. This generalizes earlier formulas by Siegel and Weil \cite{Siegel1935,Siegel1951,Weil1964,Weil1965}, thus known as Siegel--Weil formulas. As already mentioned, the Siegel--Weil formula in turn underpins the result that the ``Narain average = Poincar\'e series'' of \cite{Maloney:2020nni, Afkhami-Jeddi:2020ezh}.

Let us give some more details, relating the general formula above to the Siegel--Weil formulas of averages over unimodular lattices. For a certain choice of $\varphi$, the theta kernel becomes the theta function of an even unimodular lattice of signature $(n,\bar n)$. Then \eqref{eq:thetaLift} reduces to an average over the lattice theta functions, and it is proportional to a conventional Eisenstein series.
In the case of $n=\bar n$, the theta-lift \eqref{eq:thetaLift} leads to the formula \eqref{eq:averNarain}, 
\begin{equation}
    \int d\mu (m)\,\Theta_{\Lambda}(\Omega,m) = E_{\frac n2}(\Omega)/(\det\IM\Omega)^{n},
\end{equation}
which underpins ``Narain average = Poincar\'e series'' of \cite{Maloney:2020nni, Afkhami-Jeddi:2020ezh}, see also \cite{Ashwinkumar:2021kav}.\footnote{We learned about the connection between Howe duality and average formulas in holography from a comment in \cite{Ashwinkumar:2021kav}.} 
Another interesting case is $\bar n=0$, which is an average over positive-definite lattices. In this case, there is a finite set of such lattices  for $n=0$ (mod  $8$) and their average is
\begin{equation}
\label{eq:latticeAverMain}
    \sum_\Lambda \frac{\Theta_\Lambda(\Omega)}{|\aut \Lambda|} = m_{\frac n2}  E_{\frac n2}(\Omega),
\end{equation}
where $m_{\frac n2}=|\frac{B_n}2\, \frac{B_2}4\frac{B_4}8\cdots \frac{B_{2n-2}}{4n-4}|$ is a product of Bernoulli numbers, and $E_{\frac n2}(\Omega)$ is a holomorphic Eisenstein series. The derivation of \eqref{eq:latticeAverMain} from the theta lift can be found in section~2.2 of \cite{Li2021}.
A key step in the derivation is that the adelic integral in \eqref{eq:thetaLift} localizes to a finite sum over positive-definite unimodular lattices, using a description of the set (``genus'') of such lattices as an adelic quotient, see \cite{Hanke2011}. 

The relation~\eqref{eq:latticeAverMain} was first derived by Siegel \cite{Siegel1935}, generalizing the earlier Smith--Minkowski--Siegel mass formula \cite{Smith1868,Minkowski1885}. 
It was conjectured in \cite{Dymarsky:2020qom}, also see \cite{Henriksson:2022dml}, that the relation \eqref{eq:latticeAverMain} can be interpreted in CFT terms as an average over the ensemble of chiral code-based lattice CFTs dual to chiral Chern--Simons theory.
Note however that full set of chiral (meromorphic) CFT of central charge $n=0$ (mod 8) includes other theories (see \cite{Schellekens:1992db,Lam2019} for $n=24$), 
and it is not obvious how to perform the average over this whole set.

\subsection{Other code constructions}
Our proof in section \ref{sec:finiteFieldAndProof} covers only the case where $k=p$ is an odd prime, in which case $\Z_k$ becomes a field. A natural question to ask is what happens for a general $k$.

Extending our proof using Howe duality to the case of $p = 2$ is already non-trivial because $\F_2$ has a special structure different from that of $\F_p$ with odd $p$ -- see \cite{Heinrich2021} for a useful overview. For genus $g=1$, this was considered in \cite{Aharony:2023zit}, where the same pattern was observed: there is a unique vector in ${\mathcal H}$ invariant under both groups, which ensures that the boundary ensemble average is equal to the vacuum character's Poincar\'e series.

In fact, a preliminary study shows that whenever $k$ is square-free, the orthogonal group acts transitively on codes over $\Z_k\times \Z_k$ and there is still a unique vector in ${\mathcal H}^g$ invariant under both groups. At the same time, providing a rigorous proof in that case is a non-trivial task, in part because $\Z_k$ is no longer a field. 

When there are square factors in $k$, the action of the orthogonal group on codes splits into several orbits, resulting in several vectors in ${\mathcal H}^g$ invariant under both groups. In this case the boundary ensemble average is no longer equal to the vacuum character Poincar\'e series, indicating that non-handlebody topologies contribute to the bulk sum  \cite{Dymarsky:2024frx}. 

The mathematical statements of finite-field Howe duality apply for general fields $\F_q$, where $q = p^m$, and it is possible to consider averaging over ensembles of such codes with $m>1$. However, to provide an interpretation of this result in terms of CFTs or Chern--Simons theory would require new code-based CFT constructions. Note that $\F_{p^m}$ is not the same as the ring $\Z_{p^m}$ -- the latter emerges from the  construction of this paper with $k=p^m$, but there is no known mathematical statement similar to finite-field Howe duality in this case.

However, there are many other interesting code CFT constructions in the literature, originating from  the  appropriate Chern--Simons theory, as outlined in \cite{Barbar:2023ncl}. An immediate goal is to use the ideas of this paper to prove the equality ensemble average = Poincar\'e sum in the case of binary codes, \emph{i.e.} the case discussed in the introduction, section~\ref{sec:codeIntro}, based on \cite{Henriksson:2022dml}. This would give an all-genus proof of the genus-$g$ generalization of \eqref{eq:typeIIaverage}. 
Likewise we can consider the generalizations discussed recently in \cite{Angelinos:2024npk, Mizoguchi:2024ahp,Angelinos:2025mjj}. As the results of \cite{Angelinos:2025mjj} suggest, in all known cases, whenever the order of all elements of the ``Pauli group'' of Wilson lines is square-root free, the corresponding orthogonal group acts on the codes transitively, ensuring the uniqueness of a vector invariant under both groups.

It would be also interesting to consider ensemble averages beyond Narain CFTs, for instance when code-based CFTs are non-abelian, and apply Howe duality to that case. 
Among the possible examples are the fermionic code CFTs studied in \cite{Kawabata:2023nlt, Kawabata:2023usr, Kawabata:2023rlt,Kawabata:2025hfd}, or $SU(2)_k$ theories for $k>1$ \cite{Gannon:1994sp}. A promising case is CFTs built from the copies of the Ising characters. The connection to codes goes back to \cite{Dong1994} and was recently discussed in \cite{Moriwaki:2020ktv,Moriwaki:2021ebe} and also in \cite{Barbar:2023ncl}. It is well known that the partition function of the Ising model, which can be thought of as a code CFT of this type for the code of length $n=1$, is given by the vacuum character Poincar\'e series at any genus \cite{Castro:2011zq,Jian:2019ubz}. Generalizing that to arbitrary $n$ would pave the way to understanding holographic ensembles of large central charge CFTs of a new class.

\subsection{Other observables} 

An interesting variation to our setup is to consider ensemble averaging of observables other than the partition function, or to consider averaging over restricted ensembles of codes. As we have seen in the main text, the average over \emph{all} codes is equal to the Poincar\'e series with the vacuum character $\psi_0$ as a seed. This character is the one shared by all codes. If we want to consider an ensemble of the code CFTs based on the codes which share a particular subcode ${\mathcal C}'\subset {\mathcal C}$, the natural guess for the seed would be the state associated with that subcode, \emph{i.e.}
\begin{equation}
\label{eq:restrictedEnsemble}
    \langle  W(\psi_i)\rangle_{\mathcal C\supset \mathcal C'}\ \propto \ \sum_{\gamma}U_\gamma\, \psi_s,\qquad \psi_s=\psi_{\mathcal C'}=\sum_{c\in \mathcal C'} \psi_c.
\end{equation}
Preliminary results suggest that whenever the orthogonal group acts on this code ensemble transitively, this guess is correct. In the ensemble defined by \eqref{eq:restrictedEnsemble}, one could consider $n$-point correlation functions of local operators $\mathcal O_v$ (for $v\in \mathcal C'$) 
defined in every theory of the ensemble. 
Averaged one-point functions have been considered in \cite{Dong2000,Ashwinkumar:2021kav}. 

It would be interesting to find a bulk interpretation of \eqref{eq:restrictedEnsemble}. It is tempting to say that the restricted ensemble is holographically dual to a model similar to CS-gravity of this paper,  with the new CS theory obtained by a partial gauging of the original theory \eqref{CSaction}. In an appropriate limit this might be analogous to the setup of \cite{Chandra:2022bqq}, where the boundary theory computed by averaging over OPE data was shown to match 3d gravity plus additional matter (conical defects) in the bulk.

\subsection{Going beyond 2d code CFT ensembles}

The mathematical framework of this paper suggests generalizations beyond 2d code CFTs. Here we discussed the equivalence between averaging over a moduli space equipped with an action by an orthogonal group and summing over modular images. A natural question is whether there is a similar story with the groups reversed -- a setup where the symplectic group  acts on the conformal manifold, while the mapping class group is orthogonal. In fact, as was briefly mentioned in \cite{Dymarsky:2024frx}, following \cite{Witten:1995gf, Verlinde:1995mz}, such an example is given 
by 4d Maxwell theory with $g$ $U(1)$ gauge fields placed on a four-dimensional manifold $M_4$. The $S$-duality group of this theory is $Sp(2g)$ while the mapping class group of $M_4$ is (a subgroup of) $O(n,\bar n)$. It would be very interesting to understand possible averages of such  theories arising from Howe duality.
Upon reversing the role of the symplectic and orthogonal groups, one may expect to find that the partition function is a type of Eisenstein series for the orthogonal groups. This sort of construction appears naturally in string theory -- if the 2d CFT is on a fluctuating background, one should integrate over all  possible worldsheets, including an integral over the complex structure $\tau$. For string theory on a compactified background, this introduces the modular forms of the orthogonal T-duality group -- see \cite{Obers:1999um}, as well as \cite{Berg:2022feq} for a recent generalization to ``massive'' automorphic forms.

The oscillator representation of the Heisenberg--Weyl group was developed with the quantization of bosons in mind \cite{Segal1959,Segal1961,Shale1962}. It is well known that there is a parallel story for the quantization of fermions -- see for instance chapter 32 of \cite{Woit:2017vqo}. The phase space of the bosonic theory has an antisymmetric form which is preserved by $Sp(n)$ -- for fermions this becomes a symmetric form which is preserved by $SO(n)$. It would be interesting to try to construct the analogous duality for free fermions. The resulting construction might also involve codes, perhaps building on the recent results of \cite{Kawabata:2025hfd}.

It would also be interesting to see if other cases of Howe duality might be relevant for physics. The set of all possible reductive dual pairs has been determined \cite{Howe1979} and forms a relatively small set given in  \eqref{eq:ListHowedualities}. Reference \cite{Basile:2020gqi} helpfully gives a construction of the oscillator representation in a number of examples. Among these are cases where the dual pairs are formed by $GL(n)$ and $GL(m)$, or by $U(n)$ and $U(m)$. 
One place where these structures might appear is in random matrix theory, where these groups play a role as the set of invariances of the elements in a matrix integral. Since matrix integrals frequently appear in various branches of high-energy physics, it would be interesting to try to identify dual pairs, and in particular to see if Howe duality is at work in other examples of holographic correspondences with gravity and ensemble averages, such as the matrix integral related to JT gravity \cite{Saad:2019lba}.

An ultimate goal would be to generalize the idea of ensemble averaging to CFTs that are not free, and in particular those which do not have an extended symmetry algebra. For Narain/lattice CFT, all conformal data is encoded by the lattice, but in general one has to consider averages over both the spectrum and OPE coefficients, see \cite{Chandra:2022bqq,Belin:2023efa} for work in this direction. 
So if there is indeed a space of all 2d CFTs with a well-defined measure, transformations between two different theories must be much more complicated objects, and it is not at all obvious that any of the structure of this paper should apply directly. These transformations then need to ``localize'' on the individual CFTs and known moduli spaces. This would be reminiscent, but immensely more complicated, to how the integral over the genus of positive-definite lattices localizes to a discrete sum in the derivation of Siegel's lattice average in section~\ref{eq:SiegelFormulas}. 

\subsection{Holography and quantum information}
As a byproduct of formulating the statement of holographic duality \eqref{eq:abstractSW} in terms of group representation theory, we arrived at the formulation in terms of families of quantum stablizer states, given by \eqref{holrelcodes}. This formulation provides an intriguing interpretation to \eqref{eq:abstractSW} in terms of quantum information theory, and resonates with the lore that quantum codes of various kind must play a crucial role in holography \cite{Almheiri:2014lwa,Pastawski:2015qua}. It is given that quantum stabilizer codes, that are based on the commutative (abelian) groups $S$, is the artifact of the abelian nature of fusion rules of the bulk CS theory. More complicated, and interesting models of TQFT gravity would call for a generalization of this simple construction. It  nevertheless would be very interesting to generalize the approach of quantum codes to describe the holographic duality in more general setting. 

On a related note, we point out a curious similarity between the  identity \eqref{holrelcodes} of this paper and the results of \cite{Gross:2021ixc} evaluating averages over random stabilizer states. It would be very interesting to place both of these results under the same framework  governed by the Howe duality over finite fields.

\acknowledgments

We would like to thank 
Meer Ashwinkumar, Christian Copetti, Gabriele Di Ubaldo, Matthias Gaberdiel, Bob Kinghton, Chao Li, Alexander Maloney, Dalimil Maz\'a\v{c}, Eric Perlmutter, Dipendra Prasad, Brandon Rayhaun, Eric Reins, Jakub Vo\v{s}mera, and Masahito Yamazaki for interesting conversations and comments on our work, and we are particularly grateful to Anne-Marie Aubert, Henri Darmon, and Shu-Yen Pan for crucial correspondences and clarifications concerning the mathematical literature.

A.D.~acknowledges support by the NSF under  grant  2310426. 
B.M.~is supported by the Gloria and Joshua Goldberg Fellowship at Syracuse University and NSERC (Canada), with partial funding from the Mathematical Physics Laboratory of the CRM.
The work of J.H. has received funding from the European Research Council (ERC) under grant agreement 853507. 

The authors thank the Bootstrap workshop 2022 for hospitality, where this project was initiated. 
The authors thank the Institut Pascal at Universit\'e Paris--Saclay for the hospitality during the program \emph{Speakable and Unspeakable in Quantum Gravity}, ``Investissements d’avenir'' ANR-11-IDEX-0003-01.

\appendix
\addtocontents{toc}{\protect\setcounter{tocdepth}{1}}

\section{The Heisenberg group and the oscillator representation}
\label{app:HeisenbergReview}

A key observation behind  this paper is that Wilson lines of the $U(1)^n \times U(1)^n$ Chern--Simons theory on 3d manifold with a boundary form a faithful representation of the Heisenberg group $H_{2N+1}$ where $2N = 4ng$. Here we will review some mathematical generalities of the Heisenberg group and its representations. The discussion is drawn from a variety of introductory sources including the wonderful textbook \cite{Woit:2017vqo} and notes \cite{Woitnotes} and two comprehensive reviews about the applications of dual pairs in physics \cite{Rowe:2012ym, Basile:2020gqi}.  

\paragraph{The Heisenberg group.} We shall start by considering the field $\R$, which is simpler and more commonly considered, and then we will explain the situation for finite fields. Consider a quantum particle, free or with some potential, in $N$ spatial dimensions. It is described by position and momentum operators which satisfy canonical commutation relations of the form
\begin{align}
    [ Q_i, Q_j] \ = \ 0 \, ,\qquad [ P_i, P_j] \ = \ 0 \, , \qquad [ P_i, Q_j] \ = \ C \delta_{ij} \, .
\end{align}
The elements $\left\{P_1,  \ldots , P_N, Q_1,  \ldots  Q_N, C \right\}$, where the central element $C$ will be chosen to be $-i$, form the Heisenberg Lie algebra $\mathfrak{h}_{2N + 1}$. The algebra elements can be identified with $(N + 2) \times (N + 2)$ nilpotent matrices. For example, for $N = 1$, we have 
\begin{align}
    Q = \begin{pmatrix}
        0 & 1 & 0 \\
        0 & 0 & 0 \\
        0 & 0 & 0
    \end{pmatrix} \, , \qquad P = \begin{pmatrix}
        0 & 0 & 0 \\
        0 & 0 & 1 \\
        0 & 0 & 0
    \end{pmatrix} \, , \qquad 
    C = \begin{pmatrix}
        0 & 0 & 1 \\
        0 & 0 & 0 \\
        0 & 0 & 0
    \end{pmatrix},
\end{align}
so a general element given by $h = a\, P + b\, Q + c\, C$ is identified with 
\begin{align}
    h = \begin{pmatrix}
        0 & a & c \\
        0 & 0 & b \\
        0 & 0 & 0
    \end{pmatrix},
\end{align}
and the commutation relations become the matrix commutator
\begin{align}
    \left[ \begin{pmatrix}
        0 & a & c \\
        0 & 0 & b \\
        0 & 0 & 0
    \end{pmatrix} , \begin{pmatrix}
        0 & a' & c' \\
        0 & 0 & b' \\
        0 & 0 & 0
    \end{pmatrix} \right] \ = \ \begin{pmatrix}
        0 & 0 & a b' - b a' \\
        0 & 0 & 0 \\
        0 & 0 & 0
    \end{pmatrix} .\end{align}
Moving now to $N$ dimensions, $a$ and $b$ get replaced by $N$-vectors, denoted by $\bf a, \bf b$ and a general element $\mathbf{a} \cdot P + \mathbf{b} \cdot Q + c\, C$ is identified with an $(N + 2) \times (N + 2)$ matrix.
These element can be exponentiated to form the Heisenberg group\footnote{The redefinition $d = c + \frac{1}{2} \textbf{a} \cdot \textbf{b}$ leads to another common form of the Heisenberg group where the matrix representation is simpler, but the relation to the symplectic group introduced later is less straightforward.} 
\begin{align}
    \exp \begin{pmatrix}
        0 & \mathbf{a}^T & c \\
        0 & 0 & \mathbf{b} \\
        0 & 0 & 0
    \end{pmatrix} \ = \ \begin{pmatrix}
        1 & \mathbf{a}^T & c + \frac{1}{2} \mathbf{a} \cdot \mathbf{b} \\
        0 & I_n & \mathbf{b} \\
        0 & 0 & 1
    \end{pmatrix},
\end{align}
which satisfies the multiplication law 
\begin{align}
    \begin{pmatrix}
        1 & \mathbf{a}^T & c + \frac{1}{2} \mathbf{a} \cdot \mathbf{b} \\
        0 & I_N & \mathbf{b} \\
        0 & 0 & 1
    \end{pmatrix}  
    \begin{pmatrix}
        1 & \mathbf{a'}^T & c' + \frac{1}{2} \mathbf{a'} \cdot \mathbf{b'}\\
        0 & I_N & \mathbf{b'} \\
        0 & 0 & 1
    \end{pmatrix} = \begin{pmatrix}
        1 & \mathbf{a}^T + \mathbf{a'}^T & c + c' + \frac{1}{2}(\mathbf{a} \cdot \mathbf{b'} - \mathbf{b} \cdot \mathbf{a'}) \\
        0 & I_N & \mathbf{b} + \mathbf{b'} \\
        0 & 0 & 1
    \end{pmatrix}.
\end{align}
This can also be represented abstractly with the law
\begin{align}
        ( \mathbf{a}, \mathbf{b}, c)(\mathbf{a}', \mathbf{b}', c') = (\mathbf{a} + \mathbf{a}', \mathbf{b} + \mathbf{b}', c + c' + \frac{1}{2}(\mathbf{a} \cdot \mathbf{b'} - \mathbf{b} \cdot \mathbf{a'}) ) \, .
\end{align}

The Heisenberg group plays a central role in quantum mechanics, with the Hilbert space being an irreducible representation. Note that this group is not a symmetry of the theory, and it does not commute with any particular operator. 

One of the most important facts about the representation theory of the Heisenberg group is the \textbf{Stone--Von Neumann theorem}, which states that, once you fix the dimension and central element, every irreducible representation of the Heisenberg group $H_{2N+1}$ is unitarily equivalent. This means that we will not need to worry about different representations. However there are a number of different spaces on which the Heisenberg algebra acts, and the equivalent representations may appear quite different. For example, Schr\"odinger's formulation of quantum mechanics in terms of wavefunctions satisfying a differential equation can be stated in terms of a representation of the Heisenberg group on the space $L^2(q)$ of square-integrable functions, \textit{i.e.}~wavefunctions, on position space: 
\begin{align}
    Q = q, \qquad P = -i \frac{\partial}{\partial q} \, .
\end{align}
When exponentiated, $Q$ acts on the wavefunctions by multiplying by a coordinate-dependent phase and $P$ generates a translation. This gives rise to the \textbf{Schr\"odinger representation} of the Heisenberg group. Considering $N$ dimensions, a general element $( \mathbf{a}, \mathbf{b}, c)\in H_{N+2}$ acts on the wavefunction $\psi(\textbf{q}) \in L^2(\bf q)$ by
\begin{align}
    \pi_S(\mathbf{a}, \mathbf{b}, c) \psi(\textbf{q}) = e^{-i c} e^{i \frac{1}{2} \mathbf{a} \cdot \mathbf{b}} e^{-i \mathbf{a} \cdot \mathbf{q}} \psi(\mathbf{q} - \mathbf{b}).
\end{align}
One can define a similar action on other functional spaces such as the space of Schwartz functions $\mathcal{S}(q)$ or its dual, the space of tempered distributions $\mathcal{S}'(q)$. 

\paragraph{Symplectic automorphisms group.}
The commutation relations of the Heisenberg algebra can be written in terms of the antisymmetric form
\begin{align}
\label{SS}
    S((\mathbf{a}, \mathbf{b}), (\mathbf{a}', \mathbf{b}')) = \mathbf{a} \cdot \mathbf{b}' - \mathbf{b} \cdot \mathbf{a}' \, ,
\end{align}
where  vectors $(\mathbf{a}, \mathbf{b})$ and $(\mathbf{a}', \mathbf{b}') \in \R^{2N}$. Namely, 
\begin{align}
    \bigg[ \sum_i \left( a_i P_i + b_i Q_i \right) , \sum_j\left( a'_j P_j + b'_j Q_j \right) \bigg] = - i S((\mathbf{a}, \mathbf{b}), (\mathbf{a}', \mathbf{b}')).
\end{align}
Likewise, we can use this to express the Heisenberg group multiplication law
\begin{align}
\label{eq:HeisenbergGroupMultiplicationAppendix}
        ( \mathbf{a}, \mathbf{b}, c)(\mathbf{a}', \mathbf{b}', c') = (\mathbf{a} + \mathbf{a}', \mathbf{b} + \mathbf{b}', c + c' + \frac{1}{2}S((\mathbf{a}, \mathbf{b}), (\mathbf{a}', \mathbf{b}')) ) \, .
\end{align}
This means that the antisymmetric form can essentially be used to define the Heisenberg group. As a result, the linear transformations of $\mathbb{R}^{2N}$ which leave the Heisenberg group invariant are the same as those which leave the form invariant. The latter is the symplectic group $Sp(2N, \mathbb{R})$, the group of the canonical transformations of quantum mechanics. This leads to a statement which will be important later: 
\textit{the symplectic group is the group of (outer) automorphisms of the Heisenberg group}.
\noindent The automorphism group also contains the Heisenberg group (modulo its center), which acts on itself by conjugation. The full automorphism group is the semi-direct product of these two groups, sometimes referred to as the Jacobi group $Sp(2N, \R) \ltimes H_{2N+1}$, but we will mainly be concerned with the symplectic part. An element
\begin{align}
    g = \begin{pmatrix}
        A & B \\ C & D
    \end{pmatrix} \in Sp(2N, \mathbb{R})
\end{align}
acts on an element of the Heisenberg group by
\begin{align}
\label{eq:SpOnHeisenbergGroup}
    g\, (\mathbf{a}, \mathbf{b}, c) \ = \ (A \mathbf{a} + B \mathbf{b}, C \mathbf{a} + D \mathbf{b} , c)\,.
\end{align}

\paragraph{Metaplectic group.}
The spaces carrying representations of the Heisenberg group also transform under the symplectic group, but this is a projective representation. Consider a unitary representation $\pi$ of $H_{2N+1}$. Then the automorphisms $g \in Sp$ map this to a new representation $g: \pi \to \pi_g$ where 
\begin{align}
    \pi_g (\mathbf{a}, \mathbf{b}, c) = \pi g\, (\mathbf{a}, \mathbf{b}, c)
\end{align}
By the Stone--von Neumann theorem, these are unitarily equivalent, \textit{i.e.}
\begin{align}
    \pi_g = U_g \pi U_g^{-1} \, .
\end{align}
The operators $U_g$ are unique up to a phase and hence provide a projective representation. 
\begin{align}
    U_{g_1} U_{g_2} = e^{i \theta(g_1, g_2)} U_{g_1 g_2} \,.
\end{align}
One can try to redefine $U_g$ trivialize the phase $\theta$ in the multiplication law -- however, this can only be achieved up to a sign, and the $U_g$ truly form a projective representation of the symplectic group.\footnote{This is an important difference between the cases of real fields, discussed here, and finite fields, discussed in the main text. In the latter case, this phase can be trivialized and the irreducible representation of the Heisenberg group is a true representation of the symplectic group.}

In fact, the matrices $U_g$ form a true representation of the double cover of the symplectic group $\widetilde{Sp}(2N, \mathbb{R})$, which is called the metaplectic group. The metaplectic group is not a matrix group; it has no faithful finite-dimensional representations. However its minimal faithful representation, the so-called oscillator representation, or Weil representation, has a natural definition in terms of the raising and lowering operators. 

\paragraph{Oscillator representation (Weil representation).}
The reader familiar with quantum mechanics will expect that the Heisenberg group $H_{2N+1}$ also has a natural interpretation in terms of raising and lowering operators (or \textit{oscillators}), which are related to the generators of the Heisenberg algebra by 
\begin{align}
    a_j = \frac{1}{\sqrt{2}} \left( P_j - i Q_j \right) \, , \qquad a^\dagger_j = \frac{1}{\sqrt{2}} \left( P_j + i Q_j \right) \, ,
\end{align}
and satisfy $[a_i, a_j^\dagger] \ = \ \delta_{ij}$. With these we can define the infinite dimensional \emph{oscillator representation} $\omega$, which is the minimal faithful representation of the metaplectic group. The oscillator representation is
\begin{align}
    \label{eq:OmegaAsFock}
    \omega \ = \ \spn_\mathbb{C} \left\{ a^\dagger_{i_1} \,  \ldots  \, a^\dagger_{i_k} | 0 \rangle \, \middle| \ k \in \mathbb{N} \right\}  ,
\end{align}
where we define the ground state $|0 \rangle$ by $a_i  | 0 \rangle = 0 $. For instance, in the Schr\"odinger representation, the ground state is given by $\psi_0 = e^{-\frac{q^2}{2}}$. Note that we are abusing notation by using the term ``oscillator representation'' to refer to the space carrying the representation, rather than the representation itself.  

The space $\omega$ of \eqref{eq:OmegaAsFock} is essentially the Fock space. 
It carries a representation of the symplectic algebra $sp(2N, \mathbb{R})$. In terms of oscillators we define
\begin{align}
    K^{ij} = a^\dagger_i a^\dagger_j \, ,  \qquad K^{i}{}_j = \frac{1}{2} \{ a^\dagger_i, a_j \} \, , \qquad K_{ij} = - a_i a_j \, .
\end{align}
These operators satisfy the canonical commutation relations for $sp(2N, \mathbb{R})$, with $K^i{}_i$ (no summation over $i$) forming the Cartan subalgebra.

The oscillator representation naturally carries a unitary irreducible representation of the Heisenberg algebra, which can be exponentiated to form a representation of the Heisenberg group. Consider $h(\mathbf{\alpha}, \mathbf{\beta}, \gamma) =  e^{\alpha^i a_i + \beta^i a^\dagger_i + \gamma} \in H_{2N+1} $.
Then the product can be computed using the Baker--Campbell--Hausdorff formula, yielding
\begin{align}
    h(\alpha, \beta, \gamma)h(\alpha', \beta', \gamma') = e^{(\alpha^i + \alpha^i) a_i + (\beta^i + \beta'^i) a^\dagger_i + (\gamma + \gamma') + \frac{1}{2} (\alpha^i \beta'_i - \beta^i \alpha'_i)} 
\end{align}
in agreement with the multiplication law of the Heisenberg group. Exponentiation of the symplectic algebra turns out to be double-valued, and the symplectic group acts projectively on $\omega$.

\paragraph{Howe duality.}
The formalism of oscillator representation allows for the introduction of dual pairs. For a symplectic vector space $W$ with an associated group $Sp(2N, \mathbb{R})$, a \emph{dual pair} is a pair of subgroups $G$ and $H$ which (a) are mutually centralizing and (b) act reductively on $W$. A group $G$ acts reductively on $W$ if it breaks $W$ into a finite sum of $G$-invariant subspaces. In terms of matrices, reductive subgroups are block diagonal, with each block being an irreducible representation of the subgroup.

Howe duality is essentially a correspondence between the representations of each group in a dual pair. The way this works is the following: the oscillator representation is irreducible under the entire metaplectic group, but its restriction to the subgroup $G \times H$ is reducible. As a result it can be decomposed as
\begin{align}
    \label{eq:pairing}
     \omega|_{G \times H} \ = \ \sum_{\pi_G \in \irr G} \pi_{G} \otimes \theta(\pi_G) \, ,
\end{align}
where $\pi_{G}$ is a representations of $G$ and
\begin{align}
\label{theta}
     \theta:\irr G \longrightarrow \irr H
\end{align}
Each representation of $G$ appears only once in the sum so it occurs in $\omega|_{G \times H}$ with multiplicity equal to $\dim (\theta(\pi_{G}))$. The invertible map \eqref{theta} between the irreducible representations of $G$ and $H$ is called theta correspondence.

\subsection{Finite fields}
\label{app:Heisenberg_finite}

Much of the above discussion for $\R$ applies also to finite fields.\footnote{This discussion partially follows \cite{Neuhauser2002}. For original work, see \cite{Gerardin1977}.} Let us consider $\F_p$ for prime $p> 2$. Then the Heisenberg group $H_{2N+1}(\F_p)$ has a multiplication
\begin{align}
    \label{eq:finiteHeisenberg}
        ( \mathbf{a}, \mathbf{b}, c)(\mathbf{a}', \mathbf{b}', c') = (\mathbf{a} + \mathbf{a}', \mathbf{b} + \mathbf{b}', c + c' + 2^{-1}(\mathbf{a} \cdot \mathbf{b'} - \mathbf{b} \cdot \mathbf{a'}) ) \, .
\end{align}
identical to the case for $\R$ except with $1/2$ replaced with the inverse of $2$ in the field. Here we have $\mathbf{a}$, $\mathbf{b}$, $\mathbf{a}'$, $\mathbf{b}' \in \F_p^{n}$ and $c$ and $c' \in \F_p$.

The finite-field Heisenberg group also obeys a version of the Stone--von Neumann theorem. For any group $G$ and representation $\rho$, the restriction of the representation to the center $\rho(Z(G))$ must be one-dimensional by Schur's lemma. Therefore in an irreducible representation, $z \in Z(G)$ acts by multiplication by a (complex) scalar $\chi (z)$. The function $\chi$ is called the central character of $\rho$. In the case of the Heisenberg group, the center is $(\mathbf{0}, \mathbf{0}, c) \simeq \F_p$. The statement of uniqueness is that the representations of the Heisenberg group are uniquely (up to unitary equivalences) determined by the value of the central elements in the representation, \textit{i.e.} by the choice of central character. There are $p$ central characters $\chi(z) = \xi^z$, where $\xi$ is a $p^\text{th}$ root of unity. 

As before, the isometries of the symplectic form $\langle (\mathbf{a} , \mathbf{b} ) ,  (\mathbf{a} , \mathbf{a}) \rangle = (\mathbf{a} \cdot \mathbf{b'} - \mathbf{b} \cdot \mathbf{a'}) \in \F_p $ serve as outer automorphisms of the Heisenberg group. These automorphisms from the symplectic group $Sp(2n, \F_p)$, defined as the matrices preserving the form $J = \left(\begin{smallmatrix}
    0 & I \\
    -I & 0
\end{smallmatrix}\right)$. An element $g = \left(\begin{smallmatrix}
    A & B \\
    C & D
\end{smallmatrix}\right) \in Sp(2n, \F_p)$ acts on the Heisenberg group by
\begin{align}
    \label{eq:appaction}
    g (\mathbf{a} , \mathbf{b}, c ) \mapsto (A \mathbf{a} + B \mathbf{a} , C \mathbf{a} + D \mathbf{b}, c) \, .
\end{align}
The automorphisms of the Heisenberg group also include the Heisenberg group, acting on itself via conjugation. The full automorphism group, $Sp \ltimes H$, is called the Jacobi group. 

\paragraph{Oscillator representation and Howe duality.}
The oscillator representation is constructed using a representation $\eta$ of the Heisenberg group, along with the action of the symplectic group in~\eqref{eq:appaction}. We can then define $\eta_g (\mathbf{a} , \mathbf{b}, c ) = \eta g \,(\mathbf{a} , \mathbf{b}, c )$. By the Stone--von Neumann theorem, this is unitarily equivalent to $\eta$:
\begin{align}
    \label{eq:finiteosc}
    \eta_g = U_g \eta U_g^{-1} \, .
\end{align}
Now $U_g$ forms a representation of the symplectic group with dimension $p^N$-- this is known as the oscillator representation. It is naturally a projective representation, but (unlike in the case for local fields) the phases can be chosen so that it is a genuine (non-projective) representation. We make no distinction between the two (projective / genuine) representations in our notation, though they differ by a phase in the definition of $U_g$.

To see why $\eta$ has dimension $p^N$, we can consider how the Heisenberg group acts on the space of functions $\psi : \F_p^N \to \C$. We have
\begin{align}
    \eta(\mathbf{a} , \mathbf{b}, c ) \psi(\mathbf{q}) = \chi\left(c- 2^{-1} \mathbf{a} \cdot \mathbf{b}  +  \mathbf{a} \cdot  \mathbf{q} \right) \psi(\mathbf{q} -\mathbf{b}),\quad \mathbf{q}\in \F_p^N \, .
\end{align}
So $\eta$ is a homomorphism on the space of functions, which over a finite field is a vector space of dimension $p^N$. This argument uses the representation on the space of function fields, but of course it is a general fact that all irreps with a non-trivial center have exactly this dimension (see \cite{Garrettnotes} for another argument). 

The space of dimension $p^N$ (here, the space of functions) which carries a representation of the Heisenberg and symplectic groups is what we call the oscillator representation, denoted $\mathcal X$ in our paper. For us, it is spanned by the set $\psi_{c_1,\ldots, c_g}$ for codewords $c_i$ of length $2n$,
\begin{equation}
    \mathcal X=\mathrm{span}_\C\left\{\psi_M|M=(c_1,\ldots,c_g)\in \F_p^{N}\right\},
\end{equation}
where $N = 2 n g$. 

The oscillator representation $\mathcal X$ is an irrep of the Heisenberg group, but actually breaks into two irreps of the symplectic group. On the space of functions, these representations are the even and odd functions, which are preserved by the symplectic group. These subspaces have dimensions $(p^N + 1)/2$ and $(p^N - 1)/2$, respectively and correspond to charge-conjugation even (odd). 
We can see this at work in the case where $p = 3$, $n = 1$, $g = 1$, presented in section~\ref{sec:example}. There we find that the oscillator representation is $3^2 = 9$ dimensional, and it breaks into symplectic irreps of dimension $5$ and $4$.
More generally, the group $Sp(2N,\F_q)$ has, apart from the trivial representation, four ``minimal'' representations with dimensions \cite{Gurevich2016}
\begin{equation}
\label{eq:specialirreps}
\frac{q^N-1}2,\frac{q^N-1}2,\frac{q^N+1}2,\frac{q^N+1}2.
\end{equation}
Apart from some small values of $q$ or $N$, all other representations have significantly larger dimensions. 

\section{Classical groups over finite fields}
\label{app:groupsFiniteFields}

Here we collect a few simple facts about classical groups over finite fields. These facts can, for instance, be found in \cite{Atlas1985}. 
We consider the finite field $\F_q$ with $q=p^m$ elements for $p$ an odd prime, implying that the characteristic is $p$. $\F_q$ can be seen as a vector space over $\F_p$ with dimension $m$. In the main text we consider $m=1$, although here we give some statements for general $q=p^m$. 

Lie groups over finite fields are defined to mimic the definitions over $\R$ or $\C$. The symplectic group $Sp(2g,\F_q)$ is defined as the set of matrices that preserve an antisymmetric bilinear form $J$,
\begin{equation}
    Sp(2g,\F_q)=\left\{
    M \in GL(2g,\F_q) \middle| M^TJ M=J
    \right\},
\end{equation}
where $GL(n,\F_q)$ is the group of invertible $n\times n$ matrices over $\F_q$. For fixed $g$, all symplectic groups constructed with different antisymmetric $J$ (of maximal rank) are isomorphic. 

The group $Sp(2N, \F_p)$ is generated by three types of elements: 
\begin{align}
    u(b) = \begin{pmatrix}
        I & 0 \\ b & I
    \end{pmatrix} \, , \quad s(a) = \begin{pmatrix}
        a & 0 \\ 0 & a^{* -1}
    \end{pmatrix}  \, , \quad J = \begin{pmatrix}
    0 & I \\
    -I & 0
\end{pmatrix}
\end{align}
where $b = b^T$ and $a \in GL(2n, \F_p)$.
It has 
\begin{equation}
\label{eq:orderSp}
|Sp(2N,\F_q)|=q^{N^2}\cdot(q^2-1)(q^4-1)\cdots(q^{2N}-1)
\end{equation}
elements. 

The orthogonal group is likewise defined as the set of matrices in $GL(m,\F_q)$ that preserve a symmetric bilinear form $\eta $ of dimension $m$. For odd $m$, this group is unique up to isomorphisms. For even $m=2n$, however, there are two inequivalent groups,
\begin{equation}
    O^\epsilon(2n,\F_q)=\left\{
    M \in GL(2n,\F_q) \middle| M^T\eta  M=\eta
    \right\}
\end{equation}
where $\epsilon=\pm$ is the Witt index of the form $\eta$. We are interested in, c.f.~\eqref{eq:H5innerproduct},
\begin{equation}
\label{eq:defEtaApp}
    \eta=\begin{pmatrix}
        0&I\\I&0
    \end{pmatrix}
\end{equation}
and thus $\epsilon=+$.\footnote{We have that $\epsilon=+$ if there is a basis in which $\eta=\left(\begin{smallmatrix}
    0&1\\1&0
\end{smallmatrix}\right)^{\oplus n}$; in the case of $\epsilon=-$, there is a basis in which $\eta=\left(\begin{smallmatrix}
    0&1\\1&0
\end{smallmatrix}\right)^{\oplus (n-1)}\oplus W$ for some $W$ which depends on $p$ mod 4.} 
The orthogonal group $O^\epsilon(2n,\F_q)$ has 
\begin{equation}
\label{eq:orderO}
|O^\epsilon(2n,\F_q)|=2(q^n-\epsilon)q^{n(n-1)}\cdot (q^2-1)(q^4-1)\cdots (q^{2n-2}-1)
\end{equation} elements, where $\epsilon=\pm1$.

\section{\texorpdfstring{Phase ambiguity for $\boldsymbol{p=3}$ (mod $\boldsymbol 4$)}{Phase ambiguity for p=3 (mod 4)}}
\label{sec:p3mod4}

In section~\ref{sec:qudits}, we noted that in order for the oscillator representation to be a regular (i.e.~not a projective) representation of $\mathcal S=Sp(4gn,\F_p)$, generators of the form $\left(\begin{smallmatrix}
    (A^{-1})^T & 0\\ 0 &A
\end{smallmatrix}\right)$ must act with an extra phase factor of $(\det A)^{(p-1)/2}$. In terms of the code variables (basis of the oscillator representation),
\begin{eqnarray}
\label{SA-app}
S_A=\left(\begin{array}{cc}
(A^{-1})^T & 0\\
0 & A
\end{array}\right),\qquad U_A: \psi_{c_1,\ldots c_g} \mapsto (\det A)^{\frac{p-1}2g}\,\psi_{Ac_1,\ldots, Ac_g}.
\end{eqnarray}
Here $A\in GL(2n,\F_p)$ is a general invertible matrix.  The orthogonal group $O^+(2n,\F_p)$ discussed in the main text should be seen as a subset of such $A$.  
Now for $p=3$ (mod $4$) and $g$ odd, the representation $U_A$ of $A$ is not the same as the representation $\pi(\mathsf g)$ of $O^+(2n,\F_p)$, defined in \eqref{eq:actionOgroup} to be
\begin{equation}
    \pi(\mathsf g) \psi_{c_1,\ldots c_g}= \psi_{\mathsf gc_1,\ldots,\mathsf gc_g}.
\end{equation}
This means that $U_{\mathsf g}=(\det \mathsf g)^{\frac{p-1}2g}\,\pi(\mathsf g)$. 

To connect with the notations in the literature (e.g. \cite{Neuhauser2002}), we write these relations in the Schr\"odinger picture\footnote{In physics terms, $f(x)$ is the wavefunction $f(x)=\langle x|\psi\rangle$. For finite fields, $x\in \text{Mat}_{2n\times g}(\F_p) \simeq \F_p^N$, and we can go between notations using $\psi_{c_1,\ldots c_g}(x)=\prod_{i,I}\delta_{c_{Ii}x_{Ii}}$.}
\begin{align}
    [\pi(\mathsf g) f ](x)&= f(\mathsf g^{-1}x)\, ,
\\
    [U_{\mathsf g} f ](x)&= (\det \mathsf g)^{\frac{p-1}2g}f(\mathsf g^{-1}x)\,,
\end{align}
where $f(x) \in L^2(\F_p^{2ng}) $.
Now while the first of these expressions has a more natural action, it is the second that is compatible with the oscillator representation being a genuine representation of $Sp(4ng,\F_p)$, see \cite{Neuhauser2002}. In other words, if we have $G\times H=O^+(2n,\F_p)\times Sp(2g,\F_p)$ as a subgroup of $\mathcal S=Sp(4gn,\F_p)$, and $\omega$ is the oscillator representation, we then have the restrictions
\begin{equation}
    \omega|_{G}=U=(\det )^{\frac{p-1}2g}\otimes \pi,
\end{equation}
that is $\omega(\mathsf g)=U_{\mathsf g}$ for $\mathsf g\in G\subset\mathcal S$. 

This extra phase factor  $(\det \mathsf g)^{\frac{p-1}2g}$ (the tensor product with a power of the determinant rep) is sometimes not made explicit in the literature, see however equation (1.4) of \cite{Ma2022}.\footnote{We thank Shu-Yen Pan for correspondence.} It also introduces an ambiguity in the nomenclature of the singlet representation: the case we need for our proof in the main text is where $\pi$ acts trivially, $\pi(\mathsf g)=\mathbf 1$;  for odd $\frac{p-1}2g$ this would correspond to the case where $U_{\mathsf g}$ acts as the determinant representation $U_{\mathsf g}=\det(\mathsf g)\mathbf 1$.

\section{Higher-genus modular transformations and MacWilliams identities}
\label{app:MacWilliamsHigherGenus}

Just as the genus-1 enumerator polynomials of even self-dual codes  are invariant under the $S$ and $T$ transformations \eqref{eq:MacWilliamsS} and \eqref{eq:MacWilliamsT}, which correspond to the genus-$1$ modular group, the genus-$g$ enumerator polynomial are invariant under transformations of the genus-$g$ modular group $Sp(2g,\Z)$.
Here we give explicit form of the generators of  $\mathrm{Sp}(2g,\F_p)$ acting on genus-$g$ codeword blocks (Chern--Simons wavefunctions), or equivalently, variables entering enumerator polynomials.
All relations here in fact generalize to groups defined over $\Z_k$ for general $k$.

Recall first the explicit expression  of the genus-$1$ codeword block is 
\begin{equation}
\Psi_c(\tau,\xi,\bar \xi)=\frac1{ |\eta(\tau)|^{2n}} \sum_{v\in \Lambda_c} e^{i \pi \tau p_L^2-i \pi \bar \tau p_R^2+2\pi i (p_L \cdot \xi -p_R\cdot \bar \xi) +\frac\pi{ 2\tau_2}(\xi^2+\bar\xi^2)},
\end{equation}
where
\begin{eqnarray}
\label{lambdac}
\Lambda_c:=\left\{\left.\frac{c+p\, u}{\sqrt{p}}\, \right|\, u=(\vec{a},\vec{b}),\,
\vec{a},\vec{b}\in \mathbb{Z}^n\right\},\quad c\in\mathscr{D}=(\F_p \times \F_p)^n,
\end{eqnarray}
and 
\begin{eqnarray}
\label{ptov}
    \left(
\begin{array}{c}
p_L+p_R\\
p_L-p_R\\
\end{array}\right)=\sqrt{2}\, {\cal O} v,\quad v\in \Lambda_{c}. 
\end{eqnarray}
Here ${\cal O}\in O(n,n,\R)$ is an  ``embedding'' parameter, a fixed arbitrary orthogonal matrix that preserves $\eta$ \eqref{etamatrix}.

We now define the higher-genus version of these,
\begin{align}
\label{Psifugacity}
\Psi_{c_1,\ldots, c_g}(\Omega,\xi,\bar \xi) =\frac1{  \Phi(\Omega)} 
\sum_{v_I \in \Lambda_{c_I}} 
e^{i\pi\, v_I^T \Pi_{IJ} v_J+2\pi i (p^I_L \cdot \xi^I-p^I_R \cdot \bar \xi^I )+ \pi\, \Xi_{IJ} ( \xi^T_I \xi_J +\bar \xi^T_I  \bar \xi_J) },
\\
\Pi_{IJ}^{ij}=(\RE(\Omega)_{IJ} \eta_{ij} +i \IM(\Omega)_{IJ} g_{ij}),\quad \Xi^{IJ}=\frac12(\IM(\Omega))^{-1}_{IJ},\quad g={\mathcal O}^T {\mathcal O}\,.
\end{align}
Again, ${\mathcal O}\in O(n,n,\mathbb{R})$ is the 
embedding parameter.
The indices $I,J$ run from 1 to $g$, while the indices $i,j$ run from $1$ to $2n$. The $i,j$ indices span the target space, where Narain lattice (or codewords) live.
The sum in \eqref{Psifugacity} goes over all possible vectors $v_I$ from $\Lambda_{c_I}$ and the relation between $p_L^I, p_R^I$ and $v_I$ is given by \eqref{ptov}.

There are 3 types of modular  generators acting on the modular parameter $\Omega$, which is a symmetric $g\times g$ matrix with positive-definite imaginary part, and $2n \times g$ matrices $\xi$ and $\bar \xi$. 
We call these generators $T$, $S$ and $A$. 
The generator $T_{ij}$ is a symmetric integer-valued $g \times g$ matrix which acts as 
\begin{eqnarray}
T:\Omega \rightarrow \Omega'&=&\Omega+T,\quad \xi'=\xi,\quad \bar \xi'= \bar \xi,
\\
\Psi_{c_1,\ldots, c_g}(\Omega',\xi',\bar \xi')&=&\Psi_{c_1,\ldots, c_g}(\Omega,\xi,\bar \xi)\, e^{\frac{i \pi}p (c_I,c_J) T_{IJ}}.
\end{eqnarray}
The corresponding $Sp(2g,\F_p)$ group element is $\left(\begin{smallmatrix}
    1& T \\0 & 1
\end{smallmatrix}\right)$.

The generator $S_L$ at the position $L$ acts as follows, 
\begin{eqnarray}
\Omega \rightarrow (\Omega')_{ij}&=
\left\{\begin{array}{cr}
\Omega_{IJ}-\Omega_{IL}\Omega_{LJ}/\Omega_{LL},& I,J\neq L,\\
\Omega_{IL}/\Omega_{LL}, & J=L,\\
\Omega_{LJ}/\Omega_{LL}, & I=L,\\
-1/\Omega_{LL}, & I=J=L,
\end{array}\right.\\
\xi \rightarrow \xi'_i&= \left\{ 
\begin{array}{cr}
\xi_I-\xi_L \Omega_{LI}/\Omega_{LL}, & I\neq L,\\
\xi_L/\Omega_{LL}, & I=L,
\end{array}
\right. 
\\
\bar\xi \rightarrow \bar\xi'_i &= \left\{ 
\begin{array}{cr}
\bar\xi_I-\bar \xi_L{\bar\Omega}_{LI}/\bar{\Omega}_{LL}, & I\neq L,\\
\bar\xi_L/\bar{\Omega}_{LL}, & I=L,
\end{array}
\right. \nonumber
\\
\Psi_{c_1\dots c_L \dots  c_g}(\Omega',\xi',\bar \xi')&= \frac1{ p^n} \sum_{c' \in {\mathscr D}} \Psi_{c_1\ldots c' \ldots  c_g}(\Omega,\xi, \bar \xi)  e^{-{2\pi i\over p} (c_L, c')}.
\end{eqnarray}
This transformation corresponds to the $Sp(2g)$ matrix 
\begin{equation}
   \begin{pmatrix}
       A&-B\\B & A
   \end{pmatrix} ,
   \qquad 
   A=I-Q,\quad 
   B=Q,
\end{equation}
where $Q_{IJ}=0$ unless $I=J=L$, when $Q_{LL}=1$.

Finally, there are general linear transformations  $A$, with the action given by
\begin{align}
A:\Omega \rightarrow \Omega'&=A^T\, \Omega\,  A,\quad \xi\rightarrow \xi'=A^T\,\xi,\quad \bar\xi\rightarrow \bar\xi'=A^T\,\bar\xi,\\
\Psi_{c_1\dots c_g}(\Omega',\xi',\bar\xi')&=\Psi_{c'_1\dots c'_g,\xi, \bar\xi}(\Omega),\qquad 
c'_I=\sum_J A_{IJ} c_J.
\end{align}
The $Sp(2g)$ element is $\left(\begin{smallmatrix}
    A^T & 0 \\ 0 & A^{-1}
\end{smallmatrix}\right)$.

\section{Explicit examples}
\label{app:examples}

In this appendix, we present additional examples that generalize the simplest non-trivial case of $[png]=[311]$ of section~\ref{sec:example}. We begin by presenting some commands in GAP \cite{GAP} that are useful for working out the examples. Then we give complete examples of the theta correspondence for the cases $[png]=[312],[511],[321]$, and partial results for the case $[png]=[322]$.

\subsection{Method and manipulations in GAP}
\label{app:GAP}

First we explain how to perform some of the computations behind the examples. In summary, we need to exhibit a finite-dimensional reducible representation of the group $G\times H$ on the space $\mathcal X$ of dimension $p^{2ng}$ whose basis are the vectors $\psi_{c_1,\ldots, c_g}$. We take $G=O(n,n,\F_p)=O^+(2n,\F_p)$ (the meaning of $+$ was explained in appendix~\ref{app:groupsFiniteFields}) and $H=Sp(2g,\F_p)$. These groups can be generated in GAP by specifying the quadratic/bilinear form. Here we give the commands for $[png]=[311]$. The orthogonal group and its character table are generated by the following commands:\footnote{In GAP, double semicolon suppresses the output. We found that for the case $p=5$, $n=2$, $g=1$ there are some issues in executing the code given here. Note also that in GAP, as in ATLAS \cite{Atlas1985}, the orthogonal group is denoted ``GO'' (general orthogonal).}\\

\gapline\texttt{p := 3;;}

\gapline\texttt{n := 1;;}

\gapline\texttt{g := 1;;}

\gapline\texttt{etaMatG := One(GF(p)) * MatrixByBlockMatrix(BlockMatrix([[2,1,\\ \phantom{xxxx} IdentityMat(n)], [1,2,IdentityMat(n)]],2,2));;}

\gapline\texttt{LoadPackage("forms");}

\noindent \texttt{true}

\gapline\texttt{formG := QuadraticFormByMatrix(etaMatG);;}

\gapline\texttt{gpG := GO(1,2*n,p,formG);}

\noindent \texttt{GO(+1,2,3)}

\gapline \texttt{Order(gpG);}

\noindent \texttt4

\gapline \texttt{charTabG := CharacterTable(gpG);;}

\gapline \texttt{Display(charTabG);}

\noindent [\emph{output omitted}]
\\

\noindent To write the character table in matrix form, one can use
\\

\gapline \texttt{List(Irr(charTabG),ValuesOfClassFunction);}

\noindent \texttt{[ [ 1, 1, 1, 1 ], [ 1, -1, -1, 1 ], [ 1, 1, -1, -1 ], [ 1, -1, 1, -1 ] ]
}
\\

\noindent This corresponds to table~\ref{tab:charTabSL2}, right. 

 The symplectic group and its character table is generated by
\\

\gapline \texttt{etaMatH := One(GF(p)) * MatrixByBlockMatrix(BlockMatrix([[2,1,\\ \phantom{xxxx} (p-1)*IdentityMat(g)],  [1,2,IdentityMat(g)]],2,2));;}

\gapline \texttt{formH := BilinearFormByMatrix(etaMatH);;}

\gapline \texttt{gpH := Sp(2*g,p,formH);}

\noindent \texttt{SL(2,3)}

\gapline \texttt{charTabH := CharacterTable(gpH);;}
\\

\noindent Next, we want to extract the character table and the representatives of the conjugacy classes. This is done with the following commands:\footnote{It appears to us that the algorithm to produce the character table and the associated data is not deterministic, implying that each time the code is run it produces a different but equivalent character table (some rows and columns permuted). To avoid the recomputation of the character table, the commands below are written to take as argument the same \texttt{charTabG} as computed in the previous step.}\\

\gapline \texttt{ccrepresentativesG := List(ConjugacyClasses(charTabG),\\\phantom{xxxx} Representative);;}

\gapline \texttt{ccsizesG := List(ConjugacyClasses(charTabG), Size);;}
\\

\noindent and likewise for \texttt{charTabH}.

We performed the rest of the computations in Mathematica. We imported the expressions for the character table, the representatives of the conjugacy classes and the sizes of the conjugacy classes. With these expressions at hand, it is possible to exhibit the representation on the space $\mathcal X$ in the following way.
\begin{enumerate}
    \item Using the generators of the groups\footnote{For the orthogonal group, we use the same generators as GAP, namely those given in \cite{Taylor1987}. For the symplectic group, we use the ones defined in appendix~\ref{app:MacWilliamsHigherGenus}.} and working mod $p$, the groups $G$ and $H$ can be generated by combining the generators until no new group elements are found. This gives an explicit list of group elements alongside a non-unique way to construct each group element from the generators.
    \item Using the higher-genus MacWilliams identities for the modular group, and the simple group action \eqref{eq:actionOgroup} for the orthogonal group, 
    the representation matrices of the generators $\rho(a_i)$ on $\mathcal X$ can be constructed. For instance, at $[png]=[311]$, this gives equations~\eqref{eq:genus1repS}--\eqref{eq:genus1repT} of the main text. 
    \item For each conjugacy class, we can find its representative in the list determined in step 1, and thus the way to construct it from the generators $a_i$. These give matrices for the representatives in the representation on $\mathcal X$. 
    \item Finally, the decomposition of $\mathcal X$ into irreps of the groups can be found by using orthogonality of the characters (rows of the character table),
    \begin{equation}
        \mathcal X=\sum_i n_ir_i, \qquad n_i=\frac1{|G|}\sum_j d_j \overline{\chi_{ij} }\Tr(\rho(h_j)),
    \end{equation}
    where $d_j$ is the size of the conjugacy class $j$, $\overline{\chi_{ij} }$ is the complex-conjugate of the entries of the character table, and $h_j$ is the representative in the class $j$.
\end{enumerate}
Note that for a direct product of groups, all representation theory becomes the ``outer product'' of the representation theory of the respecting group.

\subsubsection{Example at genus $g=2$ }

We now consider the case $[png]=[312]$. The set of codes is the same as in the example in the main text, namely the two codes whose genus-1 full-weight enumerators are given by \eqref{eq:codesP3N1}. The Poincar\'e series at genus 2 has $(p+1)(p^2+1)$ terms, \emph{i.e.}\ $40$ at $p=3$.

The space $\mathcal X_+^{[312]}$ has dimension $3^4=81$,\footnote{Comparing with \eqref{eq:specialirreps} ($N=4$), it is a direct sum of two irreps of the larger symplectic group $Sp(8,\F_3)$ with dimensions $40$ (charge-conjugation odd) and $41$ (charge-conjugation even) respectively.} and following the steps above, we find that under the product group $O^+(2,\F_3)\times Sp(4,\F_3)$ it has a decomposition,
\begin{equation}
\label{eq:decompX312}
    \mathcal X^{[312]}\big|_{O\times Sp}= \pi_1\otimes(r_1+r_{24})+ \pi_1'\otimes(r_1+r_{15}) + \pi_1''\otimes r_{20}''+ \pi_1'''\otimes r_{20}''',
\end{equation}
where the representations of $O^+(2,\F_3)$ are the same as those discussed in the section~\ref{sec:example} (see table~\ref{tab:charTabSL2}), and the representations of $Sp(4,\F_3)$ will be discussed shortly. We find that the singlet $\pi_1\otimes r_1$ appears only once, as required for  our main result in section~\ref{sec:proof}.

Next, we can compare this result with the prediction \eqref{eq:ThetaMapUnipotent} for the theta map for unipotent representations. In section~\ref{sec:example}, we identified the unipotent irreps  $\pi_{[1],\emptyset}=\pi_1$ and $\pi_{\emptyset,[1]}=\pi_1'$ of $O^+(2,\F_3)$
and using \eqref{eq:ThetaMapUnipotent} we see that
\begin{equation}
   \Theta\left(\pi_{[1],\emptyset}\right)=r_{[2],\emptyset}+\rho_{[1],[1]} \qquad  \Theta\left(\pi_{\emptyset,[1]}\right)=r_{[2],\emptyset}+r_{[1,1],\emptyset},
\end{equation}
and using the dimension formula \eqref{eq:dimensionIrrep} we confirm that $\dim \rho_{[1],[1]}=24$ and $\dim r_{[1,1],\emptyset}=15$.

\begin{table}
\centering 
{\small
\begin{tabular}{|l|c|c|c|c|c|c|l|}
\hline
$Sp_{4,3}$& $[1]$ & $[1]'$ & $[90]$ & $[4320]$ & $[6480]$ & $[40]$ & $\cdots$
\\\hline
$ r_{1}  $  &  $  1  $  &  $  1  $  &  $  1  $  &  $  1  $  &  $  1  $  &  $  1 $  & \multirow{14}{*}{} \\\hline    $
 r_{4}  $  &  $  4  $  &  $  -4  $  &  $  0  $  &  $  0  $  &  $  0  $  &  $  2 \zeta -\zeta ^2 $  & \\\hline    $
 r_{4}'  $  &  $  4  $  &  $  -4  $  &  $  0  $  &  $  0  $  &  $  0  $  &  $  2 \zeta ^2-\zeta  $  & \\\hline    $
 r_{5}  $  &  $  5  $  &  $  5  $  &  $  -3  $  &  $  1  $  &  $  -1  $  &  $  \zeta ^2-2 \zeta  $  & \\\hline    $
 r_{5}'  $  &  $  5  $  &  $  5  $  &  $  -3  $  &  $  1  $  &  $  -1  $  &  $  \zeta -2 \zeta ^2 $  & \\\hline    $
 r_{6}  $  &  $  6  $  &  $  6  $  &  $  -2  $  &  $  -1  $  &  $  0  $  &  $  -3 $  & \\\hline    $
 r_{10}  $  &  $  10  $  &  $  10  $  &  $  2  $  &  $  1  $  &  $  0  $  &  $  2 \zeta ^2+5 \zeta  $  & \\\hline    $
 r_{10}'  $  &  $  10  $  &  $  10  $  &  $  2  $  &  $  1  $  &  $  0  $  &  $  2 \zeta+5 \zeta^2  $  & \\\hline    $
 r_{15}  $  &  $  15  $  &  $  15  $  &  $  -1  $  &  $  -1  $  &  $  -1  $  &  $  6 $  & \\\hline    $
 r_{15}'  $  &  $  15  $  &  $  15  $  &  $  7  $  &  $  0  $  &  $  1  $  &  $  -3 $  & \\\hline    $
 r_{20}  $  &  $  20  $  &  $  -20  $  &  $  0  $  &  $  0  $  &  $  0  $  &  $  -7 $  & \\\hline    $
 r_{20}'  $  &  $  20  $  &  $  20  $  &  $  4  $  &  $  1  $  &  $  0  $  &  $  2 $  & \\\hline    $
 r_{20}''  $  &  $  20  $  &  $  -20  $  &  $  0  $  &  $  0  $  &  $  0  $  &  $ -5 \zeta-8 \zeta^2  $  & \\\hline    $
 r_{20}'''  $  &  $  20  $  &  $  -20  $  &  $  0  $  &  $  0  $  &  $  0  $  &  $  -5 \zeta ^2-8 \zeta  $  & \\\hline    $
 r_{20}''''  $  &  $  20  $  &  $  -20  $  &  $  0  $  &  $  0  $  &  $  0  $  &  $  \zeta -5 \zeta ^2 $  & \\\hline    $
 r_{20}'''''  $  &  $  20  $  &  $  -20  $  &  $  0  $  &  $  0  $  &  $  0  $  &  $  \zeta ^2-5 \zeta  $  & \\\hline    $
 r_{24}  $  &  $  24  $  &  $  24  $  &  $  8  $  &  $  0  $  &  $  0  $  &  $  6 $  & \\\hline    $
 r_{30}  $  &  $  30  $  &  $  30  $  &  $  -10  $  &  $  -1  $  &  $  0  $  &  $  3 $  & \\\hline    
$\vdots$ & \multicolumn{6}{c|}{$\cdots$} & $\ddots$\\\hline
\end{tabular}
}
\caption{Truncated character table for the group $Sp(4,\F_3)$ with $51840$ elements, displaying the two conjugacy classes with one element ($[1]$ and $[1]'$), three additional conjugacy classes whose size is unique among the conjugacy classes, and one additional conjugacy class necessary to distinguish the representations $r_{20}''$ and $r_{20}'''$. The whole character table contains $34$ conjugacy classes. $\zeta=e^{2\pi i/3}$.}\label{tab:charTabSp4}
\end{table}

Due to the non-uniqueness of the output of GAP, we need to give more details to characterize the representations of $H=Sp(4,\F_3)$. A truncated version of the complete character table for $H$ is given in table~\ref{tab:charTabSp4}. 
Representatives in the displayed conjugacy classes are $\1 \in I[1]$, $
2\1 \in [1]'$, and
\begin{equation}
\ \left(
\begin{smallmatrix}
 1 & 1 & 0 & 1 \\
 2 & 2 & 2 & 0 \\
 0 & 2 & 1 & 2 \\
 1 & 0 & 1 & 2
\end{smallmatrix} \right)\in [90], \ \left(
\begin{smallmatrix}
 2 & 2 & 2 & 2 \\
 2 & 2 & 0 & 1 \\
 0 & 2 & 0 & 1 \\
 0 & 1 & 2 & 2
\end{smallmatrix} \right) \in [4320], \ \left(
\begin{smallmatrix}
 1 & 1 & 2 & 0 \\
 0 & 0 & 1 & 2 \\
 2 & 2 & 1 & 1 \\
 2 & 0 & 0 & 1 
 \end{smallmatrix} \right)\in [6480], \ \left(
\begin{smallmatrix}
2 & 1 & 2 & 1 \\
 2 & 0 & 1 & 2 \\
 1 & 1 & 0 & 1 \\
 1 & 1 & 2 & 2 
 \end{smallmatrix} \right)\in [40].
\end{equation}
This data is enough to unambiguously identify the irreps in \eqref{eq:decompX312}.

\subsection{Example for prime $p=5$}
We now consider the case $[png]=[511]$.
At length $1$ and prime $p$, there are always only two codes. At $p=5$, they contain five elements each, and have full-weight enumerators,
\begin{equation}
\label{eq:codesP5N1}
 W_{\mathcal C_1}=\psi_{00}+\psi_{10}+\psi_{20}+\psi_{30}+\psi_{40}, \qquad  W_{\mathcal C_2}=\psi_{00}+\psi_{01}+\psi_{02}+\psi_{03}+\psi_{04}, \qquad 
\end{equation}
The Poincar\'e series contains six terms, as given by \eqref{eq:genus-1-average}, and upon expanding it agrees with $\frac12( W_{\mathcal C_1}+ W_{\mathcal C_2})$. 
The space $\mathcal X^{[511]}$ has dimension $25$, and decomposes under the product $G\times H=O^+(2,\F_5)\times SL(2,\F_5)$ as
\begin{equation}
\label{eq:decompX511}
    \mathcal X^{[511]}\big|_{O\times SL} = \pi_1\otimes \left(r_1+ r_5\right) + \pi_1'\otimes r_1 + \pi_1''\otimes r_3+ \pi_1'''\otimes r_3'+\pi_2\otimes r_6,
\end{equation}
where the last term constitutes the charge-conjugation odd subspace of dimension $12$. Again, we see that $\pi_1\otimes r_1$ only appears once. The identification with bipartitions for unipotent irreps is the same as in the example $p=3$ in the main text, where now the dimension formula \eqref{eq:dimensionIrrep} shows that $\dim r_{\emptyset,[1]}=5$. 
\begin{table}
\centering 
{\small
\begin{tabular}{|c|ccccc|}
\hline
$O_{2,5}^+$& $[1]$ & $[2]$& $[2]'$& $[2]''$& $[1]'$
\\\hline
$\pi_1 $  &  $   1  $  &  $ 1  $  &  $ 1  $  &  $ 1  $  &  $ 1 $\\
$\pi_1'  $ &  $ 1  $  &  $ -1  $  &  $ -1  $  &  $ 1  $  &  $ 1 $\\
$\pi_1''   $  &  $ 1  $  &  $ -1  $  &  $ 1  $  &  $ -1  $  &  $ 1 $\\
$\pi_1'''  $  &  $ 1  $  &  $ 1  $  &  $ -1  $  &  $ -1  $  &  $ 1$ \\
$\pi_2$     &  $ 2  $  &  $ 0  $  &  $ 0  $  &  $ 0  $  &  $ -2$\\\hline
\end{tabular}
}
\caption{Character table for the group $O^+(2,\F_5)$. }\label{tab:chartabON25}
\end{table}

\begin{table}
\centering 
{\small
\begin{tabular}{|c|c|c|c|c|c|c|c|c|c|}
\hline
$SL_{2,5}$&$[1]$ & $ [ 12]$ & $ [ 12]'$ & $ [ 1]'$ & $ [ 12]''$ & $ [ 12]'''$ & $ [ 20]$ & $ [ 20]'$ & $ [ 30]$
\\\hline
$r_{1}$ &1 & 1 & 1 & 1 & 1 & 1 & 1 & 1 & 1 \\\hline $r_{2}$ &
 2  & $-\xi ^4-\xi  $ & $-\xi ^3-\xi ^2 $ & $-2 $ & $\xi ^4+\xi  $ & $\xi ^3+\xi ^2 $ & $-1 $ & $1 $ & $0$ \\\hline $r_{2'}$ &
 $2 $ & $-\xi ^3-\xi ^2 $ & $-\xi ^4-\xi  $ & $-2 $ & $\xi ^3+\xi ^2 $ & $\xi ^4+\xi  $ & $-1 $ & $1 $ & $0$ \\\hline $r_{3}$ &
 $3 $ & $-\xi ^3-\xi ^2 $ & $-\xi ^4-\xi  $ & $3 $ & $-\xi ^3-\xi ^2 $ & $-\xi ^4-\xi  $ & $0 $ & $0 $ & $-1$ \\\hline $r_{3'}$ &
 $3 $ & $-\xi ^4-\xi  $ & $-\xi ^3-\xi ^2 $ & $3 $ & $-\xi ^4-\xi  $ & $-\xi ^3-\xi ^2 $ & $0 $ & $0 $ & $-1$ \\\hline $r_{4}$ &
$4  $  &  $ -1  $  &  $ -1  $  &  $ 4  $  &  $ -1  $  &  $ -1  $  &  $ 1  $  &  $ 1  $  &  $ 0$ \\\hline $r_{4}'  $  &  
 $4  $  &  $ 1  $  &  $ 1  $  &  $ -4  $  &  $ -1  $  &  $ -1  $  &  $ 1  $  &  $ -1  $  &  $ 0$ \\ \hline$r_{5}  $  &  
 $5  $  &  $ 0  $  &  $ 0  $  &  $ 5  $  &  $ 0  $  &  $ 0  $  &  $ -1  $  &  $ -1  $  &  $ 1 $\\ \hline$r_{6} $  &  
 $6  $  &  $ -1  $  &  $ -1  $  &  $ -6  $  &  $ 1  $  &  $ 1  $  &  $ 0  $  &  $ 0  $  &  $ 0 $ \\\hline\hline
\end{tabular}
}
\caption{Character table for the group $SL(2,\F_5)$, where $\xi$ is a fifth root of unity.}\label{tab:chartabSL25}
\end{table}

To describe the irreps appearing in \eqref{eq:decompX511}, we need the character tables of the groups. 
For $O(2,\F_5)$, with 8 elements, the character table is given in table~\ref{tab:chartabON25}, and the representatives in the conjugacy classes are
\begin{equation}
\left(\begin{smallmatrix}
 1 & 0 \\
 0 & 1 \\
\end{smallmatrix}\right)
\in [1] , \quad 
\left(\begin{smallmatrix}
 0 & 2 \\
 3 & 0 \\
\end{smallmatrix}\right)
\in [2] , \quad 
\left(\begin{smallmatrix}
 0 & 1 \\
 1 & 0 \\
\end{smallmatrix}\right)
\in [2]' , \quad 
\left(\begin{smallmatrix}
 2 & 0 \\
 0 & 3 \\
\end{smallmatrix}\right)
\in [2]'' , \quad 
\left(\begin{smallmatrix}
 4 & 0 \\
 0 & 4 \\
\end{smallmatrix}\right)\in [1]'.
\end{equation}
The symplectic group $SL(2,\F_5)$ has $120$ elements. The character table is given in table~\ref{tab:chartabSL25}.
This information is enough to identify the irreps appearing in \eqref{eq:decompX511}.

\subsection{Example at length $n=2$}
Next we consider the case $[png]=[321]$. The orthogonal group $O(4,\F_3)$ has $1152$ elements. There are $8$ codes $\mathcal C_i$. These codes can be generated using the construction given in section~\ref{codesdef}, and have full-weight enumerators:
\begin{align}
W_{\mathcal C_1}&=\psi_{0 0 | 0 0}+\psi_{0 0 | 0 1}+\psi_{0 0 | 0 2}+\psi_{0 0 | 1 0}+\psi_{0 0 | 1 1}+\psi_{0 0 | 1 2}+\psi_{0 0 | 2 0}+\psi_{0 0 | 2 1}+\psi_{0 0 | 2 2},\nonumber\\ W_{\mathcal C_2}&=\psi_{0 0 | 0 0}+\psi_{0 0 | 0 1}+\psi_{0 0 | 0 2}+\psi_{1 0 | 0 0}+\psi_{1 0 | 0 1}+\psi_{1 0 | 0 2}+\psi_{2 0 | 0 0}+\psi_{2 0 | 0  1}+\psi
   _{2 0 | 0 2},\nonumber\\W_{\mathcal C_3}&= \psi_{0 0 | 0 0}+\psi_{0 0 | 1 0}+\psi_{0 0 | 2 0}+\psi_{0 1 | 0 0}+\psi_{0 1 | 1 0}+\psi_{0 1 | 2 0}+\psi_{0 2 | 0 0}+\psi_{0 2 | 1 0}+\psi_{0 2 | 2 0},\nonumber\\W_{\mathcal C_4}&= \psi_{0 0 | 0 0}+\psi_{0 0 | 1 1}+\psi_{0 0 | 2 2}+\psi_{1 2 | 0 0}+\psi_{1 2 | 1 1}+\psi_{1 2 | 2 2}+\psi_{2 1 | 0 0}+\psi
   _{2 1 | 1 1}+\psi_{2 1 | 2 2},\nonumber\\ W_{\mathcal C_5}&=\psi_{0 0 | 0 0}+\psi_{0 0 | 1 2}+\psi_{0 0 | 2 1}+\psi_{1 1 | 0 0}+\psi_{1 1 | 1 2}+\psi_{1 1 | 2 1}+\psi_{2 2 | 0 0}+\psi_{2 2 | 1 2}+\psi_{2 2 | 2 1},\nonumber\\W_{\mathcal C_6}&= \psi_{0 0 | 0 0}+\psi_{0 1 | 0 0}+\psi_{0 2 | 0 0}+\psi_{1 0 | 0 0}+\psi_{1 1 | 0 0}+\psi_{1 2 | 0 0}+\psi
   _{2 0 | 0 0}+\psi_{2 1 | 0 0}+\psi_{2 2 | 0 0},\nonumber\\W_{\mathcal C_7}&= \psi_{0 0 | 0 0}+\psi_{0 1 | 1 0}+\psi_{0 2 | 2 0}+\psi_{1 0 | 0 2}+\psi_{1 1 | 1 2}+\psi_{1 2 | 2 2}+\psi_{2 0 | 0 1}+\psi_{2 1 | 1 1}+\psi_{2 2 | 2 1},\nonumber\\W_{\mathcal C_8}&= \psi_{0 0 | 0 0}+\psi_{0 1 | 2 0}+\psi_{0 2 | 1 0}+\psi_{1 0 | 0 1}+\psi_{1 1 | 2 1}+\psi
   _{1 2 | 1 1}+\psi_{2 0 | 0 2}+\psi_{2 1 | 2 2}+\psi_{2 2 | 1 2}.
\end{align}
The average written explicitly is
\begin{equation}
    \overline{ W}=\sum_{i=1}^8 W_{\mathcal C_i}=\frac34\psi_{00|00}+\frac14\sum_{(\alpha_1\alpha_2)\cdot(\beta_1\beta_2)=0\text{ (mod $3$)}}\psi_{\alpha_1\alpha_2|\beta_1\beta_2},
\end{equation}
where the sum has $33$ terms. 
We can now explicitly check that this expression agrees with the Poincar\'e series
\begin{align}
\nonumber
   \overline{ W}&=\frac34(\psi_{00|00}+\rho(S)\psi_{00|00}+\rho(T)\rho(S)\psi_{00|00}+\rho(T)^2\rho(S)\psi_{00|00})
    \\\nonumber&=
    \frac34\big(\psi_{0|0}^{\otimes 2}+\frac19[\psi _{0|0}+\psi _{0|1}+\psi _{0|2}+\psi _{1|0}+\psi
   _{1|1}+\psi _{1|2}+\psi _{2|0}+\psi _{2|1}+\psi
   _{2|2}]^{\otimes 2}
   \\
   \nonumber
   &\quad +\frac19[\psi _{0|0}+\psi _{0|1}+\psi _{0|2}+\psi _{1|0}+\zeta\psi
   _{1|1}+\zeta^2\psi _{1|2}+\psi _{2|0}+\zeta^2\psi _{2|1}+\zeta\psi
   _{2|2}]^{\otimes 2}
   \\&\quad +\frac19[\psi _{0|0}+\psi _{0|1}+\psi _{0|2}+\psi _{1|0}+\zeta^2\psi
   _{1|1}+\zeta\psi _{1|2}+\psi _{2|0}+\zeta\psi _{2|1}+\zeta^2\psi
   _{2|2}]^{\otimes 2}\big)
\end{align}
where $\psi_{\alpha|\beta}\otimes\psi_{\alpha'|\beta'}=\psi_{\alpha\alpha'|\beta\beta'}$.

The space $\mathcal X$ has dimension $81$, and its decomposition under the product group $O(4,\F_3)\times SL(2,\F_3)$ reads
\begin{align}
    \mathcal X^{[321]}\big|_{O\times SL}&=\pi_1\otimes (r_1+r_3)+ \pi'_2\otimes r_1'+\pi_2'''\otimes r_1''+\pi_6'\otimes r_1+\pi_9'''\otimes r_3\nonumber
    \\&\quad + \pi_4'''\otimes r_2+\pi_8 \otimes r_2'+\pi_8''' \otimes \bar r_2'\,.
\label{eq:O4sl2decomp}
\end{align}
Just like in all other examples, we see that the singlet $\pi_1\otimes r_1$ appears only once in the decomposition. 

\begin{table}
\centering 
{\small
\begin{tabular}{|l|c|c|c|c|c|c|c|l|}
\hline
$O_{4,3}^+$& $[1]$ & $ [1]'$ & $[18]$ & $[144]$ & $[72]$& $[72]'$& $[72]''$ & $\cdots$
\\\hline
  $\pi_{1}$   &  $1 $ & $ 1 $ & $ 1 $ & $ 1 $ & $ 1 $ & $ 1 $ & $ 1 $  &
  \\\hline
$\vdots$ & \multicolumn{7}{c|}{$\cdots$} &\\\hline
  $\pi_{2}$   &  $2 $ & $ 2 $ & $ 2 $ & $ 0 $ & $ 0 $ & $ 0 $ & $ -2 $ & \\\hline
  $\pi_{2}'$   &  $2 $ & $ 2 $ & $ 2 $ & $ 0 $ & $ 0 $ & $ 0 $ & $ 2 $  &\\\hline
  $\pi_{2}''$   &  $2 $ & $ 2 $ & $ 2 $ & $ 0 $ & $ -2 $ & $ 0 $ & $ 0 $ & \\\hline
  $\pi_{2}'''$   &  $2 $ & $ 2 $ & $ 2 $ & $ 0 $ & $ 2 $ & $ 0 $ & $ 0 $ & \\\hline
  $\vdots$ & \multicolumn{7}{c|}{$\cdots$} &\\\hline
  $\pi_{6}$   &  $6 $ & $ 6 $ & $ -2 $ & $ 0 $ & $ 0 $ & $ -2 $ & $ 0 $ & \\\hline
  $\pi_{6}'$   &  $6 $ & $ 6 $ & $ -2 $ & $ 0 $ & $ 0 $ & $ 2 $ & $ 0 $ & \\\hline
  $\vdots$ & \multicolumn{7}{c|}{$\cdots$} &\\\hline
  $\pi_{9}'''$   &  $9 $ & $ 9 $ & $ 1 $ & $ -1 $ & $ -1 $ & $ 1 $ & $ -1 $ & 
\\\hline
$\vdots$ & \multicolumn{7}{c|}{$\cdots$} & $\ddots$\\\hline
\end{tabular}
}
\caption{Truncated character table for $O(4,\F_3)$. The whole character table contains 25 conjugacy classes.}\label{eq:charTabO4}
\end{table}

The irreps $r_i$ of $SL(2,\F_3)$ follow the notation from above and the corresponding character table was given in table~\ref{tab:charTabSL2}, left. For the group $O^+(4,\F_3)$ we now give some more information. The generators of the group as given in \cite{Taylor1987} are
\begin{equation}
\label{eq:generatorsO4plus}
    g_1=\begin{pmatrix}
        0 & 0 & 2 & 0 \\
 0 & 1 & 2 & 0 \\
 2 & 0 & 0 & 2 \\
 0 & 0 & 0 & 1 
    \end{pmatrix}, \qquad g_2=\begin{pmatrix}
       2 & 2 & 2 & 1 \\
 0 & 1 & 1 & 0 \\
 0 & 1 & 0 & 0 \\
 1 & 1 & 0 & 0
    \end{pmatrix}.
\end{equation}
Using these we construct a group of $1152$ elements. 
There are 25 conjugacy classes, and a partial character table is given in table~\ref{eq:charTabO4}. The representatives of the displayed conjugacy classes with non-unique size are
\begin{align}
    \1\in [1], \quad 2\1\in [1]', 
    \quad \begin{pmatrix}
0 & 2 & 2 & 0 \\
 0 & 0 & 2 & 0 \\
 0 & 0 & 2 & 2 \\
 2 & 1 & 1 & 1
    \end{pmatrix}\in [72], \quad \begin{pmatrix}
2 & 0 & 0 & 2 \\
 2 & 1 & 2 & 2 \\
 0 & 0 & 2 & 2 \\
 0 & 0 & 0 & 1 
    \end{pmatrix}\in [72]', \quad \begin{pmatrix}
 1 & 2 & 2 & 2 \\
 1 & 2 & 0 & 0 \\
 0 & 2 & 0 & 0 \\
 0 & 1 & 1 & 0
    \end{pmatrix}\in [72]''. 
\end{align}
This information suffices to uniquely identify the first line in \eqref{eq:O4sl2decomp}; we omit the bulky expressions needed to exactly identify the second line. 

We can also compare \eqref{eq:O4sl2decomp} with the form of the $\Theta$ map as given by \eqref{eq:ThetaMapUnipotent} for unipotent irreps. We have five such potential irreps, with the following maps
\begin{align}
    \Theta\left(\pi_{[2],\emptyset}\right)&=r_{[1],\emptyset}+r_{\emptyset,[1]}\,,& \Theta\left(\pi_{[1,1],\emptyset}\right)&=r_{\emptyset,[1]}\,,
    \\
    \Theta\left(\pi_{[1],[1]}\right)&=r_{[1],\emptyset} \,,& \Theta\left(\pi_{\emptyset,[2]}\right)&=\Theta\left(\pi_{\emptyset,[1,1]}\right)=0\,.
\end{align}
We know that $\pi_{[2],\emptyset}=\mathbf 1$, and moreover the dimension formula \eqref{eq:dimensionIrrep} gives $\dim \pi_{[1,1],\emptyset}=9$ and $\dim \pi_{[1],[1]}=6$, so based on comparison with \eqref{eq:O4sl2decomp} we identify $\pi_{[1,1],\emptyset}=\pi'''_9$ and $\pi_{[1],[1]}=\pi'_6$

\subsection{Example at $n=g=2$}
Our final example is the case $[png]=322$. The involved expressions are very large, $\dim \mathcal X=3^8=6561$, and therefore computing the complete decomposition of $\mathcal X$ is quite demanding. Instead we consider the space of invariants under $O^+(4,\F_3)$ inside $\mathcal X$, in order to evaluate the action of $\Theta(\pi_1)$, that is the theta lift of the singlet of $O^+(4,\F_3)$ as a direct sum of irreps of $Sp(4,\F_3)$. 

Using invariance under the generators \eqref{eq:generatorsO4plus}, we determine a $40$-dimensional subspace $V_{\mathrm{inv}(O)}$ of invariants under $O^+(4,\F_3)$. 
On this space, we determine the action of the generators of $Sp(4,\F_3)$, and thus the decomposition of $V_{\mathrm{inv}(O)}$ under $Sp(4,\F_3)$. We find that
\begin{equation}
    \Theta(\pi_1)=r_1+r_{15}'+r_{24},
\end{equation}
in the notation of table~\ref{tab:charTabSp4}. Thus we find that $\Theta(\pi_1)=r_1+r_{15}'+r_{24}$, confirming that $\pi_1\otimes r_1$ only appears once inside $\mathcal X$, and that $\Theta(\pi_1)$ contains exactly $1+\min(n,g)=3$ terms in agreement with \eqref{eq:detailsThetaOfId}, where $r_1=\mathbf 1$, $r_{24}=r_{[1],[1]}$ and $r_{15}'=r_{\emptyset,[2]}$. In all the previous examples, we had $1+\min(n,g)=2$, and there were two irreps in $\Theta(\pi_1)$.

\section{\texorpdfstring{``Pless--Sloane-type'' derivation of averaged full enumerator}{"Pless--Sloane-type" derivation of averaged full enumerator}}
\label{sec:Plessproof}
In this appendix we use the method pioneered by Pless and Sloane in \cite{Pless1975} to derive the averaged full enumerator at genera $g=1$ and $g=2$. The idea is to consider the orbits of the symmetry group $O(n,n,\Z_p)$ acting on individual even codewords. A different, brute force derivation for $p=g=2$ and a somewhat different code ensemble (which included not all codes, but only codes with the generator matrix in the canonical form), was given in \cite{Dymarsky:2020pzc}.
 
In what follows we assume $k=p$ is prime.  We consider even self-dual codes over $\F_p\times \F_p$. 
There are 
\begin{equation}
\label{eq:numCodes}
N(n)=\prod_{j=0}^{n-1}(p^j+1)
\end{equation}
such codes of length $n$. There are also 
\begin{equation}
\label{Nofeven}
{\mathcal N}(n)=(p^n-1)(p^{n-1}+1)+1
\end{equation}
even codewords $c$ of length $n$, i.e.~vectors in ${\mathscr D}=\F_p^{2n}$ satisfying \eqref{evenness}.
To derive that, we first consider the case of $c=(\vec{\alpha},\vec{\beta})$ with  $\vec{\alpha}\ne 0$.  There are $p^n-1$ such vectors. For each such vector $\vec{\alpha}$ we can have arbitrary vector $\vec{\beta}$ of length $n-1$. One of $n$  entries of $\vec{\beta}$ will be fixed to assure $\vec{\alpha}\cdot \vec{\beta}=0$ (mod) $p$. This entry could be not the last one, but should correspond to  $\alpha_i\neq 0$. In the end, there are $p^{n-1}$ such vectors $\vec{\beta}$. Hence we have $(p^n-1)p^{n-1}$ pairs. Next, we consider the case of zero vector $\vec \alpha$. Then any vector $\vec{\beta}$ works -- there are $p^n$ such possibilities. We thus have ${\mathcal N}(p)=(p^n-1)p^{n-1}+p^n$.

The  group $O(n,n,\Z)$, more precisely its image inside $O(n,n,\F_p)$, acts transitively on codes. 
This follows from the fact that any code can be brought to the canonical form with the generator matrix $({\mathcal I},B)$ \cite{Angelinos:2022umf}. Therefore the space of codes can be described as a coset \cite{Dymarsky:2020qom,Aharony:2023zit}
\begin{equation}
{O(n,n,\Z)\over \Gamma_0(p)}\,,
\end{equation}
where by $\Gamma_0(p)$ we denote the following subgroup of $O(n,n,\Z)$ consisting of matrices 
\begin{eqnarray}
\left(\begin{array}{cc}
A & B\\
C & D\end{array}\right) \in O(n,n,\Z)
\end{eqnarray}
with $C=0$ (mod $p$). This subgroup is a stabilizer of the particular code consisting of codewords $(\vec{\alpha},0)$ with arbitrary $\vec{\alpha}\in \F_p^n$. 

Furthermore, since $p$ is prime,  the  group $O(n,n,\Z)$ acts transitively on non-zero even codewords. 
Therefore each non-zero codeword appears in exactly 
\begin{equation}
\label{K}
K(n)= \prod_{j=0}^{n-2}(p^j+1)={N(n)(p^n-1)\over {\mathcal N}(n)-1}
\end{equation}
codes. To see that, we note that  ${\mathcal N}(n)-1$ is the number of non-zero even codewords, and  $(p^n-1)$ is the number of non-zero codewords in each code. 

The transitive action of $O(n,n,\Z)$ on codes means they should enter the average with equal coefficients. Therefore averaged full-weight enumerator at genus $g=1$, which get contribution from two orbits of even codewords -- a zero codeword, and all others --  has the following form 
\begin{equation}
\overline{W}=\psi_{ 0}+{K(n)\over N(n)} \sum_{c\neq { 0}}\psi_c=\psi_{ 0}+{1\over p^{n-1}+1} \sum_{c\neq  0}\psi_c. \label{sumovereven}
\end{equation}
Here we use ${ 0}$  to denote the zero element inside ${\mathscr D}$.
A sum here is over all non-zero even codewords. It can be represented as a sum over all $p^{2n}$ vectors $c=(\vec{\alpha},\vec{\beta})\in {\mathscr D}=\F_p^{2n}$ subject to the constraint \eqref{evenness}, which can be imposed by an additional sum, 
\begin{equation}
\label{g=1W}
\overline{W}={ \psi_{0}+p^{-n}\sum_{c\in {\mathscr D}} \sum_{r=0}^{p-1} \psi_{c}\, \, e^{2\pi i\,  {\vec{\alpha}\cdot \vec{\beta}\over p} r}\over 1+p^{1-n}},\quad c= (\vec{\alpha},\vec{\beta}).
\end{equation}

To find genus $g=2$ full-weight enumerator we would need to classify  orbits of unordered pairs of even codewords $\{c,c'\}$ under the action of $O(n,n,\Z)$.
One orbit consists of an element $\{{ 0}, 0\}$. Another two orbits include pairs of the form $\{{ 0},c\}$ and $\{c,{ 0}\}$ with all possible even non-zero $c$. Finally there are pairs of even non-zero vectors $c,c'\neq { 0}$, which split into different orbits based on the value of the scalar product $(c,c')$ (mod $p$), which is, by definition, $O(n,n,\Z)$-invariant. 
Even codes only include codewords with vanishing (mod $p$) mutual scalar product. Each even non-zero $c$ has vanishing scalar product with 
\begin{equation}
    P(n)=p^{2n-2}+p^n-p^{n-1}-1
\end{equation} 
even non-zero codewords $c'$ (including itself). 
The space of all  pairs $\{c,c'\}$ of even non-zero codewords $c,c'\neq  0$, satisfying $(c,c')=0$ (mod $p$), which consists of 
\begin{equation}
{({\mathcal N}(n)-1)P(n)}
\end{equation}
pairs, splits into two orbits. First includes pairs of collinear vectors, i.e.~$\{c,c'\}$ with $c'=k\,c\, \, {\rm mod}\, \, p$ for some $k$. There are 
\begin{equation}
({\mathcal N}(n)-1)(p-1)=(p^n-1)(p^{n-1}+1)(p-1)
\end{equation}
such pairs. 
All other 
\begin{equation}
({\mathcal N}(n)-1)(P(n)-p+1)
\end{equation}
pairs belong to the second orbit. 

A genus-2 full enumerator of any given even self-dual code includes $(p^n-1)(p^n-p)$ non-collinear pairs of even codewords. 
Therefore averaged full enumerator $\overline{W}$ includes the sum over  $\psi_{c,c'}$ with $\{c,c'\}$ being all possible pairs of non-collinear mutually orthogonal non-zero even vectors with the coefficient 
\begin{equation}
{(p^n-1)(p^n-p)\over ({\mathcal N}(n)-1)(P(n)-p+1)}={p^3\over (p^n+p)(p^n+p^2)}.
\end{equation}
Similarly the coefficient of $\Psi_{c,c'}$ for collinear $c,c'$ (when both vectors are non-zero) should be 
\begin{equation}
{(p^n-1)(p-1)\over ({\mathcal N}(n)-1)(p-1)}={1\over p^{n-1}+1}. \label{cf-1}
\end{equation}
If we  instead sum  $\psi_{r_1 c,r_2 c}$ for all even non-zero $c$ and arbitrary $0\leq r_1,r_2<p$, we should additionally divide by $(p-1)$ because each non-zero $c_1=r_1 c$ will appear $p-1$ times. 
Let us also note that in the sum $\psi_{r_1 c,r_2 c}$  over $c, r_1, r_2$, the number of terms $\psi_{ 0,c}$ and $\psi_{c, 0}$ will be $(p-1)$ for each non-zero even $c$. The number of such terms in each $\overline W$ should be $p^n-1$, hence coefficient \eqref{cf-1} is correct for terms $\Psi_{{ 0},c}$ and $\Psi_{c, 0}$ as well. We therefore find
\begin{align}
\overline{W}&={p-p^n\over p-1}\psi_{{ 0, 0}}+{1\over (p^{n-1}+1)(p-1)}\sum_{c\neq 0}\sum_{r_1,r_2=0}^{p-1} \psi_{r_1 c,r_2 c}+\\ &\quad + {{p^3\over (p^n+p)(p^n+p^2)}}\sum_{c,c'\neq 0}\sum_{r=0}^{p-1} \psi_{c,c'} {e^{2\pi i r(c,c')/p}\over p} \nonumber
\end{align}
where coefficient in front of $\Psi_{ 0, 0}$
\begin{eqnarray}
1-{{\mathcal N}(n)-1\over (p^{n-1}+1)(p-1)}={p-p^n\over p-1},
\end{eqnarray}
is chosen such that the total coefficient of $\psi_{{ 0,0}}$ is one. In the expression above the sum goes  over all non-zero even vectors $c,c'$.  The same expression can be rewritten as follows, 
\begin{align}
\nonumber
 \overline{ W}&= {{-p^{3n}\over (p-1)(p^n+p)(p^n+p^2)}} \psi_{ 0,0} +
 {{p^{n}\over (p-1)(p^n+p)(p^n+p^2)}} \sum_{\vec\alpha,\vec\beta,r,r_1,r_2}  \psi_{r_1(\vec\alpha,\vec \beta),r_2(\vec\alpha,\vec\beta)}\, e^{2\pi i {\vec{\alpha}\cdot \vec{\beta}\over p}r}\\&\quad + 
 {{1\over (p^n+p)(p^n+p^2)}} \sum_{\vec\alpha,\vec\beta,\vec\alpha',\vec\beta',r,r',l} \psi_{(\vec\alpha,\vec \beta),(\vec\alpha',\vec\beta')}  e^{2\pi i {\vec{\alpha}\cdot \vec{\beta}\over p}r} e^{2\pi i {\vec{\alpha}'\cdot \vec\beta'\over p}r'}\,  e^{2\pi i l(\vec{\alpha}\cdot \vec{\beta'}+\vec{\alpha'}\cdot \vec{\beta})/p}.\qquad \qquad 
\label{W2}
\end{align}
Here the sums over $r,r_i,l$ are from $0$ to $p-1$, and the sums over $\vec\alpha$, $\vec \beta$ etc are over $\F_p^n$.

As a consistency check, we can consider genus reduction $g=2 \rightarrow g=1$. After substituting  $\psi_{c,c'} \rightarrow \psi_c \, \delta_{c', 0}$ into \eqref{W2} it reduces to \eqref{g=1W}.

Using computer algebra we have checked for small values of $n$ and $p$ that \eqref{W2} is the Poincar\'e series of the vacuum character $\psi_{ 0,0}$.

\section{Gap of the spectrum of $U(1)$ primaries}
\label{app:gap}

For doubly-even self-dual codes over $\F_2$ it has been long known \cite{Pless1975} that the enumerator polynomial, averaged over all codes of length $n$, takes the form
\begin{align}
    \overline{W} = \frac{1}{2^{n/2} + 4} \left( 2^{n/2} (x^n + y^n) + (x + y)^n + (x + i y)^n + (x - y)^n +  (x - i y)^n \right).
\end{align}
For a single code, the enumerator polynomial is defined as $W_C = \sum_{c \in C} x^{n - w(c)} y^{w(c)}$ with $w(c)$ the Hamming weight, or $L^1$ norm, of the codeword $c$. 

These codes define chiral CFTs with $SU(2)^n$ extended symmetry algebras. When this enumerator polynomial is mapped to a code partition function, $x$ and $y$ are the (chiral) characters for $SU(2)_1$ global symmetry, with $x$ corresponding to the vacuum and $y$ corresponding to the excited state. In \cite{Henriksson:2022dml}, it was observed that this formula is a ``Poincar\'e series'' obtained by summing over modular images of the vacuum character $x^n$. At large $n$, a gap in these $SU(2)$ primaries appears, and was shown to be 
\begin{align}
    \Delta_\text{gap} \ = \ \frac{h_2^{-1}(1/2) n}{2} \ \simeq \ \frac{n}{18.177} 
\end{align}
where $h_2^{-1}$ is the inverse of the \textit{entropy function}
\begin{align}
    h_2 = x \log_2 \frac{1}{x} + (1-x) \log_2 \frac{1}{1-x} \, .
\end{align}

In the present case of non-chiral code CFTs over $\F_p \times \F_p$, the terms in the genus-1 Poincar\'e series are $I, S, ST,  \ldots , ST^{p-1}$, acting on the vacuum block $\psi_0$, giving the expression in~\eqref{eq:genus-1-average}
\begin{align}
    \label{eq:phases}
    \overline{W_\mathcal{C}} \ = \ \frac{1}{1+p^{1-n}} \left(  \Psi_0 + p^{-n}\sum_{c \in {\mathscr D}} \Psi_{c} +   p^{-n}\sum_{c \in {\mathscr D}} e^{\frac{\pi i \eta(c, c)}{p}} \Psi_{c} +  \ldots   + p^{-n} \sum_{c \in {\mathscr D}} e^{\frac{(p-1) \pi i \eta(c, c)}{p}}\Psi_{c} \right) .
\end{align}
This sum is over all vectors in $\mathscr{D}$, rather than codewords in some particular code. From the form of the sum we can see that only words $c$ with a $\eta(c, c) = 2 \alpha \cdot \beta = 0 \text{ (mod $2p$)}$ will contribute due to cancellation of phases, so we find the simplified expression
\begin{align}
        \overline  W_{\mathcal C} \ = \ \frac{1}{1+p^{1-n}} \left( \Psi_0 +   p^{1-n} \sum_{c \in {\mathscr D}'} \Psi_{c} \right) 
\end{align}
where ${\mathscr D}'$ is the set of words with $\eta(c, c) \in 2 p \Z $.

We would like to understand how many words there are in ${\mathscr D}'$ with a given weight $\Delta$. We assume that at large $n$, the phases $\eta(c,c)$ in equation \eqref{eq:phases} are randomly distributed, so that the number of codewords in ${\mathscr D}'$ with weight $\Delta$ is approximately equal to the total number of codewords with weight $\Delta$, divided by $p$.

Suppose that our codeword has $n_0$ zeros, $n_1$ ones, and so on. Then the number of such codewords is  
\begin{align}
    N(n_1,  \ldots , n_{p-1}) = \frac{(2n)!}{ n_0! n_1! n_2!  \ldots  n_{p-1}!} .
\end{align} 
The degeneracy of such a codeword appearing in the average is $\rho := N(n_1,  \ldots , n_{p-1}) / p^n$
For asymptotically large $n$, this will be small unless all $n_i$ scale with $n$. Assuming this -- and defining $n_i = 2 n a_i$ -- and using Stirling's approximation, we find
\begin{align}
    \rho \sim \exp \left( {2n \left(  - \frac{1}{2} \log p - a_0 \log a_0-  a_1 \log a_1 - \ldots -   a_{p-1} \log a_{p-1}  \right)} \right).
\end{align}
We can write this much more compactly as
\begin{align}
    \rho \sim \exp \left( 2 n \left( h_p - \frac{1}{2} \log p \right) \right),
\end{align}
where $h_p = - \sum_i a_i \log a_i $ is the \textit{Shannon entropy} of the random variables $\{ a_i \}$. It is clear that this will go to zero if the factor multiplying $2n$ in the exponent is negative, and it will be large if that factor is positive. Thus a state will have a positive degeneracy in the large $n$ limit whenever $h_p > 1/2 \log p$.

A codeword $c = (\alpha, \beta)$ defines a block $\Psi_c$ comprised of an infinite number of $U(1)^{2n}$ conformal blocks with weights (under the rigid embedding)
\begin{align}
    \label{eq:largecdim}
    \Delta_c = \frac{\alpha^2 + \beta^2}{2p} \, + (\alpha \cdot n) + (\beta \cdot m) + \frac{n^2 + m^2}{2} p,
\end{align}
with $n, m \in \Z^n$. First we note that that vacuum $(c = 0)$ block has excited $U(1)$ blocks with minimal weight $p / 2$, so we will need to take $p$ large enough to exclude these from being the lightest excited blocks. 

Each codeword has a $U(1)^{2n}$ block with minimal dimension, and these are not necessarily those with $n, m = 0$. E.g.~if $\alpha_i > p/2$ then the dimension in~\eqref{eq:largecdim} can be reduced by choosing $n_i = -1$.  The result is that the lightest $U(1)^{2c}$ primary state in the codeword block $\Psi_c$, where $c$ has $n_0$ zeros, $n_1$ ones, and so on, will have weight
\begin{align}
    \Delta(n_1,  \ldots , n_{p-1}) = \frac{n_1 + 4 n_2 + 9 n_3 +  \ldots + 4 n_{p-2} +  n_{p-1}}{2p }
\end{align}
We want to minimize the weight $\Delta$ for a fixed entropy -- or equivalently, to maximize the entropy for a fixed weight. From statistical mechanics, we know that the distribution that maximizes the Shannon entropy for a given energy is a Boltzmann distribution, 
\begin{align}
    a_k = C e^{-\mu E_k} \, ,
\end{align}
with $E_k = \min(k^2, (k-p)^2 ) $. The coefficients $C$ and $\mu$ are fixed using
\begin{align}
    \sum_{k = 0}^{p-1} a_k &= 1 \, , \qquad \sum_{k = 0}^{p-1} a_k E_k =  p \delta \, ,
    \end{align}
where $\delta = \Delta / n$. For general $p$, there is no simple expression for the constants $C$ and $\mu$, and they must be determined numerically. However for $p \to \infty$, the sums can be computed directly in terms of $\theta_3$ and its derivative, or by turning the sum into an integral. Large $p$ also corresponds to small $\mu$, leading to 
\begin{align}
    \frac{C \sqrt \pi}{2 \sqrt \mu} = \frac{1}{2} \, , \qquad \frac{C\sqrt \pi}{4  \mu^{3/2}} =  \frac{p \delta}{2}\,.
\end{align}
Therefore we find $\mu = (2p\delta)^{-1}$ and $c = (2 \pi p \delta)^{-1/2}$. The Shannon entropy in this state is given by 
\begin{align}
    h_p \ &= \ - \log C +  \mu p \delta \\
    \ &= \ \frac{1}{2} \log (2 \pi p \delta) + \frac{1}{2}\,.
\end{align}
To have a non-zero degeneracy in the large-$n$ limit, we find then that 
\begin{align}
    \log( 2 \pi e   \delta) > 0 \,.
\end{align}
This tells us that the weight where we find non-zero degeneracy  must scale as $p$, justifying our small-$\mu$ assumption, and in the end we find
\begin{align}
    \label{eq:gap}
    \Delta_\text{gap} = \frac{n }{2 \pi e}. 
\end{align}
This requires that $p/2 > \Delta_\text{gap}$, as we would otherwise find that the gap is saturated by the other $U(1)$ blocks appearing in $\Psi_0$. Equation~\eqref{eq:gap} is precisely the gap in $U(1)^{2n}$ characters found in the average over the Narain ensemble \cite{Afkhami-Jeddi:2020ezh, Maloney:2020nni}. In \cite{Aharony:2023zit} it was shown that in the limit of large $p$, the averaged code CFT torus partition function becomes that of $U(1)$ gravity while the number of codes goes to infinity. In this limit, the  code CFTs were conjectured to become dense in the Narain moduli space.
This provides a link between the code CFT ensemble and the ensemble of all  Narain theories, allowing $U(1)$ gravity to emerge as a limit of bulk CS theory  with a compact gauge group. This is precisely what we have observed here for the gap in $U(1)$ primaries. It is also worth noting that  the  gap of $U(1)$ primaries in the large-$p$ limit was previously considered in 
\cite{Angelinos:2022umf}. They considered the same class of codes but averaged with different weights, and also found the gap to be $n / (2 \pi e)$.

\let\oldbibliography\thebibliography
\renewcommand{\thebibliography}[1]{%
  \oldbibliography{#1}%
  \setlength{\itemsep}{0.5pt}%
}

\bibliographystyle{JHEP}
\bibliography{biblio}

\end{document}